\documentclass[structabstract]{aa}  

\usepackage{tabularx}
\usepackage{graphicx}
\usepackage{my_natbib}
\usepackage{txfonts}
\usepackage{longtable}

\newcommand{\hersc}{{\it Herschel}}
\newcommand{\lab}{{\it LABOCA}}
\newcommand{\APEX}{{\it APEX}}
\newcommand{\spitz}{{\it  Spitzer}}
\newcommand{\iso}{{\it ISO}}
\newcommand{\iras}{{\it IRAS}}
\newcommand{\jcmt}{{\it JCMT}}
\newcommand{\akari}{{\it AKARI}}
\newcommand{\twomass}{{\it 2MASS}}

\newcommand{\lsun}{$L_\odot$}
\newcommand{\msun}{$M_\odot$}
\newcommand{\zsun}{$Z_\odot$}
\newcommand{\mic}{$\mu$m}

\renewcommand*{\figurename}{ Fig.}
\renewcommand*{\tablename}{Table}
\newlength{\pointwidth}
\def\revisedbis{}

\begin{document}

  \title{Probing the Dust Properties of Galaxies up to Submillimetre Wavelengths }
  \subtitle{I. The Spectral Energy Distribution of dwarf galaxies using LABOCA}
  
  \author{Maud Galametz 
  	       \inst{1}~,
	       Suzanne Madden 
	       \inst{1}~,
	       Fr{\'e}d{\'e}ric Galliano
	       \inst{1}~,
	       Sacha Hony
	       \inst{1}~,
	       Fr{\'e}d{\'e}ric Schuller 
	       \inst{2}~,
	       Alexandre Beelen
	       \inst{3}~,
	       George Bendo
	       \inst{4}~,
	       Marc Sauvage
	       \inst{1}~,
	       Andreas Lundgren
	       \inst{5}~
	       \and
	       Nicolas Billot
	      \inst{6}}

	      \institute{Laboratoire AIM, CEA, Universit\'{e} Paris Diderot, IRFU/Service d'Astrophysique, Bat. 709, 91191 Gif-sur-Yvette, France,  \\
              \email{maud.galametz@cea.fr} 
              \and
             Max-Planck-Institut f\"ur Radioastronomie, Bonn, Germany 
              \and
              Institut d'Astrophysique Spatiale, Orsay, France
              \and
              Imperial College, London, UK
              \and
              ESO, Santiago, Chili
               \and
               Caltech, Pasadena, USA
             }


 \abstract
 {}
 {We are studying the dust properties of four low metallicity galaxies by modelling their spectral energy distributions. This modelling enables us to constrain the dust properties such as the mass, the temperature or the composition to characterise the global ISM properties in dwarf galaxies. }
 {We present 870 \mic\ images of four low metallicity galaxies (NGC~1705, Haro~11, Mrk~1089 and UM~311) observed with the Large \APEX\ BOlometer CAmera (LABOCA) on the Atacama Pathfinder EXperiment (\APEX) telescope. We model their spectral energy distributions combining the submm observations of \lab, \twomass, \iras, \spitz\ photometric data and the IRS data for Haro~11.}
  {We find that the PAH mass abundance is very low in these galaxies, 5 to 50 times lower than the PAH mass fraction of our Galaxy. We also find that a significant mass of dust is revealed when using submm constraints compared to that measured with only mid-IR to far-IR observations extending only to 160 \mic. For NGC~1705 and Haro~11, an excess in submillimeter wavelengths is detected when we use our standard dust SED model. We rerun our SED procedure adding a cold dust component (10 K) to better describe the high 870 \mic\ flux derived from LABOCA observations, which significantly improves the fit. We find that at least 70 $\%$ of the dust mass of these two galaxies can reside in a cold dust component. We also show that the subsequent dust-to-gas mass ratios, considering HI and CO observations, can be strikingly high for Haro~11 in comparison with what is usually expected for these low-metallicity environments. Furthermore,  we derive the star formation rate of our galaxies and compare them to the Schmidt law. Haro~11 falls anomalously far from the Schmidt relation. These results may suggest that a reservoir of hidden gas could be present in molecular form not traced by the current CO observations. While there can be a significant cold dust mass found in Haro~11, the SED peaks at exceptionally short wavelengths (36 \mic), also highlighting the importance of the much warmer dust component heated by the massive star clusters in Haro~11. We also derive the total IR luminosities derived from our models and compare them with relations that derive this luminosity from \spitz\ bands. We find that the Draine $\&$ Li (2007) formula compares well to our direct IR determinations. }
  {}


     \authorrunning{M. Galametz et al}
     \titlerunning{The SED of dwarf galaxies using LABOCA}

 \maketitle


\section{Introduction}

The understanding of the evolution of a galaxy requires knowledge of the roles of the different actors controlling the evolution of the Interstellar Medium (ISM) and the subsequent feedback on star formation activity. Despite its low fraction of the total mass of a galaxy (less than 1$\%$), dust plays a prominent role in the heating and cooling of the ISM and thus tightly influences the overall physics of a galaxy. Since dust absorbs the stellar radiation and reemits it in a wide range of wavelengths, the star formation rate (SFR) as well as other fundamental parameters of a galaxy, such as its age, can be indirectly studied through the dust emission itself. {\revisedbis The Spectral Energy Distribution (SED) of a galaxy is its spectral footprint from which we can study the physical processes taking place in the galaxy since it synthesises the contribution of all its components to the emission of the galaxy. Using this tool, we can peer into the window of the integrated history of the galaxy and disentangle the various physical actors (stars, HII regions, molecular clouds) and processes (stellar radiation, dust emission) involved~\citep[][see also $\S$ 5 of this paper]{Draine2007,Galliano_Dwek_Chanial_2008}.}

While dust hinders the interpretation of ultraviolet (UV) and optical wavelengths, in the Mid Infrared (MIR), Far Infrared (FIR) and submillimetre (submm) wavelengths, dust emission and absorption properties expose different physical environments, from the most vigorous star formation and AGN activity \citep[e.g.][]{Gordon1995, Wu2007} to the more quiescent diffuse media \citep{Bernard1996,Arendt1998}. Many processes linked to star formation such as stellar winds \citep{Hoefner2009}, supernovae shocks, photodestruction by high-mass stars etc. can also affect the spatial distribution and the local properties and abundance of the different dust components of a galaxy such as Polycyclic Aromatic Hydrocarbons \citep[PAHs,][]{OHalloran2006}, amorphous carbon grains, silicates or composite grains, manifesting themselves in the MIR to submm wavelengths. 

Studying the interplay between galaxy properties and metal enrichment is crucial to understand galaxy evolution.
The metallicity of a galaxy is deeply linked with the dust properties of the ISM and its substructures such as HII regions and molecular clouds, but just how it affects the ISM is currently poorly known. Dwarf galaxies in the Local Universe, are metal-poor galaxies, and are thus convenient laboratories to study the effects of metallicity (Z) on the gas and dust. They exhibit a wide variety of physical conditions, and their star formation properties and ISM represent the closest analogs to proto-galaxies of the early universe. Indeed, dwarf galaxies are small and may compared to high redshift galaxies which also present lower metallicities \citep{Lara_Lopez_2009}. They are also considered to be the building blocks of much larger and more metal-rich galaxies \citep[Review by][]{Tosi2003}. They also show analogies with Gamma Ray Bursts (GRB) hosts whose ISM usually exhibit moderate chemical enrichment with a median metallicity of 1/10 \zsun \citep{Chen2009}. 

They finally show evidence for older stellar populations than their metallicity suggests \citep[e.g.][]{Aloisi1998}, posing enigmatic issues for galaxy evolution models. Many studies have been carried out to grasp this apparent paradox. \citet{Lisenfeld_Ferrara_1998}  confirmed that the dependence of the dust-to-gas mass ratio (D/G) in low metallicity galaxies was a function of metallicity using \iras\ observations.  \citet{James2002}, \citet{Walter2007} and \citet{Hirashita2008} concluded likewise using \jcmt/SCUBA submm, \spitz\ MIR/FIR and \akari\ (FIR)  observations. Finally, \citet{Galliano_Dwek_Chanial_2008} observed some systematic deviations between dust abundances of very low metallicity systems and what is expected for supernova-condensed dust. 
At MIR wavelengths, low metallicity systems also show prominent differences in the dust properties compared to the more metal-rich systems. For example, PAH features are strikingly diminished as metallicities drop \citep[e.g.][]{Madden2005, Engelbracht2005, Wu2007, Engelbracht2008} compared to metal-rich galaxies, in spite of the role the smallest grains play in the energy balance of galaxies \citep{Rubin2009}.  Some studies suggest that PAH emission depends on the hardness or strengh of the illuminating radiation field  \citep{Madden2006, Engelbracht2008, Gordon2008, Bendo2008}. The consequence of lowering the metallicity of a galaxy is the decrease in dust opacity resulting in harder and stronger radiation fields. The dearth of PAHs in low metallicity galaxies has also been explained by the destructive effects of supernovae \citep{OHalloran2006,OHalloran2008} or by the delayed injection of PAHs by AGB stars \citep{Galliano_Dwek_Chanial_2008}.

Broad wavelength coverage of the MIR to submm regime is imperative to constrain the modelling of the observed SEDs, leading to a better comprehension of the dust properties of galaxies. Since \spitz\  only observes dust emission at wavelengths shorter than 160 \mic, submm data are necessary not only to enlarge the wavelength coverage at longer wavelengths to verify the dust models but also because the potential reservoir of cold grains ($\le$ 15K), which contribute to this submm flux, may account for a significant amount of mass. Only a handful of galaxies of the Local Universe have been studied using submm ground-based instruments (e.g. \jcmt/SCUBA). When submm observations of dwarf or late-type galaxies are studied, an excess in the dust SEDs is often found in the mm/submm domain \citep{ Lisenfeld2001, Bottner2003, Dumke2004,Galliano2003, Galliano2005, Marleau2006, Bendo2006}. This excess can be interpreted as very cold dust ($\le$10K), in which case more than 50$\%$ of the total dust mass of these galaxies should reside in a very cold component. {\revisedbis Cold dust is also needed to explain the break in the Gas to Dust mas ratio as a function of metallicity relaion for low-metallicity galaxies~\citep{Galliano_Dwek_Chanial_2008, Munoz2009}}. The presence of this cold dust component is still a contentious issue in the ISM community and will have important consequences on our comprehension of ISM properties of low metallicity environments. \citet{Lisenfeld2001}, \citet{Reach1995}, \citet{Dumke2004}, \citet{Bendo2006} or \citet{Meny2007} suggested that changes in dust emission properties (changes in dust emissivity or resonances related to dust impurities) should be responsible for boosting submm emission above the 15-20K thermal emission expected at these wavelengths. However, not all low metallicity galaxies show submm excess, as shown recently by the observations of the nearby Local Group Galaxy {\revisedbis IC10 {\revisedbis (Parkin et al. 2009 in prep)}, where the main} two star forming regions were isolated with ISO, \spitz\ and 850 \mic\ observations, the SEDs were modeled without invoking a very cold dust component. Moreover, \citet{Draine2007} showed that their observations of mostly metal-rich galaxies can largely be reproduced by dust models which do not account for a very cold dust component, even in their limited number of cases where submm observations are present. Studies using submm observations for a wider range of metallicity values are necessary to check the relevance of these conclusions for low metallicity environments.

In this paper, we present the first \APEX/ \lab\ 870 \mic\  observations of dwarf galaxies: 1 extended galaxy (NGC~1705) and 3 compact sources (Haro~11, UM~311, Mrk~1089) (see Table~\ref{Sample_properties}). We have combined these data with \spitz\ and/or \iras\ observations to produce global SEDs that we use to model the dust properties. The sample is small but covers a wide range of metallicities, from $\sim$1/9 \zsun\ for Haro~11 to $\sim$1/3 \zsun\  for NGC~1705. It also presents varied morphologies, size scales and characteristics: resolved or compact galaxies, disturbed and even merging environments. 

We describe the sample in $\S$ 2 and the observations and data reduction in $\S$ 3 and the images and photometry in $\S$ 4. In $\S$ 5, we present the SED modelling and discuss the results in $\S$ 6.


\section{The sample}


\begin{table*}
$$
\begin{tabular}{ccccccc}
\hline
\hline
Name & Ra (2000) & Dec (2000) & 12+log(O/H) & Distance (Mpc) & M(HI) & Apparent size \\
\hline
\noalign{\smallskip}
NGC~1705 & 04h 54' 13.5''  & -53$^{\circ}$ 21' 40'' & 8.46 $^{(a)}$ & 4.7 $^{(a)}$ & 5.1 $\times$ 10$^{7~(1)}$ & 1.9' $\times$ 1.4'\\
Haro~11 &  00h 36' 52.5'' & -33$^{\circ}$ 33' 19'' & 7.9 $^{(b)}$ & 92 $^{(c)}$ & $\sim$ 10$^{8~(2)}$ & 0.5' $\times$ 0.5' \\
Mrk~1089 &  05h 01' 37.8'' & -04$^{\circ}$ 15' 28'' & 8.0 $^{(d)}$ & 59.8 $^{(d)}$ & 2.7 $\times$ 10$^{10~(3)}$ & 0.61' $\times$ 0.23' \\
UM~311& 01h 15' 34''  & -00$^{\circ}$ 51' 32'' & 8.3 $^{(d)}$ & 21.3 $^{(d)}$ & 2.3 $\times$ 10$^{9~(4)}$ & 0.11' $\times$ 0.11' \\

\hline
\end{tabular}
$$
 \label{Sample_properties}
 \caption{General properties of the sample: a)~\citet{Meurer1992}, b)~\citet{Bergvall_Ostlin_2002}, c)~\citet{Bergvall2000}, d)~\citet{Hopkins2002}
\newline 
The HI masses were scaled according to the distances we used. References for HI dust mass: 1) 8.9 $\times$ 10$^7$ \msun\ for D=6.2 Mpc \citep{Meurer1998}, 2) Upper limit derived from the HI detection \citep{Bergvall_Ostlin_2002}, 3) 2.1 $\times$ 10$^{10}$ \msun\ for D=53 Mpc \citep{Williams1991}, 4) 3.1 $\times$ 10$^9$ \msun\ for D=24.7 Mpc \citep{Smoker2000}.}
 \end{table*} 
 
\begin{figure*}
    \centering
    $$
    \begin{tabular}{ c c c c}
    a) & NGC1705 (H$\alpha$) &  b) & NGC1705 (8 \mic\ + H$\alpha$ contours)\\
& \includegraphics[width=8.5cm ,height=6cm]{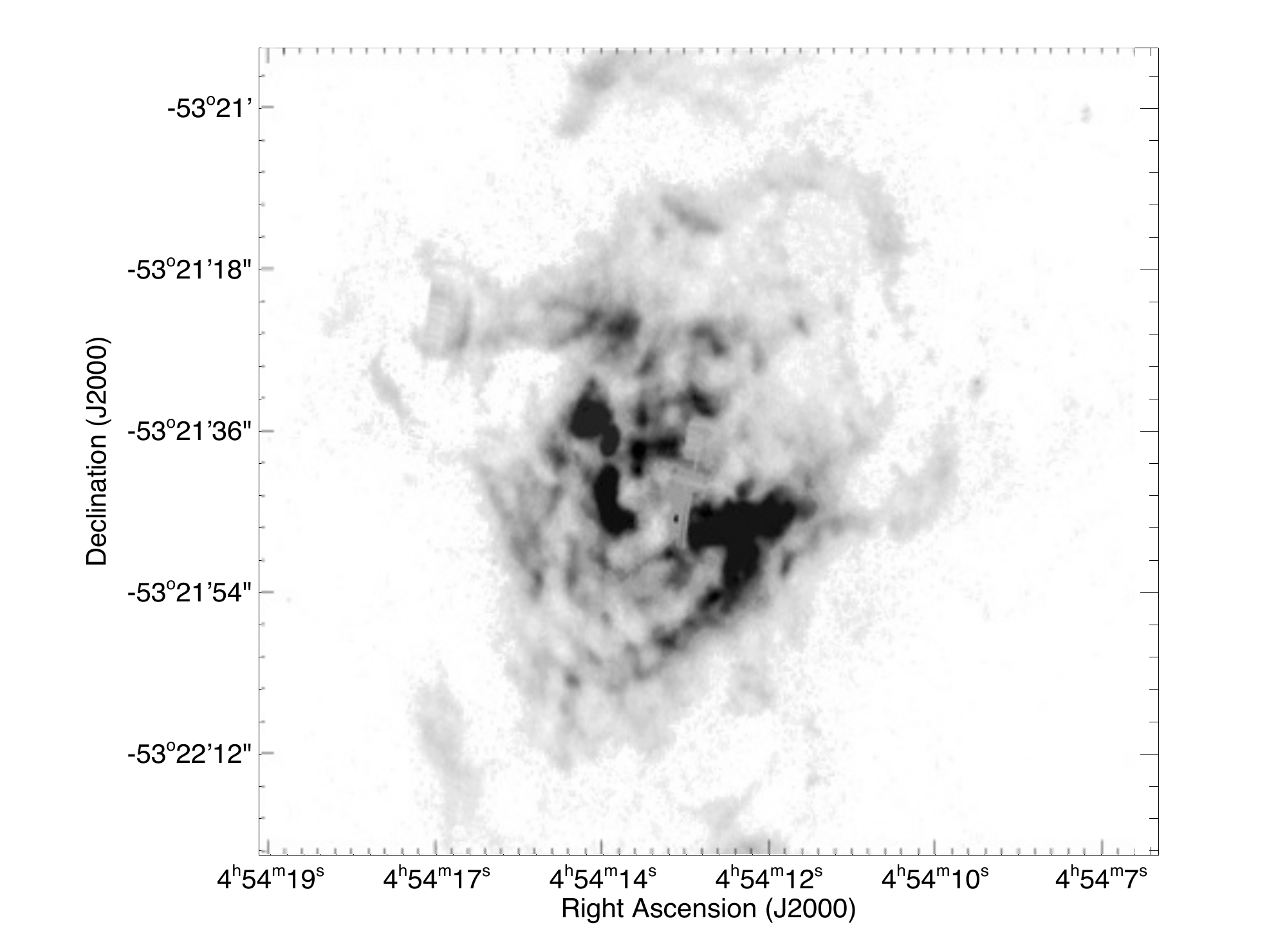} &&
      \includegraphics[width=8.5cm ,height=6cm]{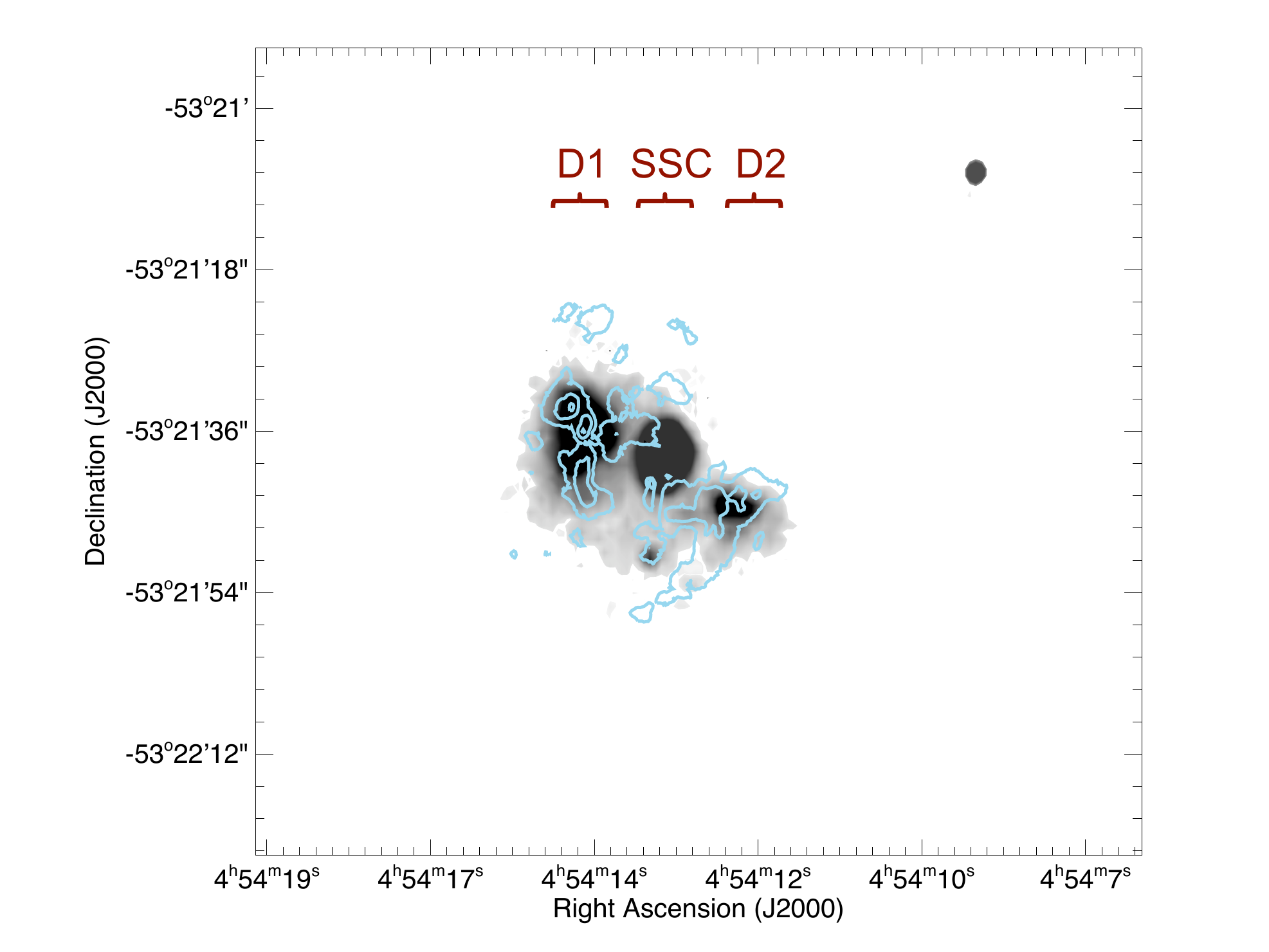} 	\\
          &&&\\
    c) & NGC1705 (HI) & d) & NGC1705 (8 \mic\ + HI contours) \\
      & \includegraphics[width=8.5cm ,height=6cm]{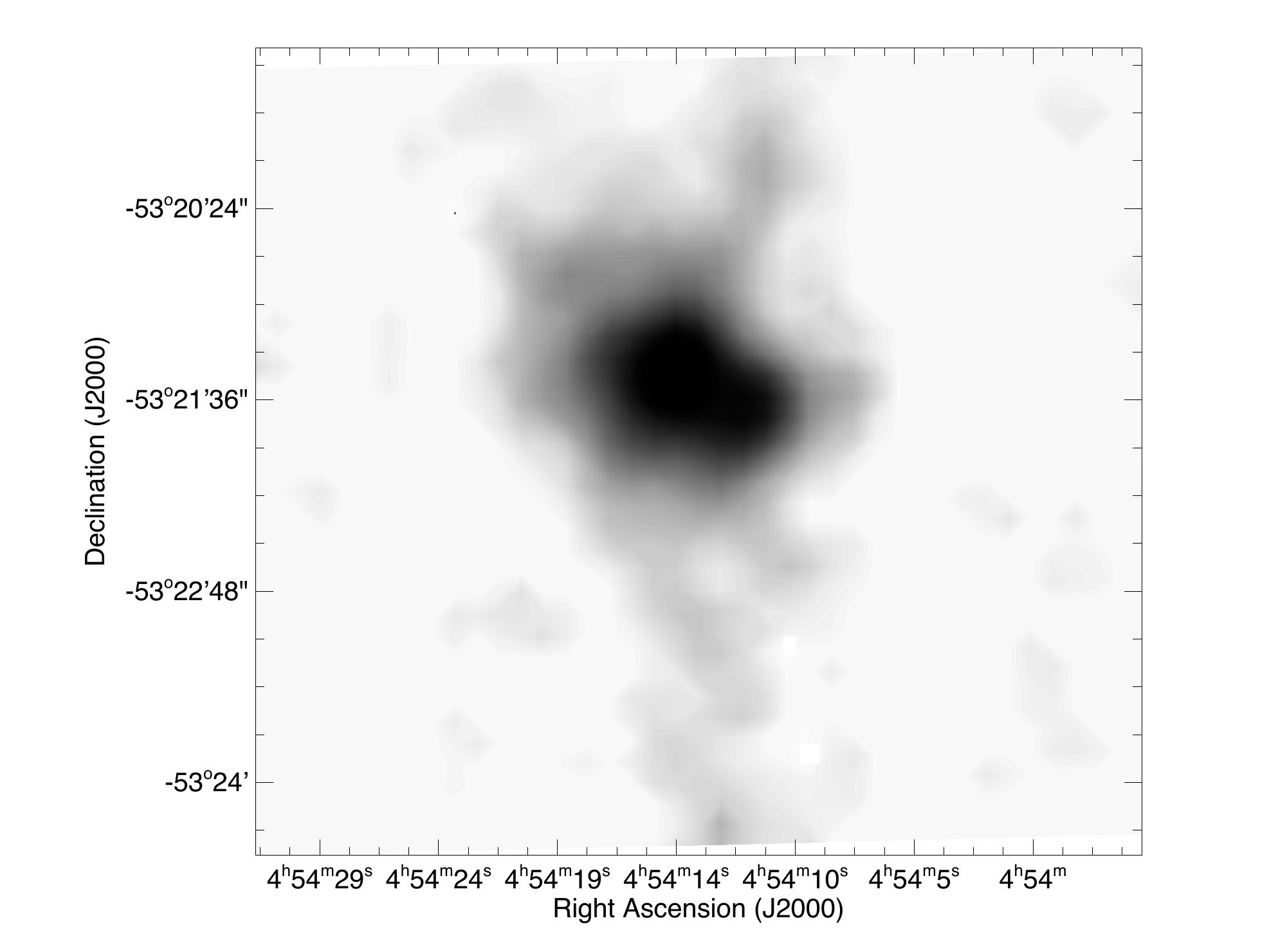} &&
      \includegraphics[width=8.5cm ,height=6cm]{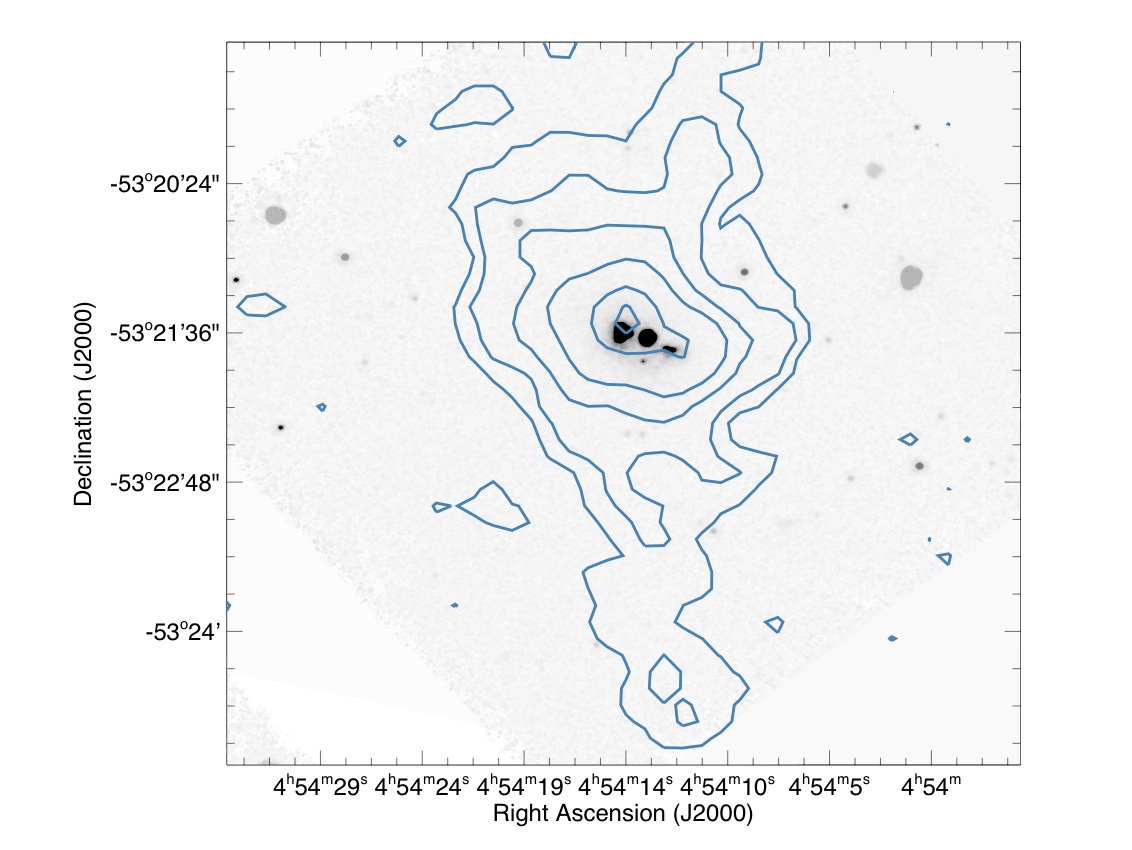} \\
         \end{tabular}
         $$
    \caption{{\it a)} H$\alpha$ image of NGC~1705. Note that the SSC is masked in the image \citep{DePaz2003}. {\it b)} NGC~1705 observed at 8 \mic\ with \spitz. H$\alpha$ contours are overlaid. Note the SSC and the two bright off-nuclear HII regions labeled D1 and D2 \citep{Cannon_NGC1705_2006}. {\it c)} HI image of NGC~1705 \citep{Meurer2006}. {\it d)} NGC~1705 observed at 8 \mic\ with \spitz. HI contours are overlaid. }
 \label{NGC1705_Halpha_HI}
\end{figure*}

 The four galaxies studied in this paper were chosen because of their diversity in morphology, distance, metallicity and star formation activity:

{\it NGC~1705 - } This is a well studied Local Group galaxy which is in the \spitz\ Infrared Nearby Galaxies Survey  \citep[SINGS;][]{Kennicutt2003}. \citet{Heckman_Leitherer_1997} found the luminosity of the galaxy in the UV to be dominated by a central bright 10$^{5}$\msun\  Super Star Cluster (SSC) also at the origin of a galactic outflow. This SSC shows similar properties to the distant gamma ray burst hosts \citep{Chen2007}.  The H$\alpha$ emission extends over the entire optical emission of the galaxy \citep{DePaz2003} while the HI emission is exceptionally extended  beyond the optical emission \citep{Meurer1998} and lie on either side of the SSC. Two off-nuclear regions called D1 and D2 (Fig.~\ref{NGC1705_Halpha_HI}) can be seen to dominate the MIR and FIR dust emission \citep{Cannon_NGC1705_2006}.  \spitz\ IRS spectroscopy reveals the PAH emission originating toward region D1 but not toward the SSC or region D2 \citep{Cannon_NGC1705_2006}. 

{\it Haro~11 - } Also known as ESO 350-IG038, this is the most distant galaxy of the sample ~\citep[92Mpc;][]{Bergvall2000}. It possesses characteristics of an extreme starburst  with L$_{FIR}$ $\sim$ 10$^{11}$ L$_{\odot}$  \citep{Sanders2003}, making it a luminous infrared galaxy (LIRG) with a high star formation rate of $\sim$ 25 M$_\odot$ yr$^{-1}$ as determined from H$\alpha$, radio continuum, FIR and hard X-ray observations \citep{Grimes2007}. {\revisedbis Broadband images of H$\alpha$ show three bright star-forming condensations} with unrelaxed kinematic structure and faint extended shell structures in the outer regions of the galaxy, all suggesting an ongoing merger \citep{Bergvall_Ostlin_2002, Ostlin1999}.  Haro~11 is a moderately strong radio source (essentially free-free continuum) with extended continuum emission at 6 and 20 cm \citep{Heisler_Vader_1995}. It is a very metal poor galaxy (Z$\sim$1/7 Z$_{\odot})$ that seems to have a very little neutral hydrogen, an unusually high ratio between blue luminosity and HI mass and little observed molecular gas  \citep{Bergvall2000}. 

{\it Mrk~1089 - } Mrk~1089 is a Wolf-Rayet (WR) galaxy \citep{Kunth_Schild_1986} and is the most luminous of the eight members of the Hickson Group 31 \citep[][]{Hickson1982}. The morphology of the group is very disturbed with tidal interactions between Mrk~1089 (HCG31 C) and the galaxy NGC~1741 (HCG31 A)(Fig.~\ref{LABOCA_images}c). The two galaxies present similar kinematics suggesting a single entity \citep{Richer2003}.  At the location of their interaction is a very strong 24 \mic\ source and high levels of infrared emission can be found throughout the whole group \citep{Johnson2007}. Mrk1089 and NGC1741 are referred to collectively as NGC 1741 in the catalog of WR galaxies of \citet{Conti1991}. Nevertherless, many surveys \citep{Rubin1990, Conti1996, Iglesias-Paramo_Vilchez_1997, Johnson2000} showed that the interacting system was undergoing a starburst which {\revisedbis was attributed to HCG31 C}. We thus decide, in this paper, to designate the interacting system compounded of HCG31 A and HCG31 C together as Mrk~1089. 

{\it UM~311 - } This compact HII galaxy \citep{Terlevich1991} is located between the pair of spiral galaxies NGC~450 and UGC~807, 2 galaxies which were once thought to be interacting but have now been demonstrated  to be physically separated \citep{Rubin_Ford_1983}. There are three very bright sources of compact HII emission between the two galaxies, UM~311 being the brightest (Fig.~\ref{LABOCA_images}d). The galaxy has been misidentified as a projected galactic star due to its quasi-stellar and compact morphology. Its H$\alpha$ luminosity and equivalent width are remarkably high for an HII galaxy \citep{Guseva1998}.

 
\section{Observations and data reduction}

\subsection{ LABOCA}

\lab\ is a multi-channel bolometer array for continuum observations at 870 \mic, built by MPIfR (Max-Planck-Institut fur Radioastronomie, Bonn, Germany) and mounted on \APEX\ (Atacama Pathfinder EXperiment), a 12-meter radio telescope of ESO, Onsala and MPIfR. The array consists of 295 channels. \lab\ on \APEX\ has a total field of view of 11.\hspace{-\pointwidth}$^{\prime}$4  $\times$ 11.\hspace{-\pointwidth}$^{\prime}$4 and a full width half maximum (FWHM) of its point spread function (PSF) $\sim$18.2''.

About 30 hours of observations were taken from November 9th 2007 to November 20th 2007. The four galaxies are smaller than the FOV of \lab. Basic spiral patterns with 4 pointings were combined with a raster mapping mode (raster-spiral) to completely fill the array (homogenous sampling of each map), to obtain enough off-source position for background substraction and to effectively execute long integration times (8.7h, 8.3h, 4h and 7.6h for Haro~11, UM~311, Mrk~1089 and  NGC~1705 respectively). {\revisedbis Flux calibration was performed} through the observations of the planets Uranus and Mars and the sources HLTAU, J0423-013, J0050-095, J0006-064, V883-ORI,  PMNJ0403-3605, PMNJ0106-4034, PKS0537-441, N2071IR. We used the BOA package (BOlometer Array Analysis Software; Schuller et al., in prep) to reduce the data. The sotware was developed through a collaboration of scientists from the MPIfR, AIfA (Argelander-Institut f{\"u}r Astronomie, Bonn, Germany), AIRUB (Astronomisches Institut der Ruhr-Universit{\"a}t, Bochum, Germany), and IAS (Institut d'Astrophysique Spatiale, Orsay, France). 

The atmospheric attenuation was determined via skydips every hour. Opacities at 638 \mic\ range from 0.103 to 0.353 throughout the observing run. 
Flat-fielding normalisation is applied by dividing the signals by the bolometre gains supplied by the observatory and thus removing the bolometre gain variations. From the 18th of October up to the 15th of November, 33 \lab\ pixels were shadowed by a plate in the beam. These pixels are removed. We also masked out stationary points and data taken outside reasonable telescope scanning velocity and acceleration limits as well as dead or noisy channels.
During the commissioning period of \lab, correlated noise was found between groups of channels sharing some parts of the electronics (amplifier boxes or cables). As our data are not all point sources, it is necessary to carefully differentiate correlated noise from extended emission. This correlated noise is subtracted from each map. Moreover, we performed a correction to suppress spikes and flatten the 1/f noise of the FFT to remove the noise caused by thermal variations. 
Finally, each reduced scan was gridded into a weighting map. These weights were built by calculating the rms of each time series which contributes to a given region of the map.

The major sources of uncertainty are the calibration uncertainty and uncertainty resulting from the background variation. We estimate the average uncertainty levels to be of $\sim$ 15 $\%$. All of the galaxies were detected with a root mean square (rms) estimated to be lower than 10 mJy/beam for three of our galaxies but as high as 15 mJy/beam for Mrk~1089. Some {\revisedbis anomalous} pixels at the edges of the final maps are the results of poor coverage at the edges of the map. The anomalous pixels are ignored in the analysis. The 870 \mic\ \lab\ images are presented in Fig.~\ref{LABOCA_images}.


\begin{figure*}
    \centering
    \begin{tabular}{c c c}
    
    a) & NGC~1705 (3.6 \mic) & NGC~1705 (8 \mic)\\
          
          & \includegraphics[width= 7.5cm ,height=5cm]{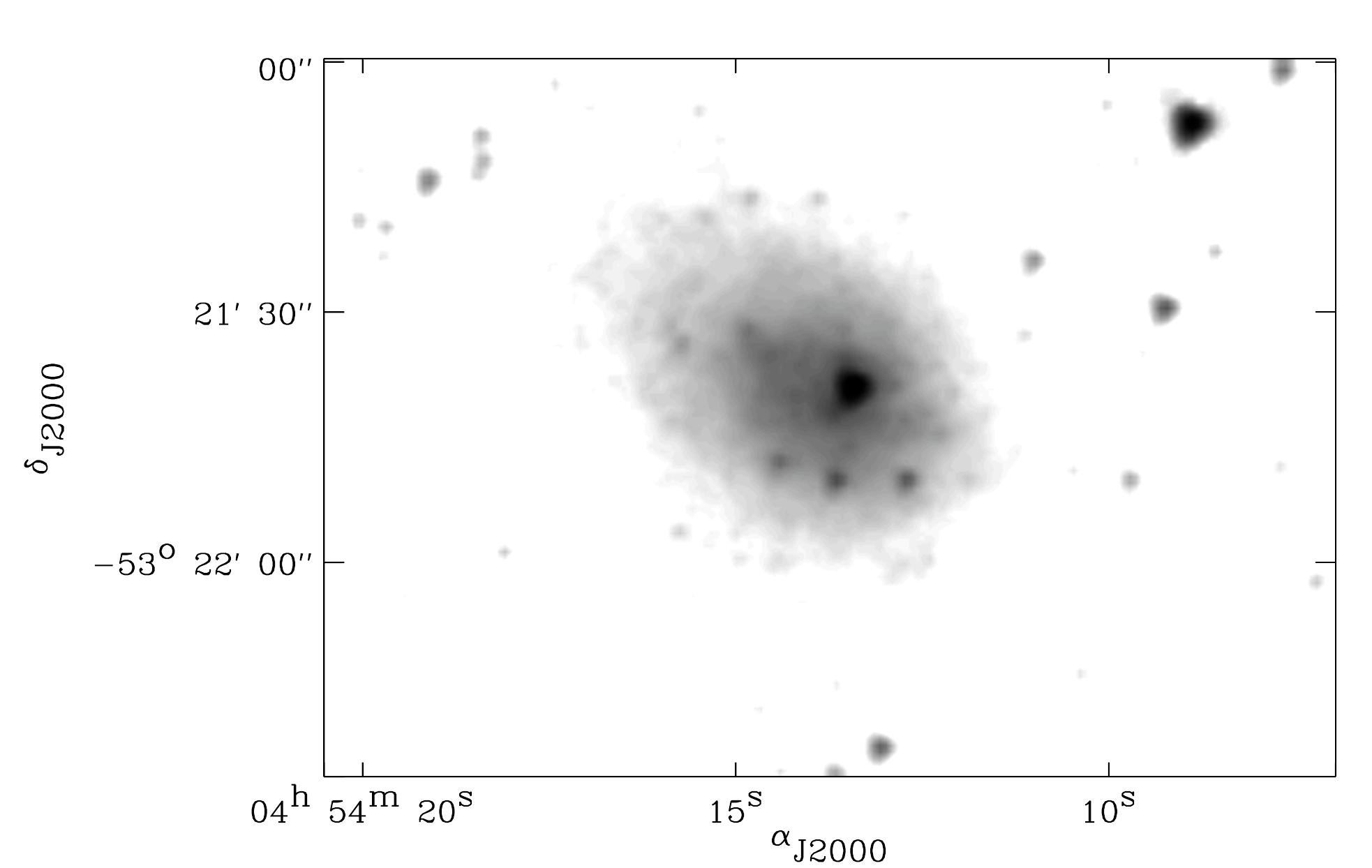} & \includegraphics[width= 7cm ,height=5cm]{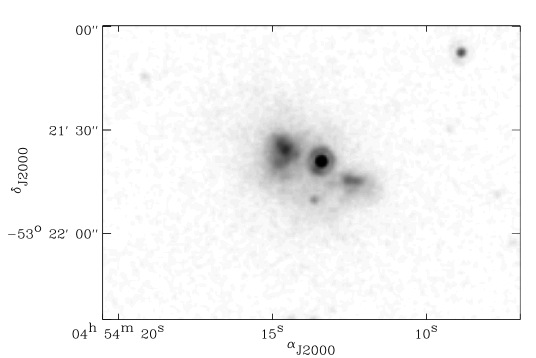}\\
           
&&\\   
&&\\  
           & NGC~1705 (24 \mic) & NGC~1705 (870 \mic)\\
          
          & \includegraphics[width= 7.5cm ,height=5cm]{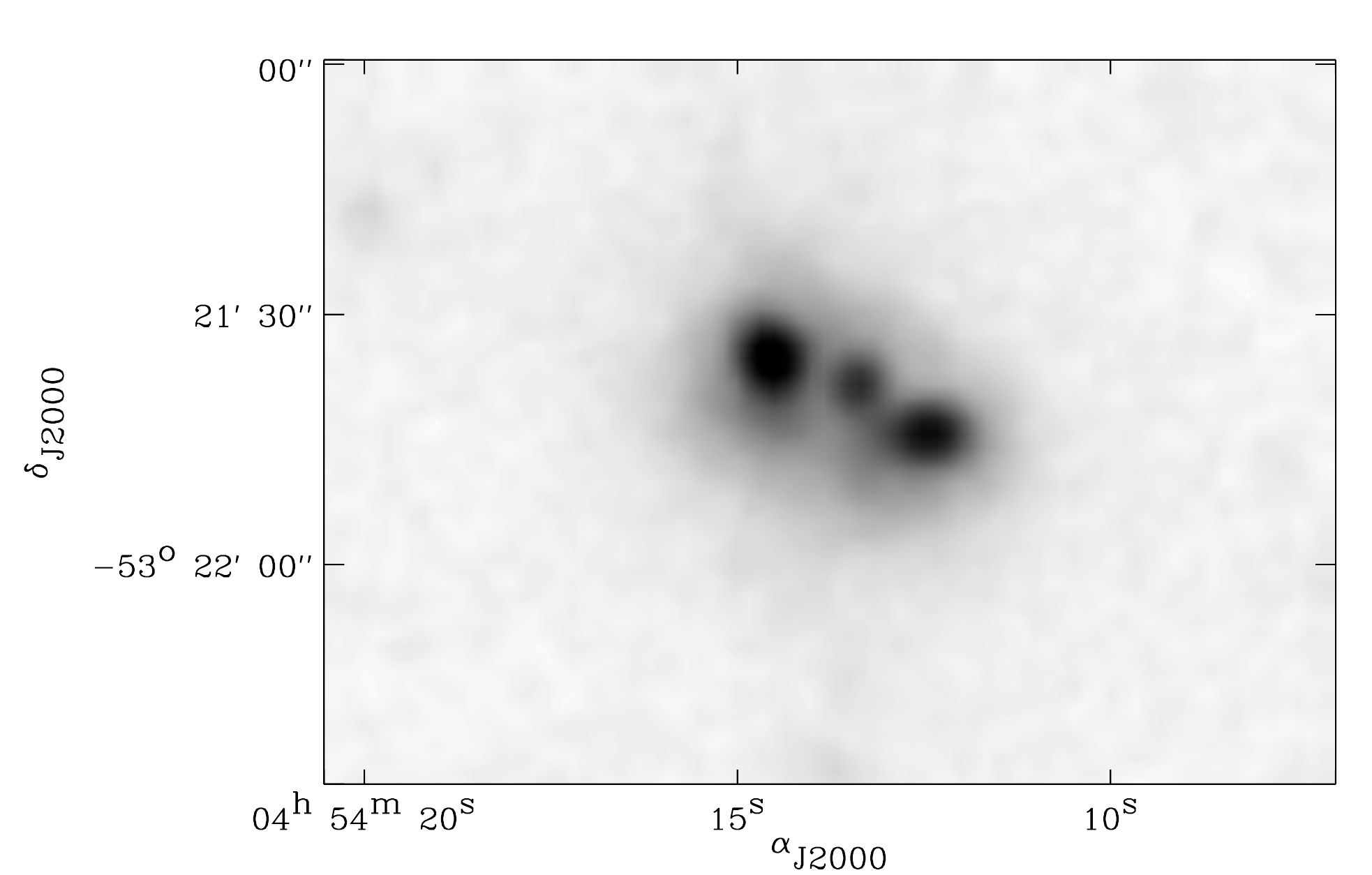} & \includegraphics[width= 7cm ,height=5cm]{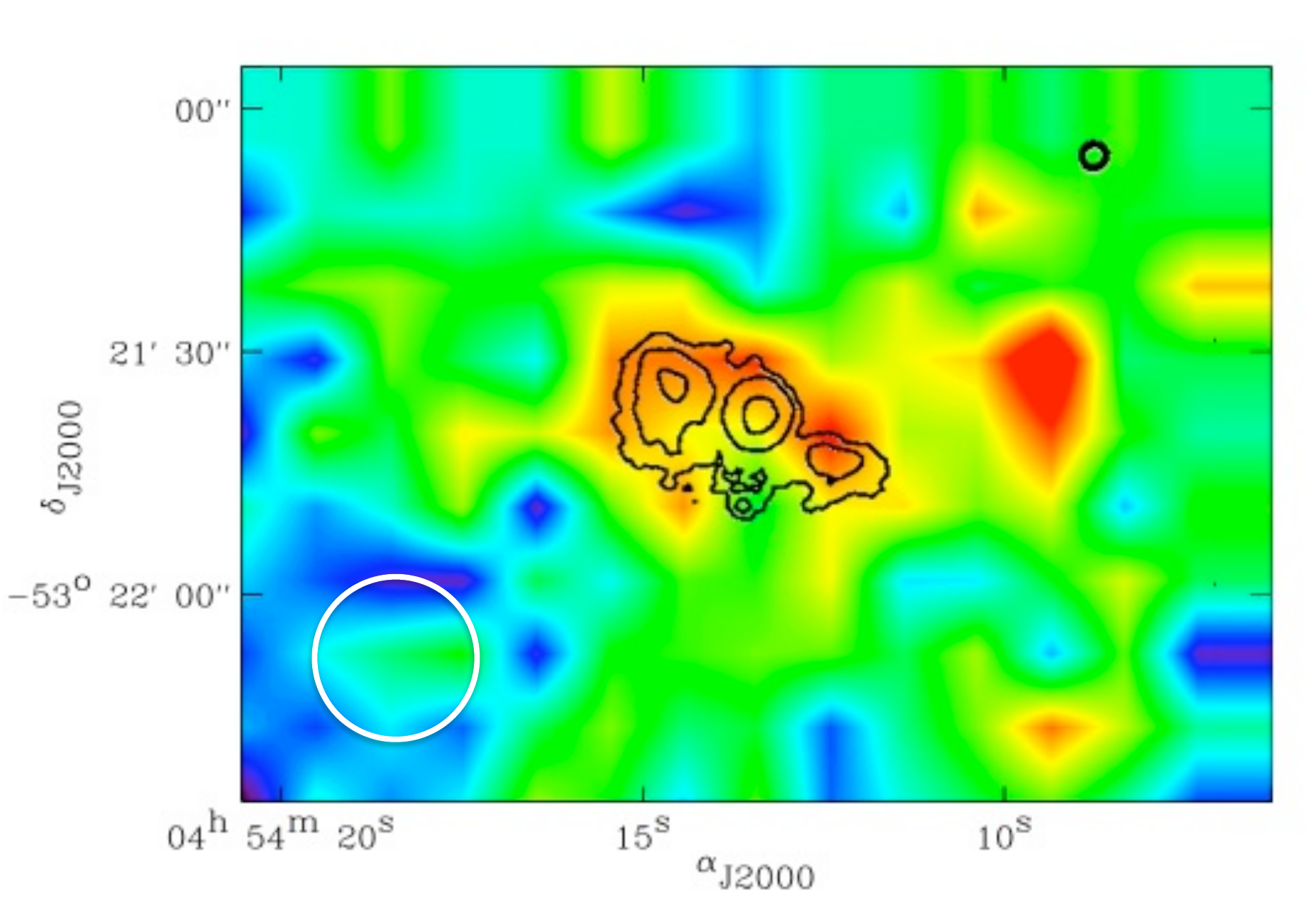}\\

           b) & Haro~11 (3.6 \mic) & Haro~11 (8 \mic)\\
           
           & \includegraphics[width= 7cm ,height=5cm]{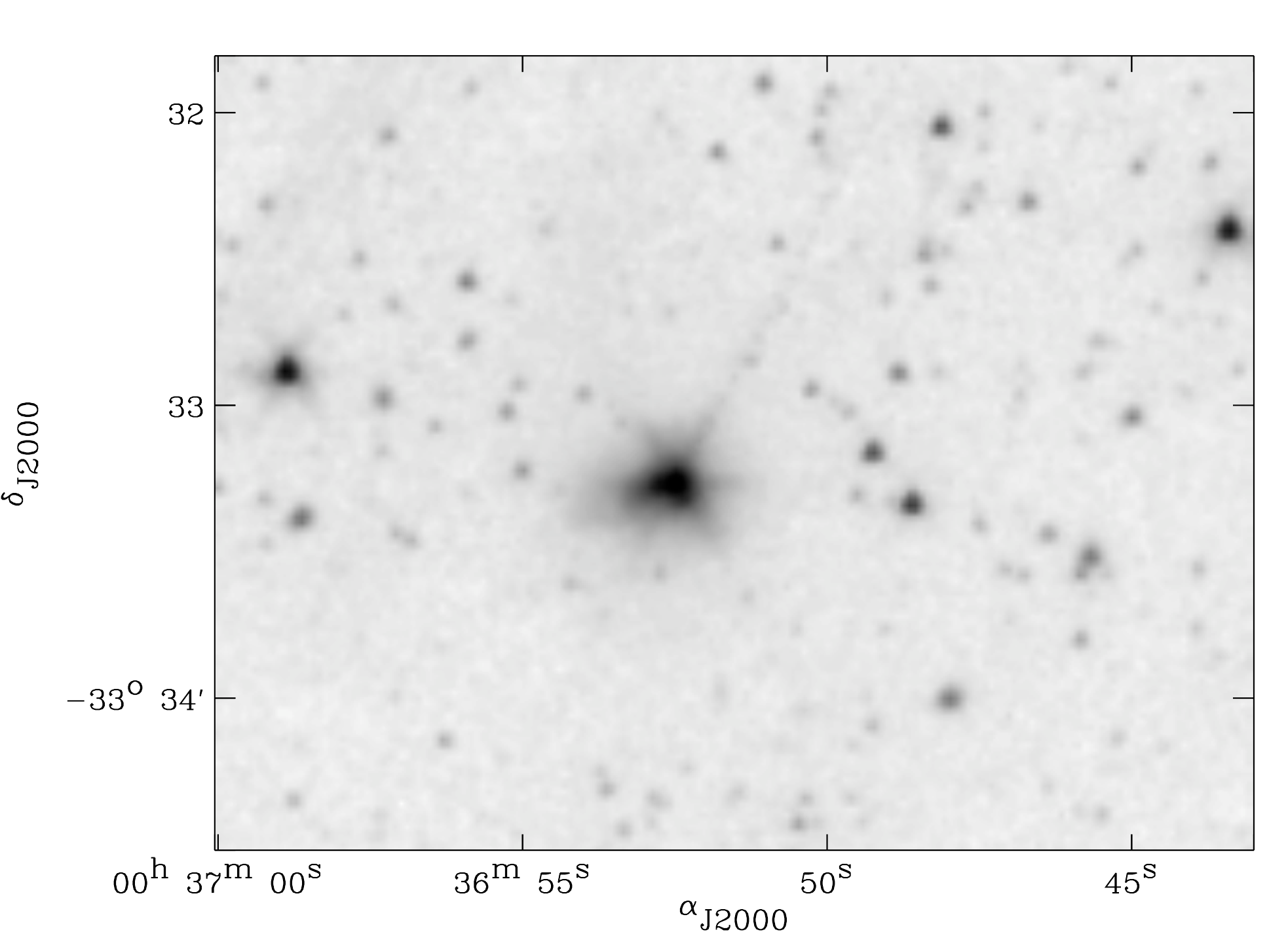} &  \includegraphics[width=7cm ,height=5cm]{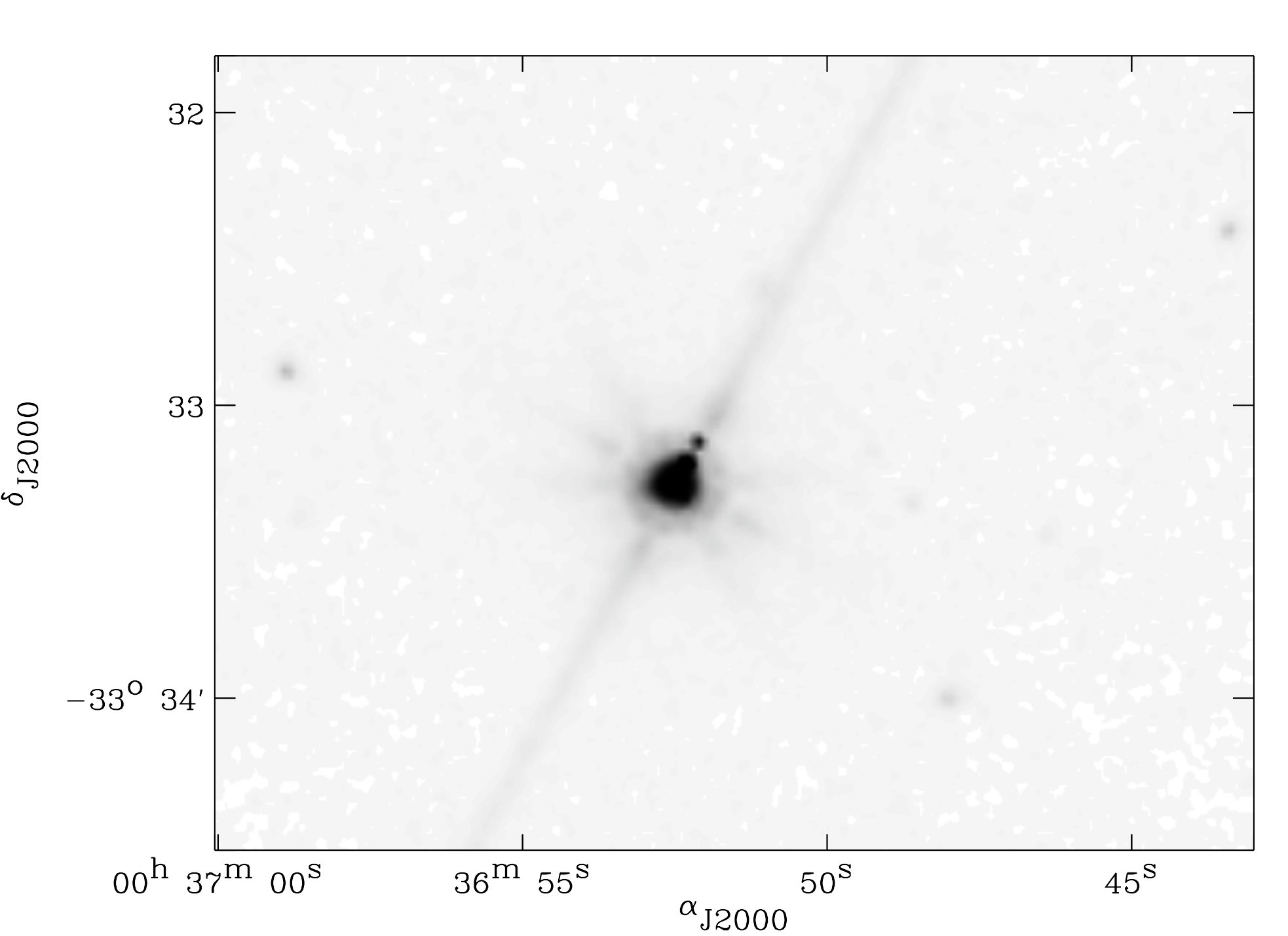} \\
&&\\
&&\\
            & Haro~11 (24 \mic) & Haro~11 (870 \mic)\\
            
           & \includegraphics[width= 7cm ,height=5cm]{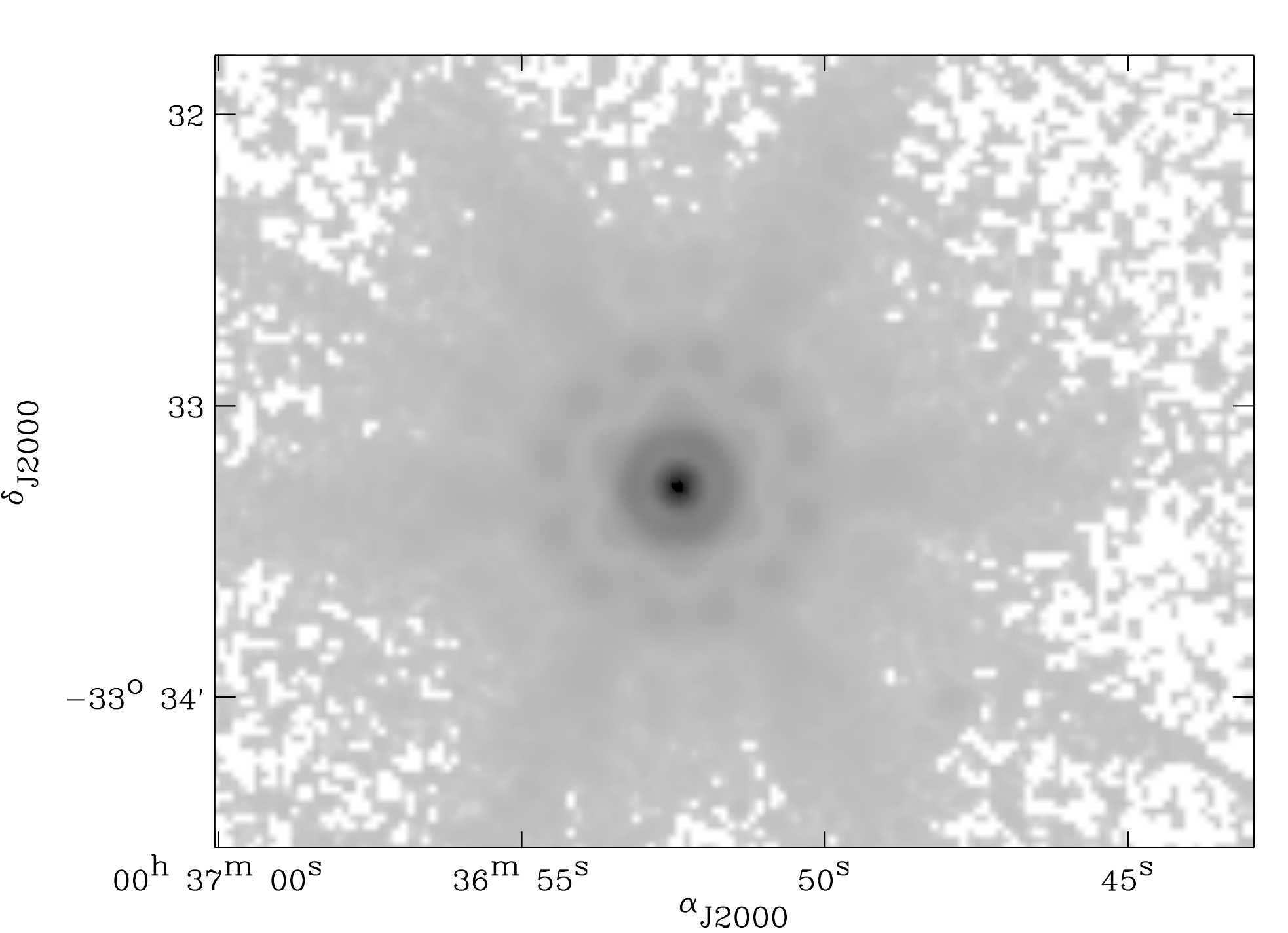} & \includegraphics[width=7cm ,height=5cm]{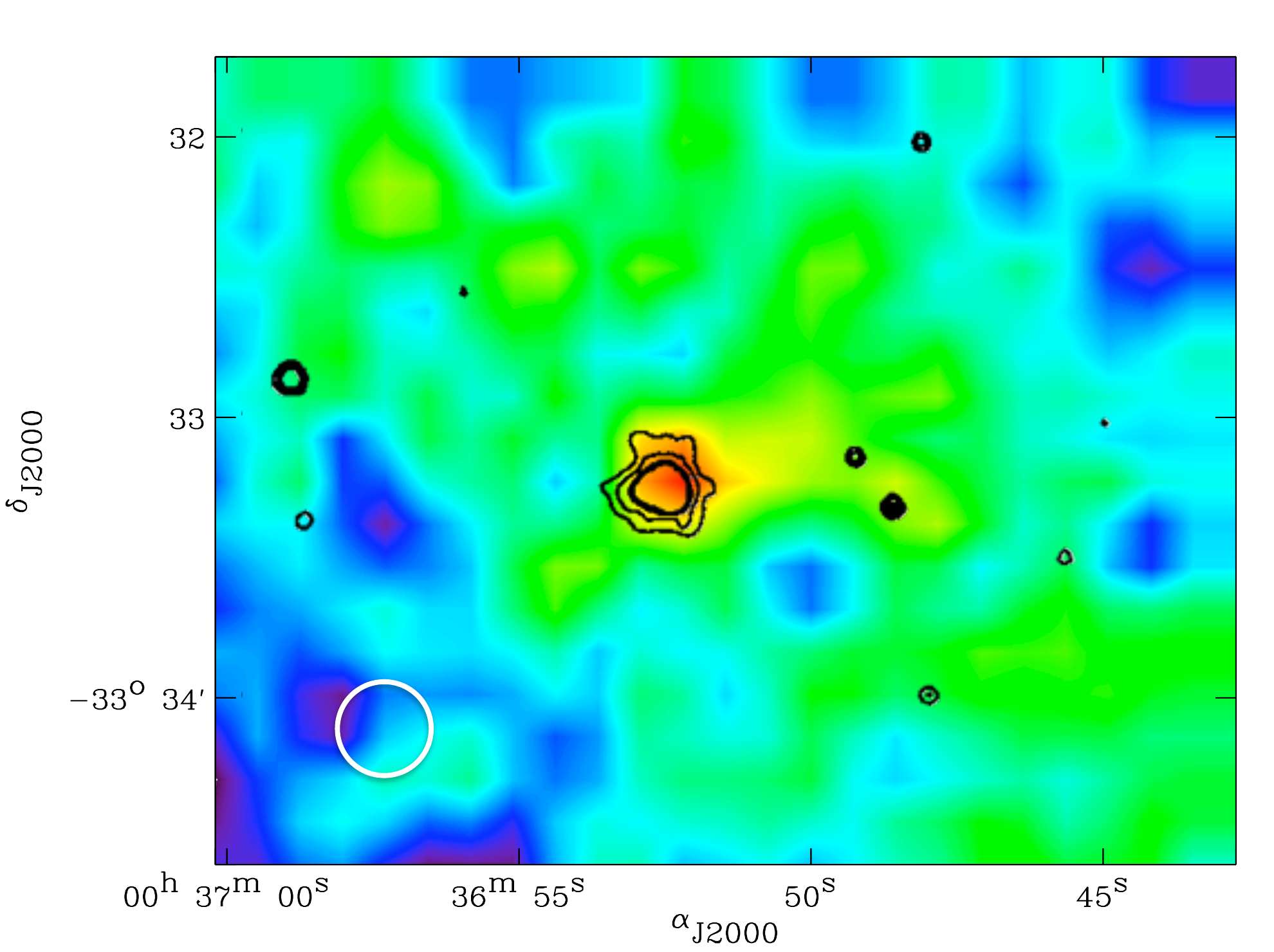}, \\
&&\\
&&\\
	      \end{tabular}
    \caption{ 3.6 \mic\ (4.5 for UM~311),  8 \mic\ , 24 \mic\  and 870 \mic\ images of our four galaxies: {\footnotesize a)} NGC~1705, {\footnotesize b)} Haro~11, {\footnotesize c)} Mrk~1089 (In the paper, the interacting system of NGC~1741 and Mrk~1089 is designated as Mrk~1089), {\footnotesize d)} UM~311 (The galaxy is identified as source number 3). North is upwards and East is to the left. The FWHM of the PSF of \lab\ is shown by a white circle.  8 \mic\ (4.5 \mic\ for Haro11) contours are overlaid on the \lab\ images: a) 0.5, 1, 2.5 MJy/sr - (b) 0.34, 0.77, 8.58 and 9 MJy/sr - c) 2, 2.5, 4.5, 8 MJy/sr - d) 4.6, 5.61, 7.8 MJy/sr. For UM~311, the photometric aperture chosen  (54'') encompasses the flux of the 3 sources at the edges of the two spiral galaxies and is marked by the dashed circle. These sources are numbered according to \citet{Moles1994}. For NGC~1705, three individual regions can be isolated: the SSC in the center and two off-nuclear HII regions on both sides of this SSC.   }
    \label{LABOCA_images}
\end{figure*}


\addtocounter {figure}{-1}
\begin{figure*}
    \centering
    \begin{tabular}{c c c}
      
          c) & Mrk~1089 (3.6 \mic) & Mrk~1089 (8 \mic)\\
	
	 & \includegraphics[width= 7.5cm ,height=5cm]{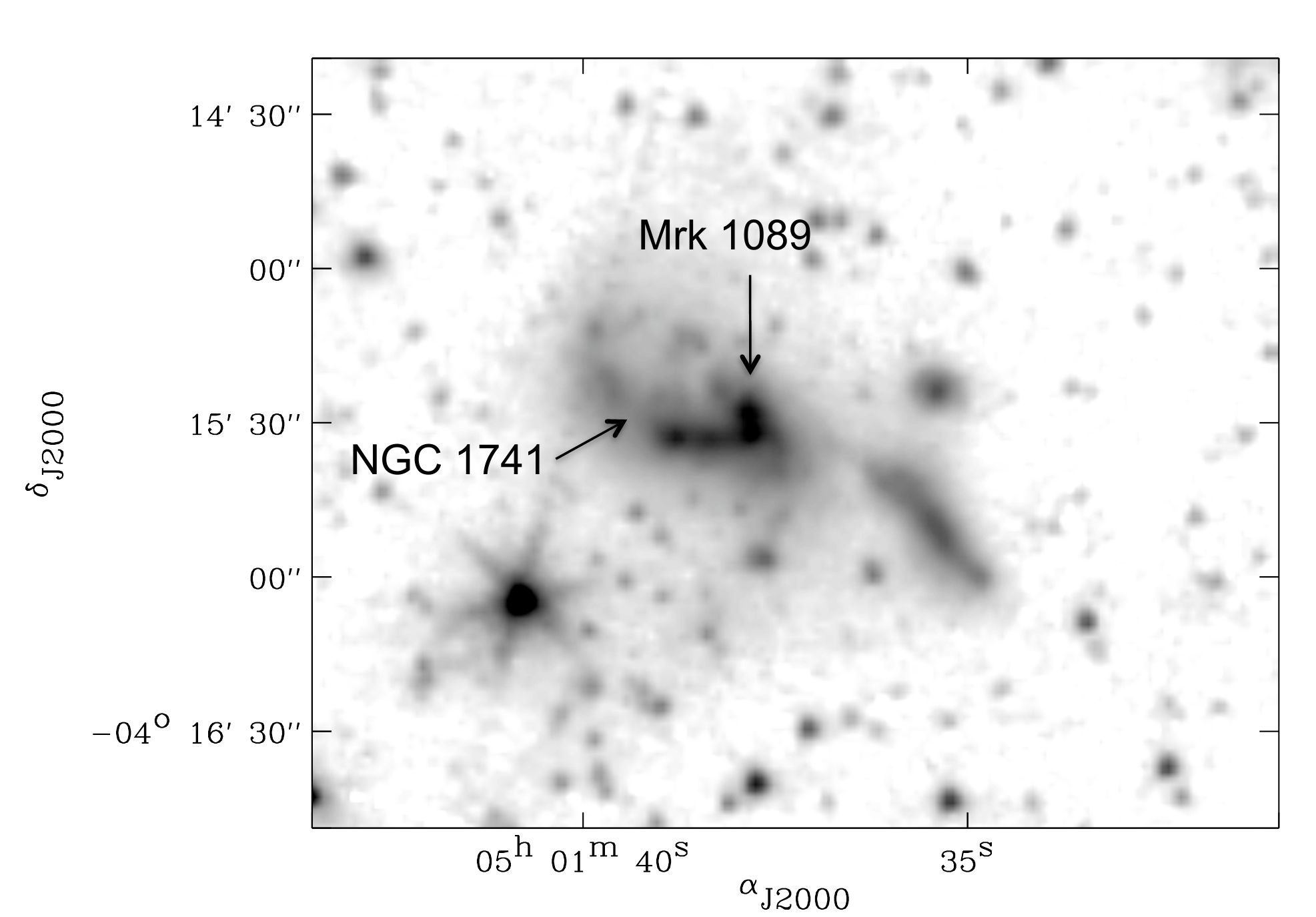} &
           \includegraphics[width= 7cm ,height=5cm]{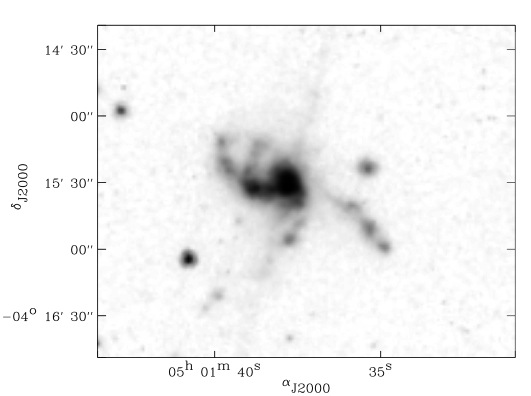} \\
&&\\   
&&\\      
           & Mrk~1089 (24 \mic) & Mrk~1089 (870 \mic)\\
	
	 & \includegraphics[width= 7.5cm ,height=5cm]{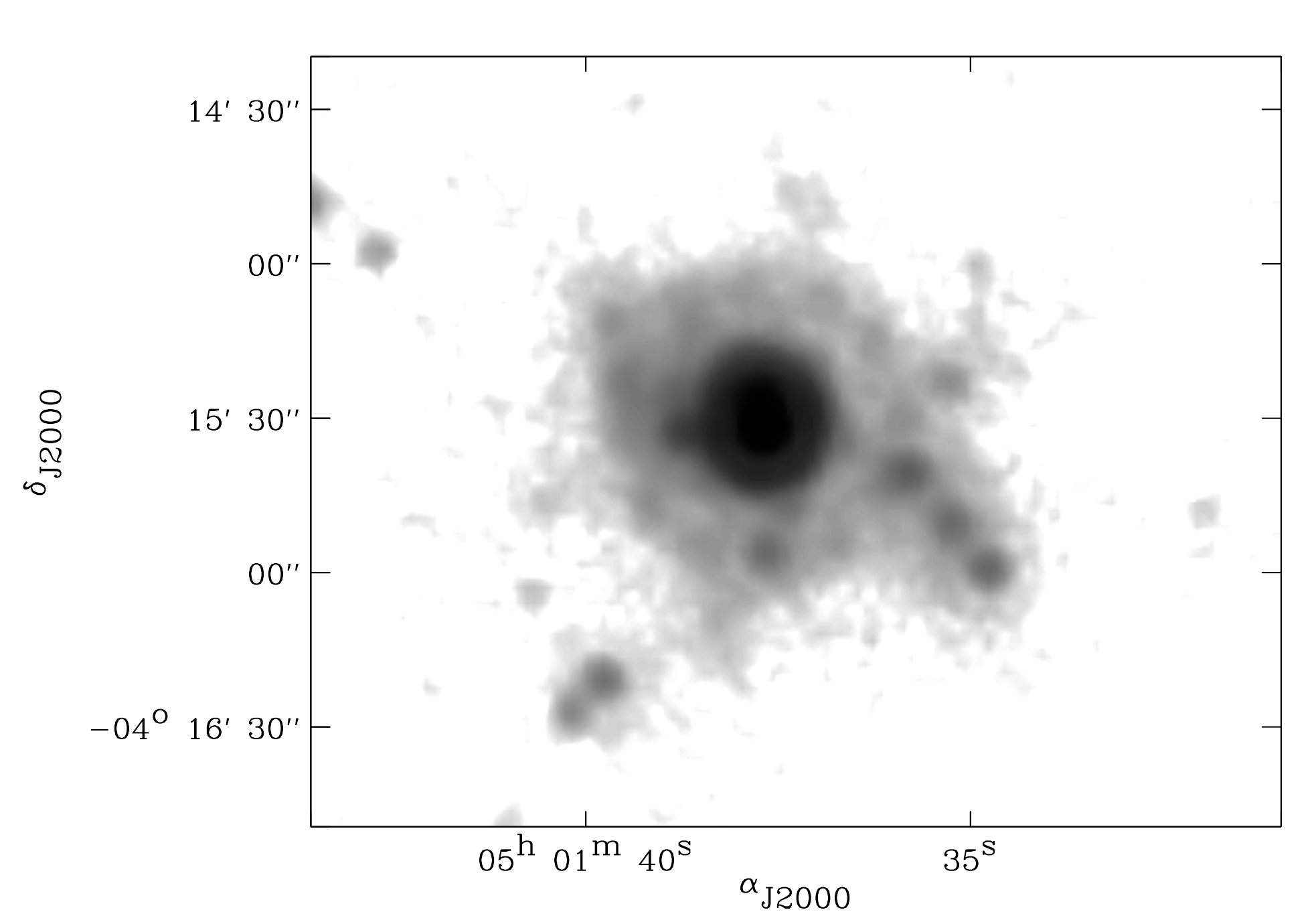} &
           \includegraphics[width= 7cm ,height=5cm]{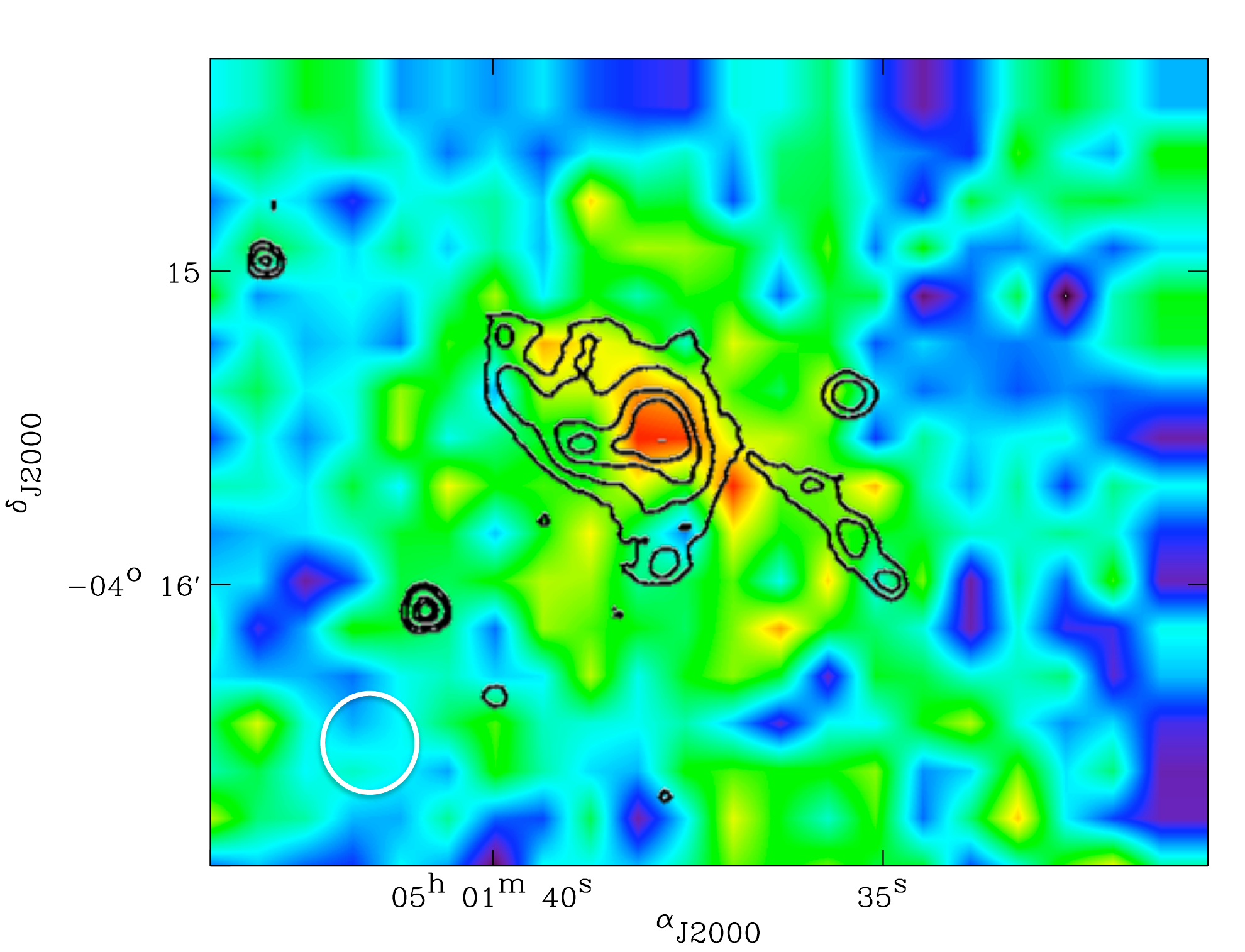} \\  
           
&&\\   
&&\\  
	d) & UM~311 (4.5 \mic) & UM~311 (8 \mic)\\

	 & \includegraphics[width= 7cm ,height=5cm]{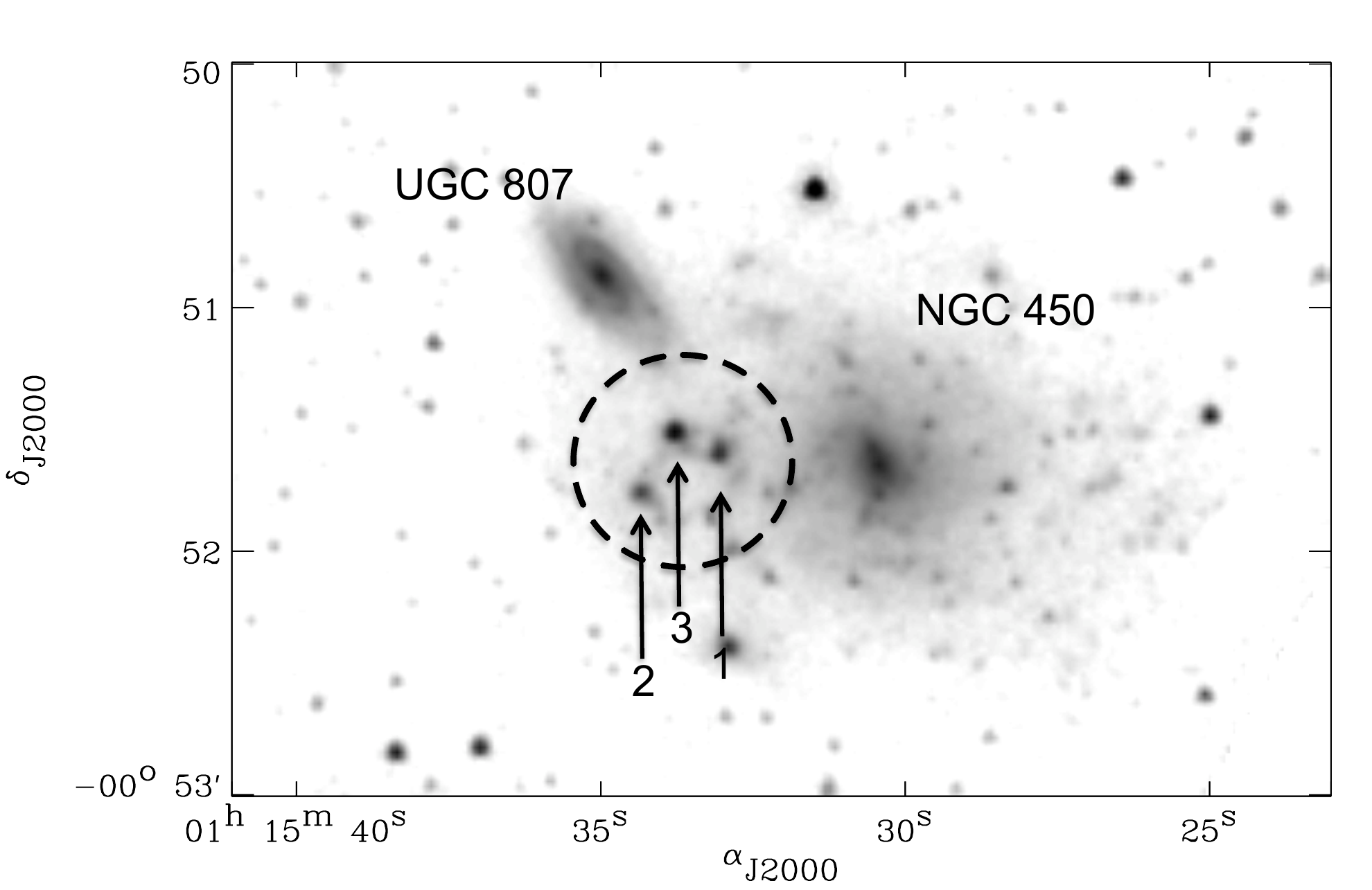} & \includegraphics[width= 7cm ,height=5cm]{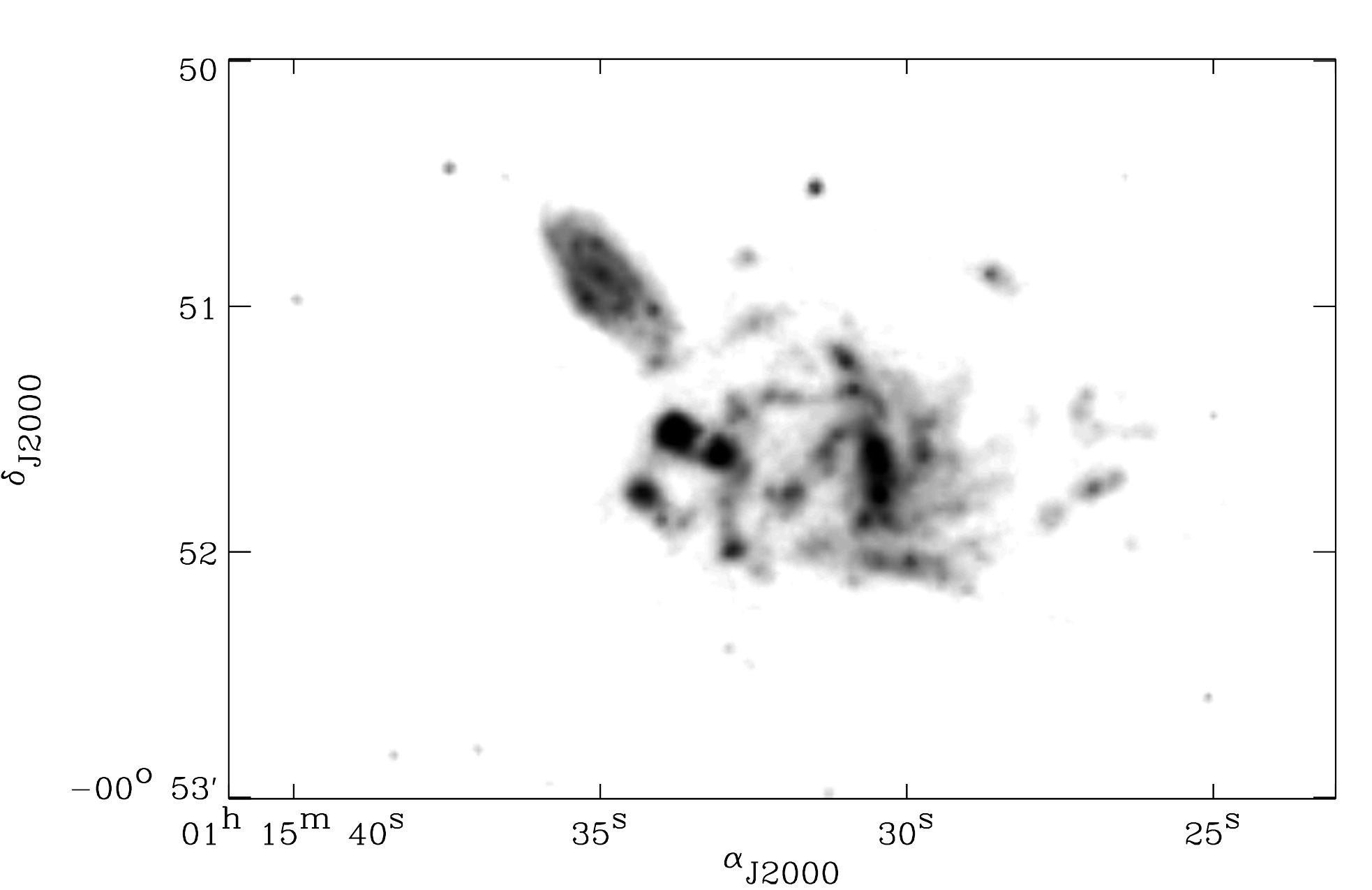} \\
&&\\
&&\\
           & UM~311 (24 \mic) & UM~311 (870 \mic)\\

	 & \includegraphics[width= 7cm ,height=5cm]{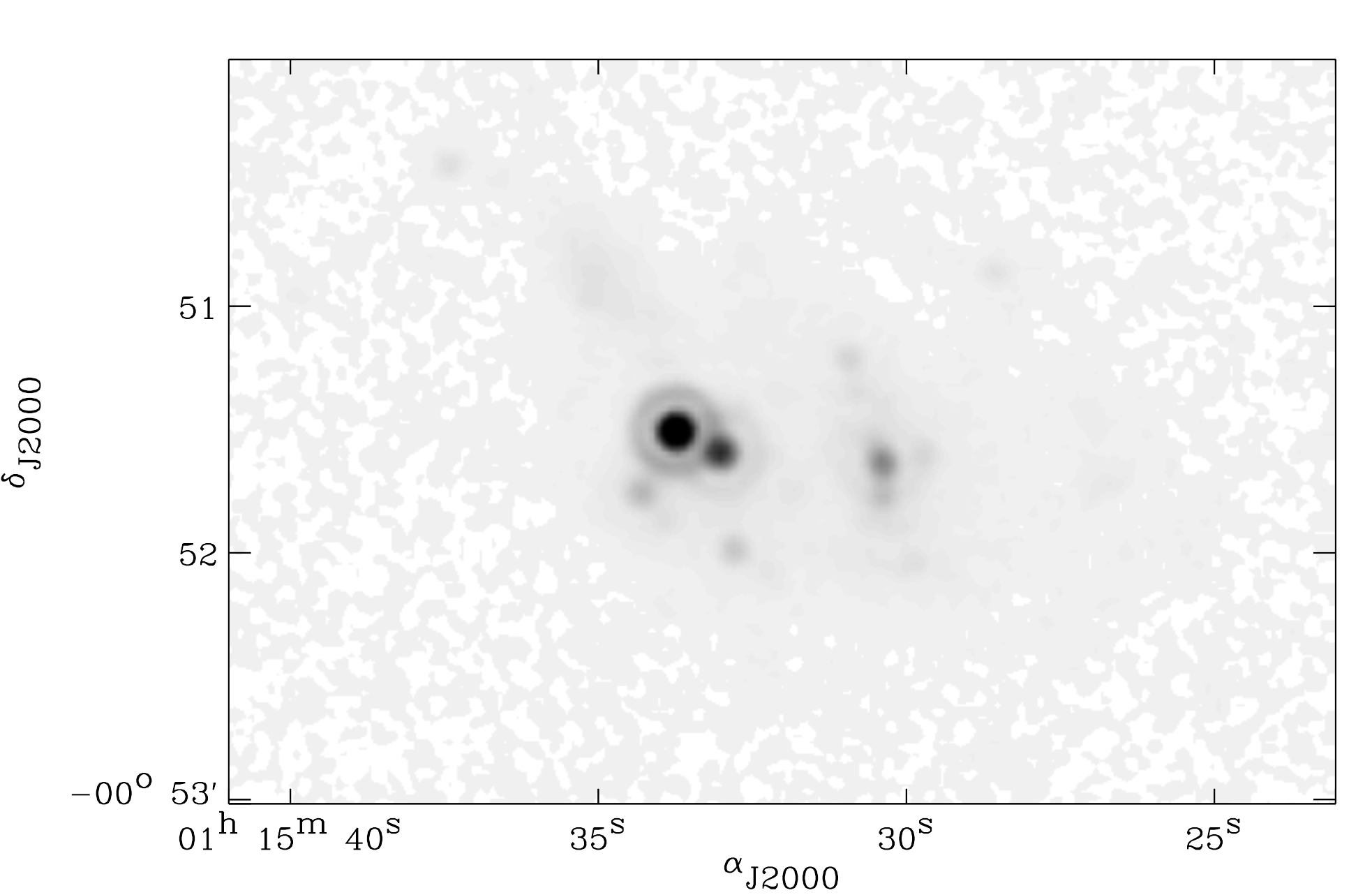} & \includegraphics[width= 7cm ,height=5cm]{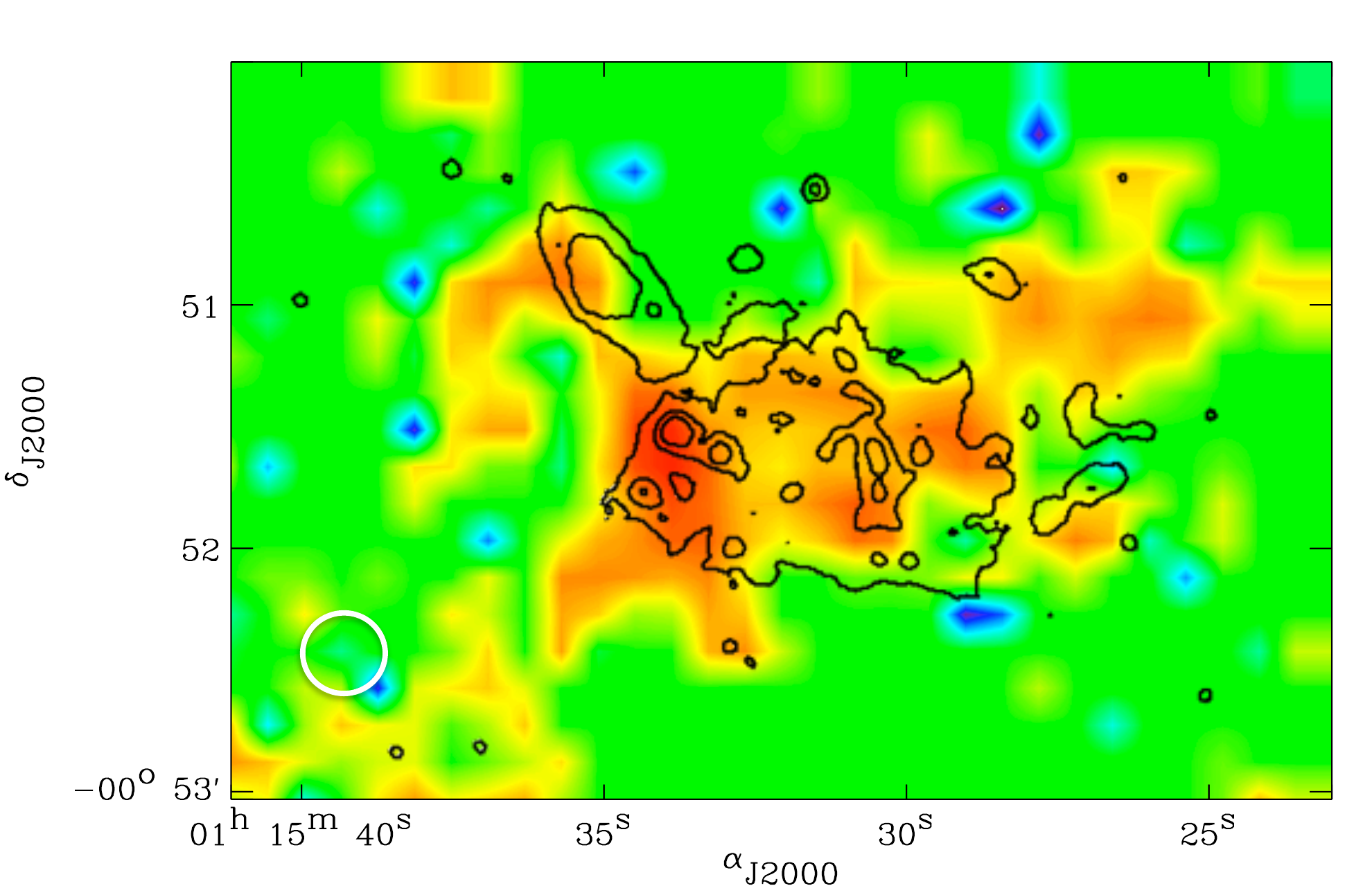} \\

                      \end{tabular}
    \caption{continued }
\end{figure*}
           

 \subsection{Spitzer data}


 \begin{table*}
 \centering
 \begin{tabular}{ccc}
\hline
\hline
&&\\
Name & AOR key \spitz/IRAC & AOR key \spitz/MIPS  \\
&&\\
\hline
&&\\
NGC~1705	&	5536000 - 5535744		&	5549056 - 5549312 - 14812416 \\
Haro~11		&	4326400		&	4344832		\\
Mrk~1089		&	11250432		&	11247360		\\
UM~311		&	10392576		&	12562176		\\
&&\\
\hline
\end{tabular}
  \caption{ AOR keys of the \spitz/IRAC and \spitz/MIPS observations.}
  \label{AORkeys}
 \end{table*} 

 \subsubsection{IRAC}
 
The IRAC bands cover 3.6, 4.5, 5.8, and 8 \mic\ with a FWHM of the PSFs of 1.8, 1.8, 2.0 and 2.2 arcsec respectively (http://ssc.spitzer.caltech.edu/documents/SOM/). {\revisedbis Two bands} are imaged in pairs (3.6 and 5.8 microns; 4.5 and 8.0 microns). IRAC provides 5.2' $\times$ 5.2' images (pixel size of 1.2'' $\times$ 1.2''). Our sources were observed with IRAC in dithering imaging mode and Mrk~1089, UM~311 and Haro~11 were obtained through the \spitz\ data archive (post-basic calibrated data). For NGC~1705, the IRAC data were obtained throught the SINGS data delivery page\footnote{Details on the data reduction of the SINGS Fifth Enhanced Data Release can be found at http://data.spitzer.caltech.edu/popular/sings/. }. Table ~\ref{AORkeys} summarizes the Astronomical Observation Request (AOR) keys of each observation. IRAC fluxes are uncertain at the $\sim$ 10$\%$ level due to systematic effects. The data we obtained were compared and verified with those published in \citet{Johnson2007} for Mrk~1089 and ~\citet{Cannon_NGC1705_2006} for NCG~1705. Images are presented in Fig.~\ref{LABOCA_images}.

 \subsubsection{MIPS}

The MIPS bands cover 24, 70 and 160 \mic\ with a FWHM of the PSFs of 6, 18 and 40" respectively (http://ssc.spitzer.caltech.edu/documents/SOM/). The galaxies were observed in MIPS scan mapping mode for NGC~1705 and MIPS Photometry/Super-Res mode for Haro~11, Mrk~1089 and UM~311. 

All of the MIPS images (see Table ~\ref{AORkeys} for AOR keys) were obtained through the \spitz\ data archive (raw data) and reduced using the MIPS Data Analysis Tools \citep{Gordon2005}, version 3.10 along with additional processing steps.  

The individual 24 \mic\ frames were first processed through a droop correction (to remove an excess signal in each pixel) and were corrected for non-linearity in the ramps.  The dark current was then subtracted. Scan-mirror-position dependent and independent flats were created in each AOR and applied to the data.  Detector pixels that had measured {\revisedbis signals superior to 2500 DNs$^{-1}$} (data numbers per second) in any frame were masked out in the following three frames to avoid latent images in the data.  Planes were fit to the zodiacal light emission in the background regions, and third order polynomials were fit to the background in each leg of each scan map. We also measured scan-mirror-position dependent residual offsets in the backgrounds. They were subtracted from the data.  For NGC~1705, we created preliminary mosaics of the data to identify transient objects (e.g. asteroids). We also performed a robust statistical analysis in which the values of cospatial pixels from different frames were compared to each other and statistical outliers to remove cosmic rays. We masked out these objects.  A final mosaic was made with data calibrated into astronomical units.  The calibration factor for the 24 \mic\ data was given in ~\citet{Engelbracht2007} as 4.54 {\small{$\pm$} }$  0.18 \times 10^{-2}$ MJy sr$^{-1}$ [MIPS instrumental unit]$^{-1}$.

In the 70 and 160 \mic\ data processing, we first fit ramps to the readouts to derive slopes. Readout jumps, cosmic ray hits and dark current  were removed. The stim flash frames taken by the instrument were used as responsivity corrections. An electronic nonlinearity correction and an illumination correction were applied. Short term variations in the signals ('drifts') were removed from all 70 \mic\ and 160 \mic\ data. To combine the photometry and scan map data for NGC~1705 without dealing with problems of background offsets, we subtracted the background from the photometry map at this stage.  {\revisedbis We corrected all pixels }affected by cosmic rays (see 24 \mic\ data treatment for details).  Then, final mosaics were built, residual backgrounds were subtracted and the data calibrated.  The 70 \mic\ calibration factors given in \citet{Gordon2007} are $702 \pm 35$ MJy sr$^{-1}$ [MIPS instrumental unit]$^{-1}$ for coarse-scale imaging and $2894 \pm 294$ MJy sr$^{-1}$ [MIPS instrumental unit]$^{-1}$ for fine-scale imaging.  The 160 \mic\ calibration factor is given by \citet{Stansberry2007} as $41.7\pm5$ MJy sr$^{-1}$ [MIPS instrumental unit]$^{-1}$.  An additional 70 \mic\ nonlinearity correction given as f{\tiny{$70$}}{\small{(true)}} = 0.581 $\times$ f{\tiny {$70$}}{\small{(measured)}}$^{1.13}$ \citep{Dale2007} was applied where the surface brightness exceeded 66 MJy sr$^{-1}$. MIPS flux uncertainties were estimated to be 10$\%$ in the MIPS 24 \mic\ band and 20$\%$ in the MIPS 70 and 160 \mic\ bands. 
{\revisedbis The fluxes calculated compared to those published in \citet{Johnson2007} for Mrk~1089 and ~\citet{Cannon_NGC1705_2006} for NCG~1705.}


\begin{figure*}
    \centering
    \begin{tabular}{p{0mm} m{8.5cm} p{0mm} m{10cm}}
      a) & \includegraphics[width=8.5cm ,height=5.5cm]{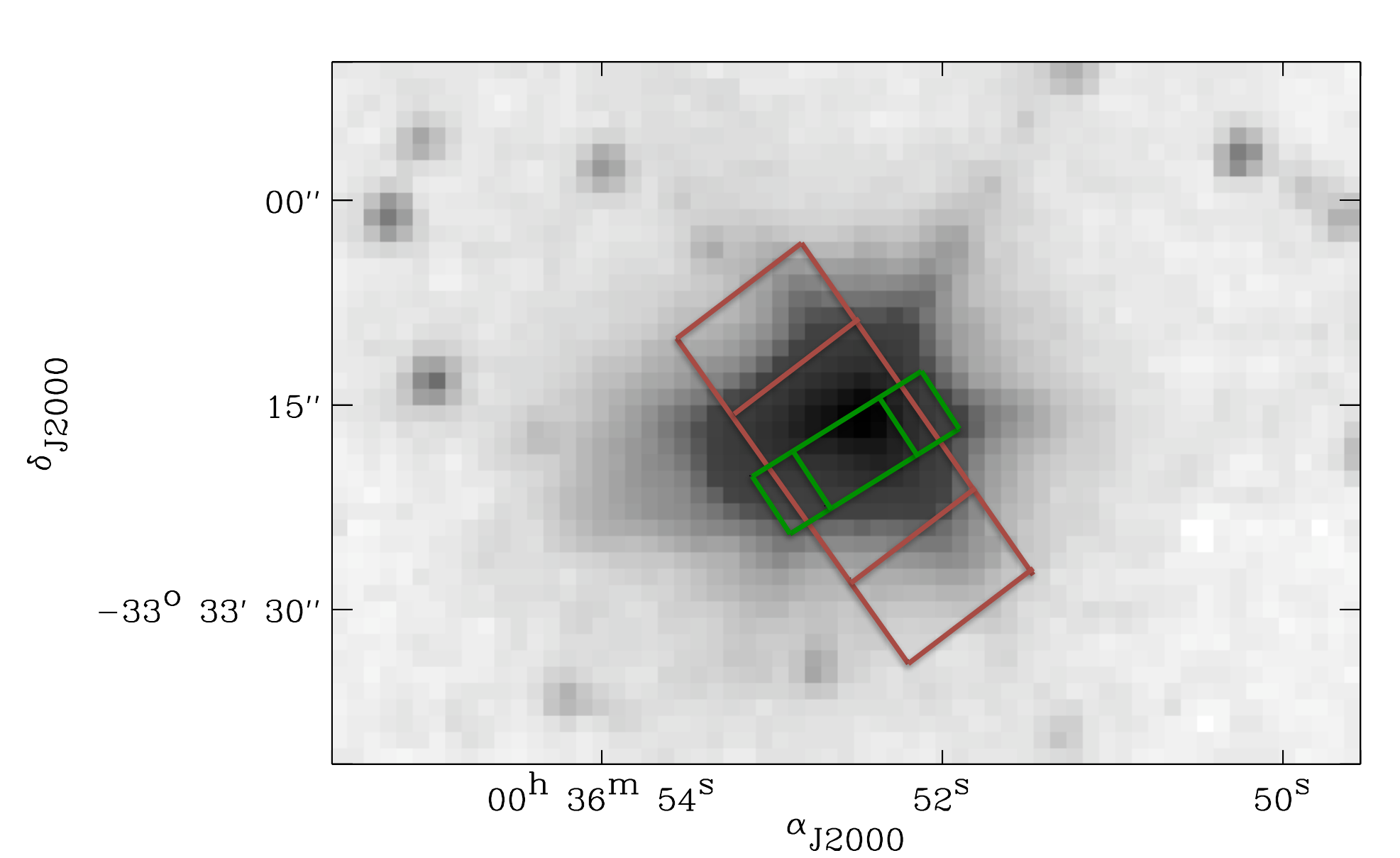}  & b) &   \includegraphics[width=8cm ,height=5.5cm]{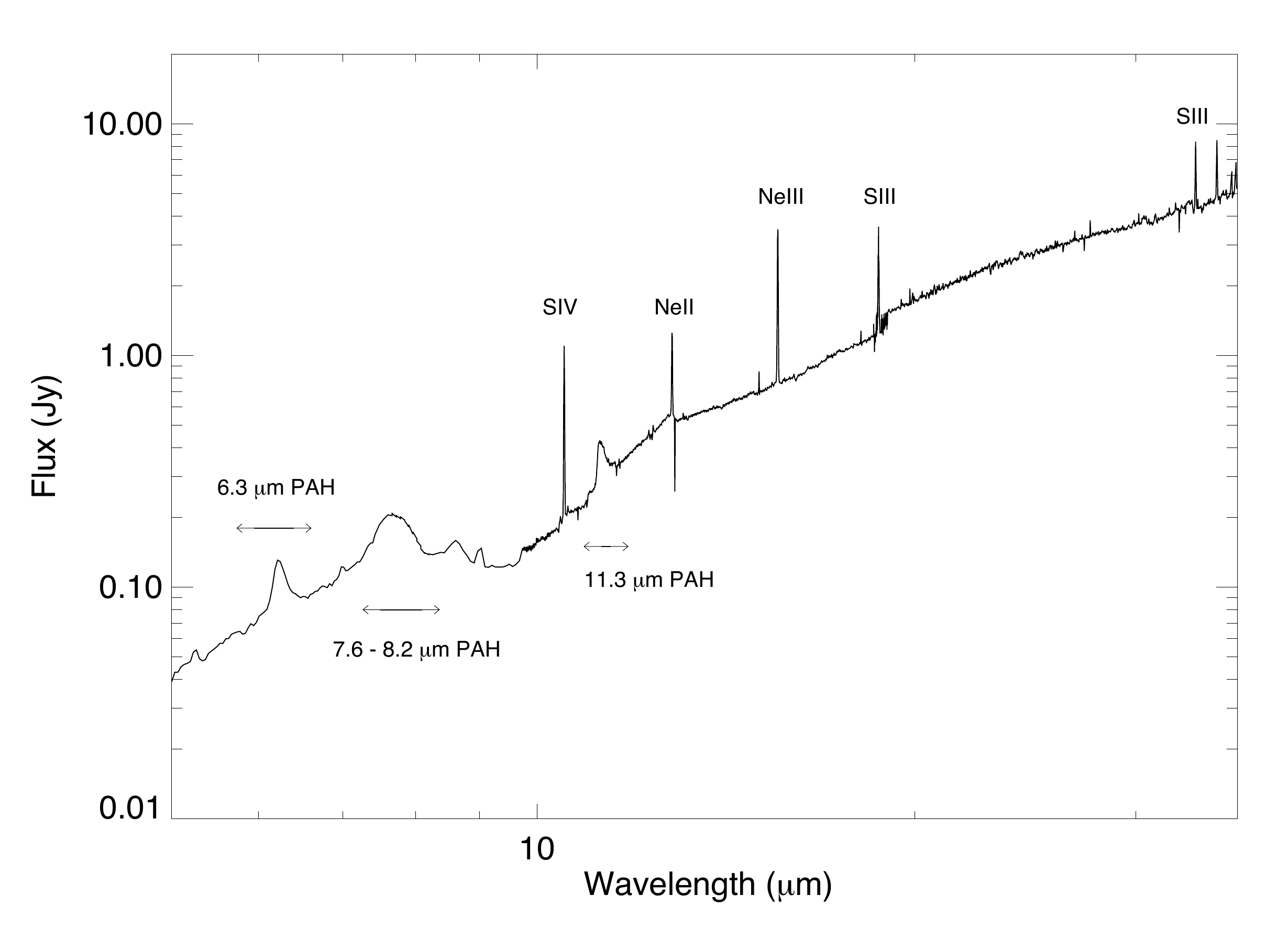} \\
     
         \end{tabular}
    \caption{{\it a) } The positions of the short-high (green) and low-high (red) slits of the IRS observations superimposed on the {\spitz}/3.6 \mic\ image of Haro~11. {\it b) } The short-low and short-high spectrum of Haro~11.   }
        \label{IRS}
\end{figure*}

\subsubsection{IRS}

For the galaxy Haro~11, we supplemented our dataset with the mid-IR spectrum from the \spitz\ Infrared Spectrograph (IRS) in order to better constrain the 5 \mic\ to 40 \mic\ range of the SED. We used public released data  (\spitz\ AOR key: 9007104) performed in staring mode. We used the Basic Calibrated Data products available in the \spitz\ archive for the short-low resolution channel (SL, 5.2-14.5 \mic; $\lambda$/$\Delta$$\lambda$$\sim$60-120) and the two high resolution channels: short-high (SH, 9.9-19.6 \mic; $\lambda$/$\Delta$$\lambda$$\sim$600) and long-high (LH, 18.7-37.2 \mic; $\lambda$/$\Delta$$\lambda$$\sim$600). IRSClean was applied to the individual files to mask out hot pixels, and the frames were coaddd together. We then used the \spitz\ IRS Custom Extraction software (Spice - 2.0.1 version) to extract the spectrum. We subtracted background contributions estimated from zodiacal models using the software SPOT (18.0.1 version - http://ssc.spitzer.caltech.edu/propkit/spot/). We finally subtracted a background estimated using zodiacal models in SPOT. The final error on the photometry derived from peak-up acquisition windows is about 15$\%$, the value recommended by the IRS data handbook. Finally, the spectrum is shifted from a redshift of z=0.021  \citep{Bergvall2000} to rest wavelengths. The position of the slits of the IRS high resolution observations is shown in the panel on the left in Fig.~\ref{IRS}. The spectrum with labels identifying the main emission features is shown in the panel on the right in Fig.~\ref{IRS}. The 6.3, 7.6, 8.2 and 11.3 \mic\ PAH features are clearly detected as well as significant emission lines. Note that NeIII/NeII ratio=3.2 (calculated on the IRS spectrum), which is an indication of the hard radiation field dominated by a young population $\leq$ 5 Myr ~\citep{Madden2006}.

 \subsection{Supplemental data}

We supplement the \spitz\ and \lab\ observations with \twomass\ J, H, K flux density estimates available on the NASA/IPAC Infrared Science Archive to describe the stellar contribution of the SEDs of the galaxies Mrk~1089, Haro~11 and NGC~1705. We assume that these \twomass\ flux densities are global measurements for the galaxies. For the UM~311 system, we perform the photometry and estimate the \twomass\ flux densities from the images also obtained from the NASA/IPAC Infrared Science Archive. 
The IRAS broadband flux densities at 12, 25, 60 and 100 \mic\ are finally obtained through the NASA/IPAC Infrared Science Archive, the IRAS Faint Source catalogue, {\revisedbis $|$b$|$ $>$ 10}, Version 2.0 ~\citep{Moshir1990}, the IRAS catalogue of Point Sources, Version 2.0 (IPAC 1986) and \citet{Dultzin-Hacyan1990}. They are added to supplement the existing data on dust emission. For Haro~11, the IRAS fluxes are consistent with the values of the IRS spectrum. For UM~311, IRAC data should probably encompass emission from both of the nearby galaxies and are thus not used in the analysis.

 \section{Images}
 \subsection{Discussion on the morphology}
  
The 870 \mic\ \lab\ maps are presented in Fig.~\ref{LABOCA_images}. The contours of the IRAC observations - 4.6 \mic\ for Haro~11 and 8 \mic\ for UM~311, Mrk~1089 and NGC~1705 - are superimposed on the \lab\ images. 

{\it NGC~1705 -} There is significant change in the galaxy morphology from NIR to FIR wavelengths. The bright 3.6 \mic\ source is observed toward the location of the SSC while two strong MIR and FIR emission peaks appear at 5.8 \mic\ and longer wavelengths, offset from the SSC and coincident with the H$\alpha$ maximum (Fig.~\ref{NGC1705_Halpha_HI} and Fig.~\ref{LABOCA_images}a). {\revisedbis These two off-SSC emitting regions} do not have bright stellar counterparts. The \spitz\ 8 \mic\ image shows the SSC in the middle of these two off-nuclear HII regions. The eastern region is the brightest source at 24 \mic, with a flux density which is two times higher than the SSC flux density at this wavelength ~\citep{Cannon_NGC1705_2006}. The three regions are also barely resolved in the 870 \mic\ images. An offset peak is detected toward the west of the centre of the galaxy. Two faint 24 \mic\ sources are possible counterparts of the emission. 

 {\it Haro 11 -} This galaxy is not resolved by \spitz\ at wavelengths greater than 4.5 \mic . It is clearly detected but barely resolved with \lab\ (Fig.~\ref{LABOCA_images}b). The final images present extended structures but this extension is not observed in the MIR images and was removed while calculating the submm flux to be conservative.

 {\it Mrk~1089 -} The 8 \mic\ image (Fig.~\ref{LABOCA_images}c) clearly shows the interaction between the galaxies NGC 1741 (East) and Mrk 1089 (West). A \spitz\ color composition image of the complete Hickson group observed with \spitz\ is described in \citet{Johnson2007}.  The merging center of Mrk~1089 dominates the emission at MIR, FIR and 870 \mic. Diffuse emission probably linked with intergalactic dust, presumably from galaxy interactions, is detected through out the whole region and was removed while performing the photometric measurements. 
 
 {\it UM~311 -} The 4.5 \mic\ image of the interacting field (Fig.~\ref{LABOCA_images}d) shows the spiral structure and arms of NGC~450 as well as the spiral  companion UGC~807 and the three bright HII regions between these two spirals, one of which is the UM~311. We adopt the same numbering as \citet{Moles1994} to describe these 3 HII sources and will call them "the UM~311 system". UM~311 is the region called 3 on the 4.5 \mic\ image.  This compact HII galaxy is the brightest source of emission at 24 \mic. In the \lab\ image, the emission peaks toward the location of UM~311. The galaxy should dominate the SED of the UM~311 system.

\subsection{Photometry}

%

\begin{table*}
 \centering
 \begin{tabular}{cp{2.6cm}cp{3mm}cp{3mm}cp{3mm}cp{3mm}c}
\hline
\hline

&&&& NGC 1705 && Haro 11 && Mrk 1089	&& UM 311 system  \\
\hline
2MASS &1.25 \mic\ flux density &(in mJy) && 52 $\pm$ 1.6	&& 13  $\pm$ 0.3  	&& 12.2 $\pm$ 0.6	&& 12.8 $\pm$ 1 \\
2MASS &1.65 \mic\ flux density &(in mJy) && 50.9 $\pm$ 2	&& 13 $\pm$ 0.5 	&& 13.9 $\pm$ 0.9	&&  12.6 $\pm$ 1 \\
2MASS &2.17 \mic\  flux density &(in mJy) && 41.1 $\pm$ 2.4	&& 13.8 $\pm$ 0.7 	&& 12.2 $\pm$ 1.2	&& 11.5 $\pm$ 1  \\
IRAC &3.6 \mic\  	flux density &(in mJy) && 26.3 $\pm$ 3	&& 22.5 $\pm$ 2 	&& 14.9 $\pm$ 2	&& ...   \\
IRAC &4.5 \mic\ 	flux density &(in mJy) && 17.9 $\pm$ 2	&& 28.9 $\pm$ 3 	&& 11.2 $\pm$ 1	&& 5.5 $\pm$ 0.5   \\
IRAC &5.8 \mic\ 	flux density &(in mJy) && 16 $\pm$ 2		&& 72.7 $\pm$ 7 	&& 33.3 $\pm$ 3	&& ...   \\
IRAC &8 \mic\ flux density &(in mJy) && 17.9 $\pm$ 2	&& 177 $\pm$ 18 	&& 92.5 $\pm$ 9	&& 35.6 $\pm$ 4   \\
IRAS  &12 \mic\ flux density &(in mJy) &&  ...			&& 417 $\pm$ 46 	&& 106 $\pm$ 21	&& ...$^*$   \\
MIPS &24 \mic\ flux density &(in mJy) && 52.7 $\pm$ 5	&& 1930 $\pm$ 193 && 460 $\pm$ 46	&& 151 $\pm$ 15   \\
IRAS &25 \mic\ flux density &(in mJy) && ...				&& 2500 $\pm$ 25 	&& 579 $\pm$ 29	&& ...$^*$  \\
IRAS &60 \mic\  flux density &(in mJy) && 868 $\pm$ 61	&& 6880 $\pm$ 41 	&& 4010 $\pm$ 40	&& ...$^*$  \\
MIPS &70 \mic\ 	flux density &(in mJy) && 1110 $\pm$ 222	&&  4190 $\pm$ 838 && 3990 $\pm$ 798 && 1480 $\pm$ 148 \\
IRAS &100 \mic\ 	flux density &(in mJy) &&  1610 $\pm$ 100	&& 4730 $\pm$ 640	&&  5430 $\pm$ 181&& ...$^*$  \\
MIPS &160 \mic\  flux density &(in mJy) && 1110 $\pm$ 222	&& 1660 $\pm$ 332 	&& ...				&& ... $^{**}$  \\
LABOCA &870 \mic\  flux density &(in mJy) && 114 $\pm$ 17	&& 40 $\pm$ 6 		&& 67.4 $\pm$ 10.1	&& 32.2 $\pm$ 7    \\
\hline
\end{tabular}
  \caption{Integrated flux densities measured with {\it {2MASS}}, \iras, \spitz/IRAC, \spitz/MIPS and \lab. The 10$\%$ non-dust contamination was subtracted from the \lab\ flux densities.  For Haro~11, no broadband data between 5 and 38 \mic\ were used in the SED modelling since we add the IRS spectrum to constrain this range. 
  \newline
  $^*  $ ~The available IRAS fluxes for UM~311 are not used as they incorporate the entire field which includes the two nearby spiral galaxies too.
   \newline
   $^{**}$ The 160 \mic\ resolution is too large to isolate the 160 \mic\ flux of the UM~311 system alone.} 
  \label{Tableflux}
 \end{table*} 

We use the function {\it aper} of the library IDL astro of the NASA Goddard Space Flight Center to perform the aperture photometry. For the {\spitz} photometric data, the background is estimated using annuli just outside the boundaries of our galaxies. For the \lab\ images, the background significantly varies in intensity throughout the entire map. We calculated the flux densities in small circles in the immediate surroundings of the galaxy and averaged them to determine a local background for each source which was removed during the photometric calculation. 

To determine the flux densities at IRAC and MIPS bands of our sources and compare data between different wavelengths, the observations are convolved and regridded to a common resolution. For NGC~1705 and Haro~11, the lowest spatial resolution is the resolution of MIPS 160 \mic\ (40''). We use convolution kernels \citep{Gordon2008} which convert a higher resolution IRAC/MIPS point-spread function (PSF) to lower resolution IRAC/MIPS PSF using Fourier transforms. For the \lab\ data, we convolved the observations with a Gaussian kernel and regridded the images to the resolution of MIPS 160 \mic. The photometric aperture chosen has a diameter of 144" for these two galaxies to encompass the entire emission of the galaxy.

As Mrk~1089 was not observed at 160 \mic,  the four IRAC bands and the MIPS 24 and 70 \mic\ observations were regridded to the resolution of \lab\ (18.2''). We choose an aperture of 72" avoiding the flux arising from the other nearby companions. However, Mrk~1089 clearly dominates the whole system at MIR-FIR wavelengths. 

For UM~311, the resolution of the MIPS bands and the proximity of the 2 nearby HII regions makes it difficult to study UM~311 alone. We decide to perform the photometry of the UM311 system (encompassing sources 1, 2, 3 in Fig.~\ref{LABOCA_images}d). The 2MASS, IRAC and MIPS 24 and 70 \mic\ images were regridded to the \lab\ resolution of 18.2" for the UM~311 system. In the 160 \mic\ image, the broad PSF may cause emission from other sources outside the region around UM~311 to bleed into that region (Fig.~\ref{LABOCA_images}d). To be conservative, we choose a 54" aperture which encompasses the UM311 system and use the 160 \mic\ observation as an upper limit when modelling the SED of this system. The procedure we use to model this system gives us solutions for the limits on the parameter space for UM~311 as if this interacting system were more distant causing these 3 sources to blend together. 
 
The IRAC measurements require aperture corrections to account for the scattering of incident light in the focal plane arrays \citep {Reach2005}. These scaling factors of 91, 94, 71, 74$\%$  are respectively applied to the final integrated flux density estimates at 3.6, 4.5, 5.8 and 8 \mic. Table~\ref{Tableflux} presents the IRAC flux densities corrected for the scaling factors. 

Finally, we compare the fluxes obtained from \spitz\ observations with the \iras\ broadband fluxes. MIPS 24 \mic\ and \iras\ 25 \mic\ as well as \iras\ 60 \mic\ and MIPS 70 \mic\ are consistent for the galaxies where both are available except for Haro~11 for which \iras\ fluxes are systematically higher than MIPS fluxes due to the fact that \iras\ fluxes were estimated using bigger aperture. The \iras\ 25 \mic\ flux is not used since we use the IRS spectrum as a constraint for that part of the SED. The \iras\ 100 \mic\ flux values seem to be high for Mrk~1089 and NGC~1705, perhaps encompassing more extended emission. We will see in $\S$5 what this implies for the SED modelling.\\

\subsection {Contamination of the 870 $\mu m$ emission from non-dust origins}

Since we are interested here in the dust emission arising in the 870 \mic\ band, we must consider and quantify possible contamination from non-dust sources and take these corrections into account when modelling the dust SEDs. The CO(3-2) line can, in principle fall within the 870  \mic\ band. While CO(1-0) has been a great challenge to detect in low metallicity galaxies \citep [c.f.][] {Leroy2005}, we can not be sure that the smaller beam, higher excitation CO(3-2) could not be present here. CO observations were attempted in NGC~1705 without positive detections   \citep{Greve1996}. In the same way, CO seems to be very faint in the Hickson Group to which Mrk~1089 belongs \citep{Yun1997}. Finally, for Haro~11, \citet{Bergvall2000} found an upper limit for L$_{CO(1-0)}$ of 10$^{29}$ W.  We can derive an upper limit to  L$_{CO(3-2)}$ from the L$_{CO(1-0)}$ estimation using the \citet{Meier2001} relations for dwarf starburst galaxies: L$_{CO(3-2)}$/L$_{CO(1-0)}$ is usually lower than 1. For the four galaxies, we conservatively estimate the CO (3-2) contribution to the 870 \mic\ band to be $\sim$ 5$\%$.

Additionally, we can expect contributions to the 870 \mic\ fluxes from radio continuum emission (synchrotron emission and/or bremsstrahlung). Three radio fluxes were estimated for Haro~11 with the NRAO VLA Sky Survey ~\citep[8.46 and 1.4 GHz - ][]{Condon1998} and the Sydney University Molonglo Sky Survey ~\citep[843 MHz - ][]{Mauch2003}. We derive the 870 \mic\ radio continuum contribution by extrapolation of the radio data tendancy ($\nu^{-1}$). We find a radio contribution of 3$\%$ of the \lab\ flux for Haro~11. The radio observations in the literature are not sufficient to constrain the expected 870 \mic\ radio contamination for NCG1705, UM~311 and Mrk~1089. We thus consider that the radio continuum contamination is of the same order for these other galaxies. Finally, we conservatively estimate the global non-dust contamination from potential CO(3-2) emission and radio continuum emission (synchrotron/bremsstrahlung) in the 870 \mic\ band to be 10$\%$ for our four galaxies, similar to the  submm contributions that \citet{Galliano2005}, for example, determine for their dwarf submm observations.


 \section{The SED modelling}
 
  The SED is a complex tool to interprete. For a macroscopic region of the ISM (HII region, cirrus), the SED summarises a wide range of physical conditions linked with the non-uniformity of its illumination and the variations of the dust composition. For an entire galaxy such as the dwarf galaxies we are studying here, the SED synthesises, on the contrary, the different components that constitute its global shape and emission: HII regions, molecular clouds, nebulae, diffuse ISM etc., that can not be studied independantly due to the lack of observational constraints. 
  
We want to construct the SEDs of our galaxy sample using the fluxes presented in Table~\ref{Tableflux} and the IRS spectrum for the galaxy Haro~11 as constraints in order to quantify elementary quantities of these galaxies such as their mass of PAHs or their total mass of dust. The stellar contribution to the SED is constrained by the 2MASS bands and IRAC 3.6 \mic. The NIR to submm wavelengths of the SEDs, signatures of the dust and physical conditions of the galaxies, are constrained by \iras, \spitz\ and our new \lab\ data (Table~\ref{Tableflux}). No radio component is taken into account as a model constraint but the radio contribution is removed from the 870 \mic\ flux (see $\S$ 4.3). 

The model we use to fit the SEDs is a simplified version (due to the smaller number of observational constraints) of the \citet{Galliano_Dwek_Chanial_2008} model. We would first like to note that a SED depends on the intensity and the hardness of the Interstellar Radiation Field (ISRF) {\revisedbis that the ISM experiences}, on the mass fraction of the dust species (silicate and carbon grains, PAHs etc.) and on the distribution of grain sizes. Thus, modelling a SED requires making {\it a priori} assumptions on the ISRF and on the global properties of the dust, assumptions that we will describe in the following paragraphs. 

We assume that the source of excitation of the dust is the ISRF which we choose to have the spectral shape of the ISRF found in the Galactic diffuse ISM \citep{Mathis1983}, although we test the influence of this spectral shape by trying a variety of forms (see $\S$ 6.4). The ISRF intensity will be scaled using a factor {\it U} \citep[defined in][]{Draine_Li_2007}, with {\it U}=1 corresponding to a normalization to the local solar neighbourhood value of 2.2 $\times$ 10$^{-5}$ W m$^{-2}$. 

We assume that the sources of IR emission are {\revisedbis dust and old stars}.  We suppose that the dust composition is homogeneous throughout the galaxy. {\revisedbis For silicates, graphites and PAHs, }we adopt the composition and size distribution of \citet{Zubko2004} (Table~\ref{Zubko2004}). \citet{Zubko2004} assume that the dust particules are PAHs, graphite and silicate grains and that the ISM has a solar abundance. The assumed optical properties of these grains are taken from~\citet[][PAHs]{Draine_Li_2007}, ~\citet[][graphites]{Laor1993} and~\citet[][silicates]{Weingartner2001}. The total mass of dust (M$_{dust}$) represents the first parameter of our model while the PAH component requires the introduction of two other parameters, the ionised PAHs-to-neutral PAH ratio (f$_{PAH+}$) and the PAH-to-total dust mass ratio (f$_{PAH}$) normalised to the Galactic value of 0.046, to be described.


\begin{table*}
 \centering
 \begin{tabular}{c|ccc}
\hline
\hline
&&&\\
 & PAHs & Graphites & Silicates \\
&&&\\
\hline
&&&\\
Minimum Size  (\mic) & 3.5 $\times$10$^{-4}$ & 3.5 $\times$10$^{-4}$ & 3.5 $\times$10$^{-4}$\\
Maximum Size  (\mic)& 5 $\times$10$^{-3}$ & 0.33 & 0.37 \\
Mass density  (g cm$^{-3}$) & 2.24 & 2.24 & 3.5\\
&&&\\
\hline
\end{tabular}
  \caption{Size range \citep{Zubko2004} and mass densities of the three dust grain components. } 
  \label{Zubko2004}
 \end{table*} 

We adopt the prescription of \citet{Dale2001} to relate the dust mass exposed to a given intensity dM$_{dust}$({\it U}) to the different heating environment intensities (U) to which dust is exposed. 
\begin{equation}
       dM_{dust}({\it U}) \propto {\it U}^{-\alpha}dU,    	\hspace{1cm}		  U_{min} \leq U \leq U_{max} 
 \label{Dale}
 \end{equation} 
 
This prescription is flexible enough to describe dense and diffuse media. The simple power law relation leads to the definition of the index $\alpha$, which represents the contribution of the different local SEDs exposed to a certain intensity U. We introduce the three quantities $\alpha$, the minimum heating intensity U$_{min}$ and the maximum heating intensity U$_{max}$ as free parameters in our modelling.

Since the stellar component contributes to the NIR part of the SED, we finally {\revisedbis add} it from the dust SED by fitting a stellar spectrum. This spectrum is synthesised using the stellar evolution code PEGASE \citep{Fioc_Rocca_1997} assuming a Salpeter Initial Mass Function (IMF). The stellar population is considered to have undergone an instantaneous burst (5 Gyr ago) and the initial metallicity is assumed to be solar (Z=Z$_\odot$). The mass of stars (M$_{oldstar}$) is introduced as a parameter of our modelling. 


In summary, the free parameters used in the modelling are:
   \[
\begin{array}{lp{0.9\linewidth}}
M_{dust}  &  total mass of dust \\
f_{PAH}    &  PAHs-to-dust mass ratio (normalised to the Galactic value) \\
f_{PAH+}  & ionised PAHs-to-total PAHs mass ratio\\
\alpha       & index describing the fraction of dust exposed to a given intensity\\
U_{min}    & minimum  heating intensity\\
U_{max}   & maximum heating intensity\\
M_{oldstar} & mass of old stars \\
\end{array}
   \]
   
The modelling is an iterative process in which we assume an initial dust grain distribution (size and composition). Optical and near-IR data are first used to constrain the stellar radiation of the model. We then compute the temperature distribution of the dust grains heated by the absorption of this stellar radiation using the {\revisedbis method} of \citet{Guhathakurta1989}. Synthesised spectra are finally computed for each of the silicates, graphites, neutral and ionized PAHs, considering that grains are stochastically heated. The sum of these discrete contributions leads to a global SED model of the galaxy. 


 
 
The interactive fitting stops when deviations from the observational constraints are minimized. The $\chi$$^{2}$ minimisation algorithm is based on the Levenberg-Marquardt methods.
To prevent $\chi^2$ from being influenced by the density of points or dominated by the highly sampled MIR spectrum in the case of Haro~11, we weight each data point depending on the density of points around its wavelength. $\Delta$ L$_{\nu}^{obs}$($\lambda$$_i$) represents the error on the luminosity at a given wavelength.
\begin{equation}
{\chi^2}=\sum_i\left(\frac{ \lambda_{i+1}-\lambda_{i-1}}{2\lambda_i}\right)\left[\frac{ L_{\nu}^{obs}(\lambda_i)-L_{\nu}(\lambda_i)}{\Delta L_{\nu}^{obs}(\lambda_i)}\right]^2
 \end{equation}
We will call this first model the "fiducial" model. 

{\revisedbis Our model presents some limits mainly linked to the assumptions we made to simplify the model. Indeed, we choose the radiation field shape of the Galaxy \citep{Mathis1983}, a profile which could be different in low-metallicity environments. The choice of a radiation field profile usually affects the PAH and small grains mass derived from the modelling. The influence of this parameter is discussed in $\S$6.4. 
Another strong assumption was in the use of the \citet{Dale2001} prescription (Eq.~\ref{Dale}). This formula directly relates the cold and the hot regions of the galaxy, an assumption which is not valid if cold dust is physically residing in different regions than star forming regions. The fact that a separate cold dust component is required for some of our galaxies (see $\S$ 6.2) may be a sign of the limit of the \citet{Dale2001} prescription.
Nevertherless, our simple model has the sufficient level of complexity (e.g. detailed dust properties) to accurately derive the global properties we want to study, especially as regards to our small number of observational constraints for each galaxy. Adding other components such as different phases, clumps or more complex geometries would have lead to an overinterpretation of the observations.}

An international conference was held (2005) to discuss other SED models. We refer the readers to the reviews of this conference in \citet{Popescu2005}.\\

\begin{figure*}
    \centering
    \begin{tabular}{ m{8cm}  m{8cm} }
      \includegraphics[width=9cm ,height=6cm]{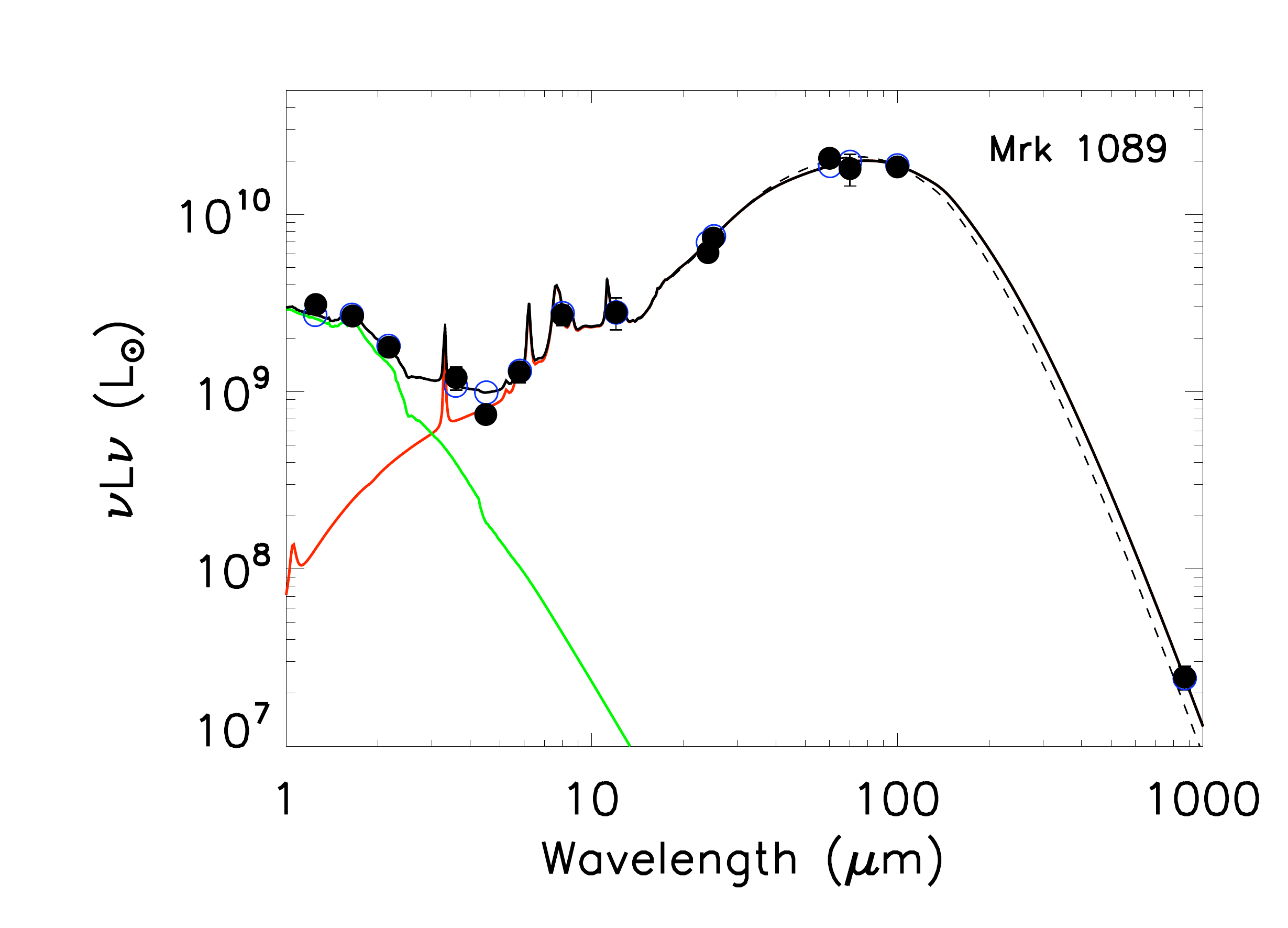} &
      \includegraphics[width=9cm ,height=6cm]{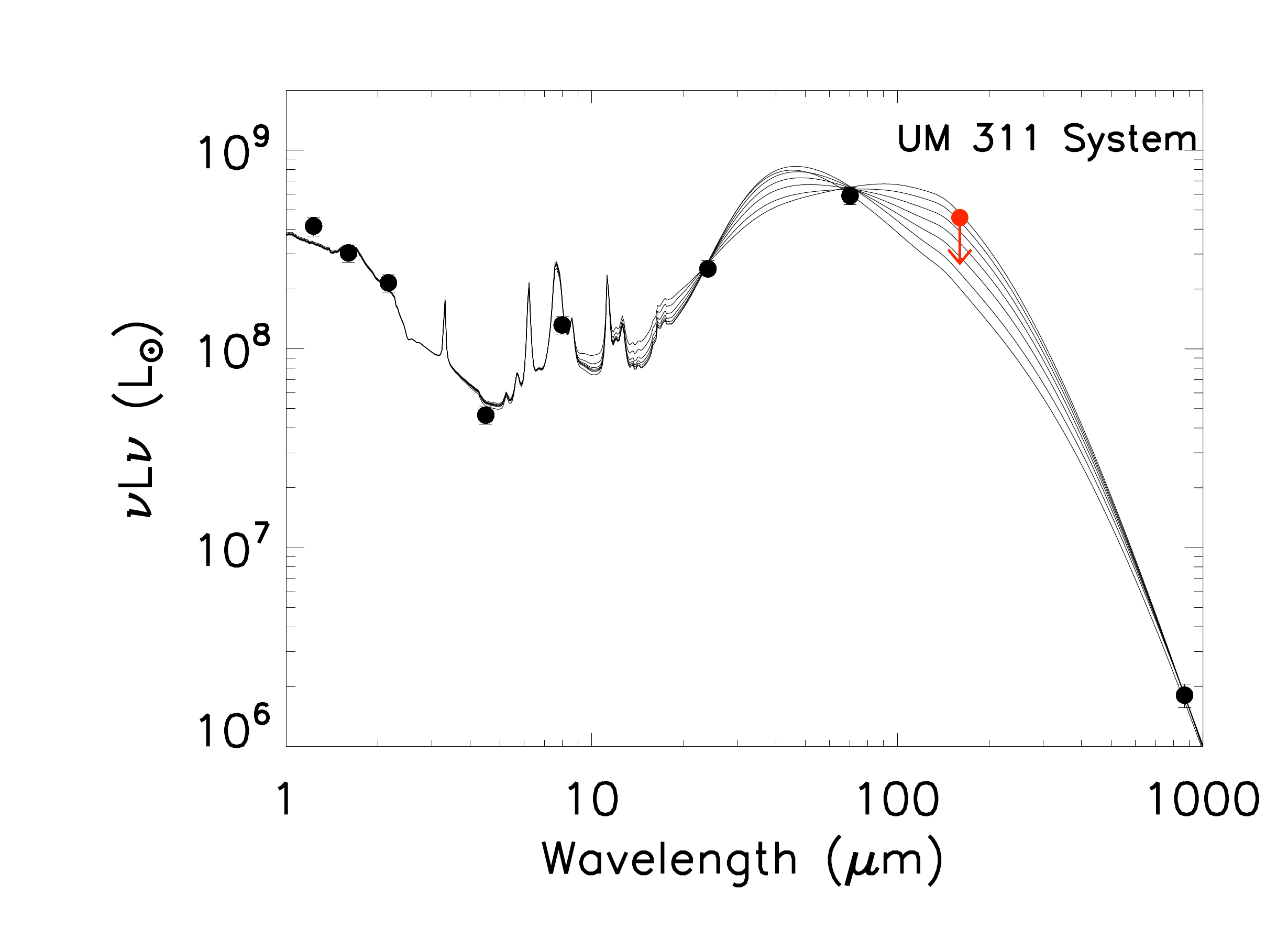} \\
       \includegraphics[width=9cm ,height=6cm]{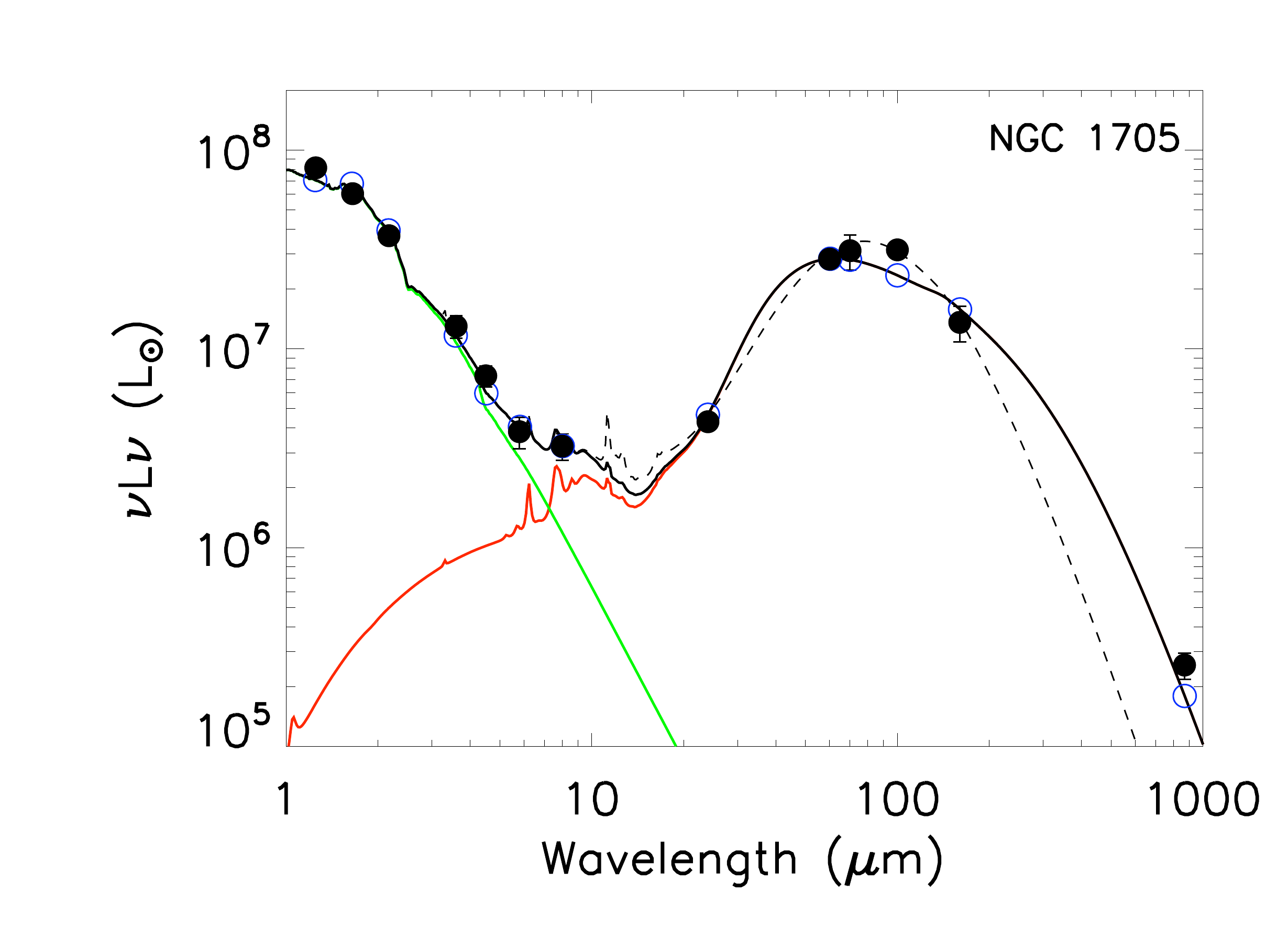} &
       \includegraphics[width=9cm ,height=6cm]{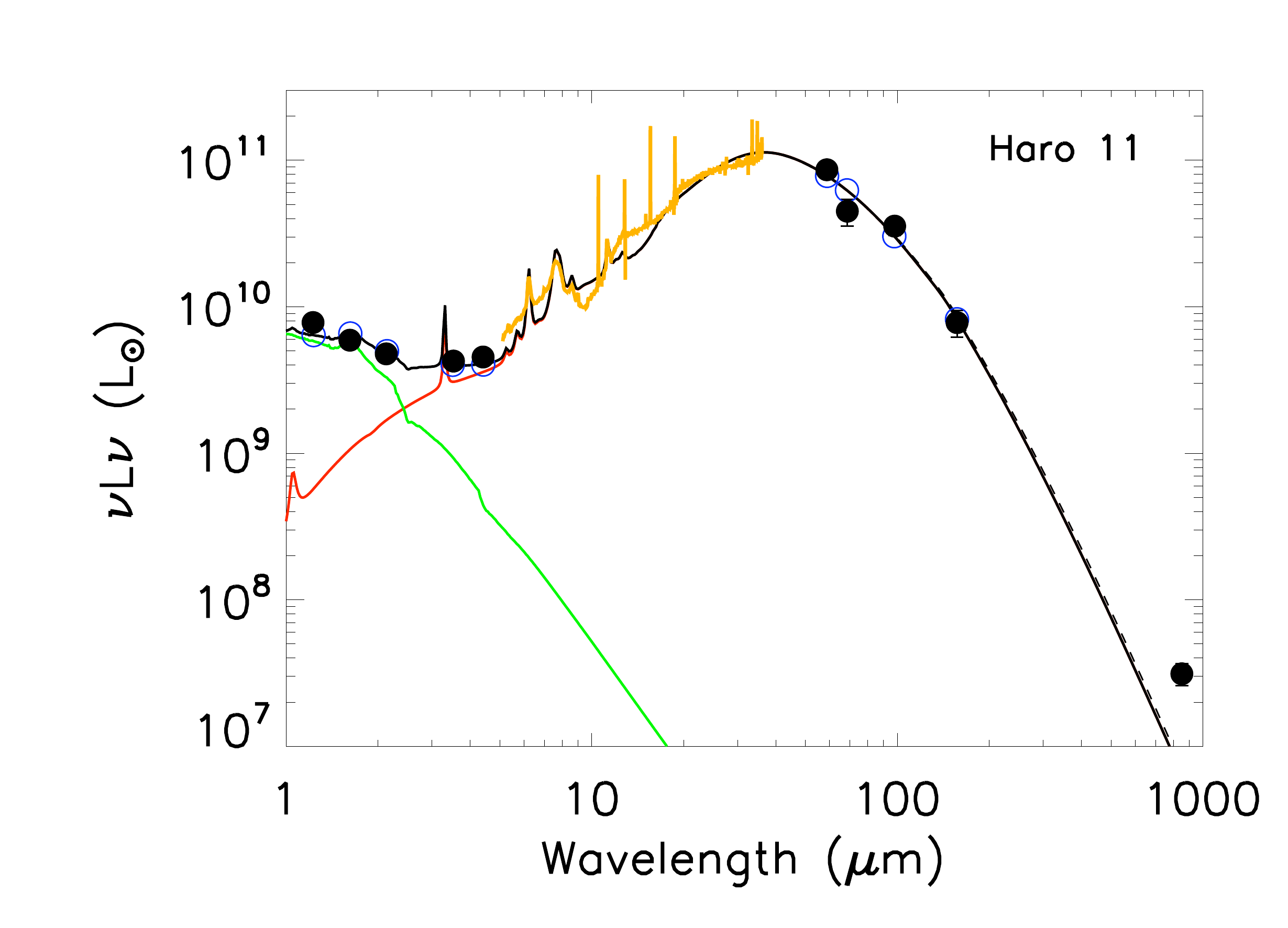} \\
         \end{tabular}
    \caption{SED models of Mrk~1089, the UM~311 system, NGC~1705 and Haro~11 using the fiducial model. The SEDs are plotted in black. Observational constraints (listed in Table~\ref{Tableflux}) are superimposed (filled circles). The green and red lines respectively distinguish the stellar and the dust contributions. The dashed black lines present the SED models of our galaxies obtained when the LABOCA constraint is not used in the modelling. The open circles represent the expected modeled fluxes integrated over the instrumental bands. When the error bars are not shown, the errors are smaller than symbols. Note that the IRS MIR spectrum used in the modelling is overlaid in orange for Haro~11. For the UM~311 system of 3 compact sources, the 160 \mic\ flux is an upper limit since it was calculated with a 40" aperture. The different SEDs represent the possible SED models that fit the observational constraints with good accuracy.}
    \label{SEDs}
\end{figure*}

\section{Analysis}

\subsection{Results of the fiducial SED modelling}

\begin{table*}
 \centering
 \begin{tabular}{c|cc|cc|cc}
\hline
\hline

& \multicolumn{2}{c|}{Haro~11}  & \multicolumn{2}{c|}{Mrk~1089} & \multicolumn{2}{c}{NGC~1705} \\

& no submm & submm &  no submm &  submm & no submm &  submm \\
\hline
& \multicolumn{2}{c|}{}  & \multicolumn{2}{c|}{} & \multicolumn{2}{c}{} \\

M$_{dust}$ (\msun)	& 6.2 $\times$ 10$^6$ &  1.7 $\times$ 10$^7$ ($\pm$ 40$\%$)       & 3 $\times$ 10$^7$  &  {\bf{5.12 $\times$ 10$^7$}} ($\pm$ 20$\%$)      &  2.9 $\times$ 10$^4$   & 1.3 $\times$ 10$^6$ ($\pm$ 50$\%$) \\
 
f$_{PAH}$   & 0.22   & 0.1      & 0.18 &  0.18 &  0.11 & 4.4 $\times$ 10$^{-2}$\\

$\alpha$ 	  &  1.88 	 &  1.73	& 2.21&  2.18    & 2.5     &1.84 \\

U$_{min}$      & 12.2    &  1.44   &  1.5	    &  0.82 & 3.68      & 9 $\times$ 10$^{-3}$\\

U$_{max}$     & 1.3 $\times$ 10$^5$      & 5.77 $\times$ 10$^4$    &  3.84 $\times$ 10$^5$ 	 &  3.5 $\times$ 10$^5$    &  1.47 $\times$ 10$^4$    & 8.67 $\times$ 10$^2$\\

 & \multicolumn{2}{c|}{} & \multicolumn{2}{c|}{} & \multicolumn{2}{c}{} \\
\hline
& \multicolumn{2}{c|}{}  & \multicolumn{2}{c|}{} & \multicolumn{2}{c}{} \\

M$_{HI}$  (\msun) & \multicolumn{2}{c|}{$\sim$ 10$^{8}$} &  \multicolumn{2}{c|}{2.7 $\times$ 10$^{10}$} & \multicolumn{2}{c}{4.1 $\times$ 10$^{7}$ $^*$} \\

M$_{dust}$/M$_{HI}$  & 6$\times$10$^{-2}$ & 0.17 & 10$^{-3}$ & {\bf 1.9$\times$10$^{-3}$} & 7$\times$10$^{-4}$ & 3.2$\times$10$^{-2}$\\

& \multicolumn{2}{c|}{}  & \multicolumn{2}{c|}{} & \multicolumn{2}{c}{} \\
\hline
\end{tabular}
 \caption{Model parameter results derived from our fiducial SED modelling of Haro~11, Mrk~1089 and NGC~1705 and dust-to-gas mass ratios implied by the dust mass. M$_{dust}$ represents the total mass of dust of the galaxy in solar mass, f$_{PAH}$ the PAH-to-total dust mass ratio, $\alpha$ the power law index describing the illuminating radiation fields, U$_{min}$ and U$_{max}$ the minimum and maximum heating intensities. Finally, M$_{HI}$ represents the HI mass and M$_{dust}$/M$_{HI}$ the dust-to-HI mass ratio. Bold numbers are the dust mass and the dust-to-gas mass ratio values of our prefered SED for the galaxy Mrk~1089. For Haro~11 and NGC~1705 where the model does not fit correctly our observed constraints, a cold 10 K dust component is added (see $\S$ 6.2).  
  \newline
$^*$  For NGC~1705, we use the HI mass associated with the aperture chosen for the dust mass determination to calculate the M$_{dust}$/M$_{HI}$. The HI  mass reported here is thus 80\% of the the total HI mass (5.1 $\times$ 10$^7$\msun) of the entire extension of HI.}
 
  \label{Tableparameters}
 \end{table*}

\begin{table*}
 \centering
 \begin{tabular}{c|c|c|c|c|c|c}
\hline
\hline
&&&&&&\\
& f$_{160}$ & 0.9 f$_{160}$ & 0.8 f$_{160}$ & 0.7 f$_{160}$ & 0.6 f$_{160}$ & 0.5 f$_{160}$\\
&&&&&&\\
\hline
&&&&&&\\
M$_{dust}$ (\msun) & 5.2 $\times$ 10$^6$ & 5.5 $\times$ 10$^6$ & 6.1 $\times$ 10$^6$ & 7.0 $\times$ 10$^6$ & 8.4 $\times$ 10$^6$ & 1.1 $\times$ 10$^7$ \\
f$_{PAH}$ & 0.34 & 0.38 & 0.39 & 0.4 & 0.41 & 0.42 \\
$\alpha$ & 2.13 & 2.08 & 2.00 & 1.93 & 1.86 & 1.79  \\
U$_{min}$ & 0.25 & 0.19 & 0.12 & 0.07 & 0.04 & 0.02 \\
U$_{max}$ & 1.6 $\times$ 10$^5$ & 5.0 $\times$ 10$^4$ & 1.8 $\times$ 10$^4$ & 10$^4$ & 6.7 $\times$ 10$^3$ & 4.9 $\times$ 10$^3$ \\
&&&&&&\\
\hline
\end{tabular}
 \caption{Model parameter results derived from our fiducial SED modelling for the UM311 system. We perform the modelling using the upper limit of the 160 \mic\ flux and decrease it in steps of 10$\%$ down to 50$\%$ to study the evolution of the model parameters. These parameters are: M$_{dust}$ the total mass of dust of the galaxy, f$_{PAH}$ the PAH-to-total dust mass ratio, $\alpha$ the power law index describing the illuminating radiation fields, U$_{min}$ and U$_{max}$ the minimum and maximum heating intensities.} 
 \label{Tableparameters_UM311}
 \end{table*}

The SED models are presented in Fig.~\ref{SEDs}, along with the observations. Haro~11 observations are displayed corrected to rest wavelengths taking into account its redshift. The Haro~11 IRS spectrum, from 5.1 \mic\ to 37.2 \mic, provides additional constraints to better describe the PAH properties as well as the slope of the rising MIR continuum. We perform SED models using the 870 \mic\ observational constraint, and then compare to the resulting parameters when we remove the 870 \mic\ data  to study the influence of the submm data on the parameters of the modelling, especially the dust mass of the galaxy (Table~\ref{Tableparameters}). To quantify the errors on our dust mass determinations, we produced a grid of 500 randomly modified observational constraints, allowing the observational fluxes to vary within their error bars, following a Gaussian distribution around their reference value. Before entering into the details of each galaxy, we can already notice that the f$_{PAH}$, which is normalized to the Galactic value, is always inferior to 1, which is consistent with the fact that low-metallicity galaxies usually show a PAH deficit compared to dustier galaxies ~\citep{Madden2005, Engelbracht2005,Engelbracht2008, Wu2007}. For NGC~1705, 11.3 \mic\ PAH emission was nevertherless detected with IRS spectroscopy in the most luminous of the two dust emission peaks (D1) but not in the second off-nuclear region nor in the SSC ~\citep{Cannon_NGC1705_2006}, so most of the PAHs should reside in D1. Finally, $\alpha$ is quite similar for the three galaxies, with Haro~11 having the smaller $\alpha$. ~\citet{Dale2005, Dale2007}  found that $\alpha$ was typically ~2.4 for SINGS galaxies while \citet{Draine2007} fixed it at 2 in their SED modelling. Our values are smaller than 2 for NGC~1705 and Haro~11. Note that {\revisedbis for a given U$_{max}$}, galaxies with smaller $\alpha$ usually contain more strongly heated regions and therefore are more likely to be associated with intense star formation activity.

 {\it Mrk 1089 -} In spite of the lack of 160 \mic\ observations, we are able to fit the observations accurately both with and without the \lab\ observations.  Notice how flat the peak of the SED is in the FIR (Fig.~\ref{SEDs}).  The slope of SINGS galaxies is positive between 60 and 100 \mic\  whereas the slopes of the dwarf galaxies here (Mrk~1089 but also in NGC~1705 and Haro~11) are flat or descending in that wavelength range. This rather flat form of the SED is not characteristic of our Galaxy nor is it common in other more metal rich galaxies, but is often seen in the SEDs of active low metallicity galaxies, such as IIZw40 and He2-10 \citep{Galliano2003, Galliano2005}, in the Large Magellanic Cloud \citep{Bernard2008} or in the SEDs of \citet{Dale2007}. There are two possible explanations for this: the lower metal abundance and the decrease in dust attenuation result in a higher overall interstellar radiation field and/or the increased abundance of very small stochastically-heated grains can inflate the continuum on the Wien side of the SED. We find a total dust mass of 5.12 $\times$ 10$^7$ \msun\ ($\pm$ 20$\%$). This mass is a factor of 1.7 larger than that we would calculate without submm constraints. This galaxy is the least affected by the submm constraint. Nevertherless, this increase is superior to the error bars estimated for this galaxy.

 {\it The UM 311 system -} The 160 \mic\ beam is too large to isolate the UM~311 system alone and, without this data, we lack sufficient observational constraints at FIR wavelengths to accurately determine the shape of the SED peak for the system. Instead we measure the 160 \mic\ flux in a bigger aperture (80") than the one chosen to measure the fluxes in the other bands (54'') and use the 160 \mic\ flux as an upper limit. To explore the range of parameter space constrained by the  160 \mic\ upper limit, we decrease the 160 \mic\ flux in steps of 10$\%$ down to 50$\%$ and find the range of shapes with associated parameters that could fit our observational constraints (see Fig.~\ref{SEDs} and Table~\ref{Tableparameters_UM311}). We also produced a grid of 200 randomly modified observational constraints to study the spread of the SED models induced by these uncertainties. These tests enable us to establish a range of possible dust masses for this galaxy system. We conclude that the dust mass of the system should reside between 5.2 $\times$ 10$^6$ and 1.1 $\times$ 10$^7$ \msun, and $\alpha$ should reside between 1.79 and 2.13 (Table~\ref{Tableparameters_UM311}). The various SED models obtained include both solutions that maximise the contribution from the hot dust with a FIR peak at about 50 \mic\  and solutions that, on the contrary, favor a large dust mass, peaking at longer wavelengths. 

We convolve the 8 \mic\ IRAC images of the UM~311 system to the resolution of MIPS 24 \mic\ and calculate the luminosities of the 3 different substructures at these wavelengths. The luminosities in $\nu$ L$_{\nu}$ of region 1 and 2 are respectively 1.5 and 2.9 times lower than that of UM~311 (region 3) at 8 \mic\ and respectively 2.6 and 8.4 times lower at 24 \mic. Thus, we obtain different 8/24 luminosity ratios for the 3 different regions, respectively 0.36 and 0.61 for region 1 and 2, and 0.21 for UM~311 which compares with the value of 0.24 obtained for the whole UM~311 system. These ratios imply that region 2, which has the lowest 24 \mic\ flux, may be relatively less active than the others, if we use the 24 \mic\ flux as a good indicator of the star formation activity of the region \citep{Helou2004,Relano2007}. On the contrary, UM~311, having the highest 24 \mic\ continuum should present the strongest star formation activity of the system. Moreover, even if the system is not clearly resolved at 70 \mic, its emission clearly peaks at the location of UM~311. This leads to the conclusion that the galaxy UM~311 should dominate the system at MIR and FIR wavelengths and the global dust SED of the UM~311 system. The galaxy should account for the major part of the hot dust mass of the system. If more distant, the 3 objects would blend, like the 3 distinct nuclei of Haro~11 which can be resolved by HST but not resolved with \spitz.
 
   {\it NGC 1705 -}  Fig.~\ref{SEDs} presents the fit which gives the lowest $\chi$$^2$ for this galaxy. The model fails to fit the \iras\ 100 \mic\ data when we introduce the LABOCA constraint but does fit this constraint when the SED is modeled with constraints up to 160 \mic. We test the influence of the \iras\ 100 \mic\ flux on the fit performing SED models with and without this data point but with the submm constraint. The global shape of the SED does not change and the dust masses derived in these 2 cases vary by less than 5\%. \citet{Cannon_NGC1705_2006}  estimated the dust mass of NGC~1705 using the models of ~\citet{Dale_Helou_2002} and ~\citet{Li_Draine_2001}, obtaining respectively (3.8 $\pm$ 1.9) $\times$ 10$^5$ \msun\ and 7 $\times$ 10$^4$ \msun\ ($\pm$ 50$\%$). The model of  ~\citet{Li_Draine_2001} combines laboratory studies and astronomical observations and fits the MIPS fluxes very well. The dust mass deduced from the ~\citet{Li_Draine_2001} model, accounting for its 50$\%$ uncertainty and the different distance they used (5.1 Mpc), compares well with our results obtained without the submm constraint: 3 $\times$ 10$^4$ \msun\ (Table~\ref{Tableparameters}). 
Including the submm \lab\ constraint, we obtain a dust mass of 1.3 $\times$ 10$^6$ \msun, which is 50 times higher than our results without the submm data. In any event, our fiducial model, even using the 870 \mic\ observations, is not satisfactory and gives a poor $\bar{\chi}$$^2$ value. For this galaxy, the uncertainty in the mass estimate obtained using the submm constraint is mostly due to the flattening FIR peak and to the elevated 870 \mic\ emission, for which the model has difficulty to find a consistent solution. Considering that the error in the dust mass reaches more than 50$\%$, we consider this first dust mass estimate obtained with our fiducial model uncertain and not satisfactory for this galaxy.
  
  {\it Haro 11 -} The SED of Haro~11 is very striking in that it peaks at very short wavelengths - 36 \mic, highlighting the extreme nature of the young starburst. This is also evidenced by the high global value of the Ne[III]/[NeII] ratio (ratio $>$ 1), already seen in other low metallicity dwarf galaxies \citep{Madden2006}, which is normally an indication of the hard interstellar radiation field dominated by a young ($<$ 5 Myr) stellar population. The peak of the IR SED compares to that of the 1/40\zsun\ galaxy SBS0335-052~\citep{Houck2004} or IC10 NW {\revisedbis (Parkin et al. 2009 in prep)}. The IRS spectroscopy provides a tight constraint on the slope of the hot MIR dust continuum emission as well as for the f$_{PAH+}$. The model does not clearly fit all the details of the IRS spectrum due to the lack of sophistication and flexibility in the dust properties. More complex modelling would be required to perfectly model the different features of the spectrum but this does not affect the following conclusions on the dust mass parameter. \citet{Engelbracht2008} previously estimated the total dust mass of the galaxy to be 6.2 $\times$ 10$^{6}$ \msun\ using \spitz\ observations and a distance of 87 Mpc. This value is consistent with the dust mass we obtain without submm constraints for a distance of 92 Mpc: 7.2 $\times$ 10$^6$ \msun. Note that when the IRS spectrum is not used for the modelling but only IRAC broabands, the dust mass is estimated to be 2 $\times$ 10$^{8}$ {\revisedbis \msun} with a poor-fit ($\bar{\chi}$$^2$ = 24). Thus the IRS spectrum greatly influences and constrains the global shape around the peak of the SED.
  
The large uncertainty in the dust mass estimated from the modeled SED (the error reaches 40 $\%$ for this galaxy) is due to the excess emission at submm wavelengths that the model does not fit (Fig.~\ref{SEDs}). Such submm excess was already observed in other galaxies using submm instruments such as SCUBA on the James Clerk Maxwell Telescope (\jcmt) \citep{Galliano2003, Galliano2005, Bendo2006, Marleau2006} and usually leads to different interpretations. \citet{Galliano2003, Galliano2005} proposed that the submillimeter excess in the dwarf galaxies that they observed could originate from 10 K dust. We decide to explore this hypotesis.
Our fiducial model does not seem to be sufficient to explain the fluxes we obtained at submm wavelengths for the two galaxies NGC~1705 and Haro~11. Two separate thermal dust models may be more successful to describe the 60-160 \mic\ and the 870 \mic\ emission separately for these two galaxies.


\begin{figure*}
    \centering
    \begin{tabular}{ m{8cm} m{8cm} }
      \includegraphics[width=9cm ,height=6cm]{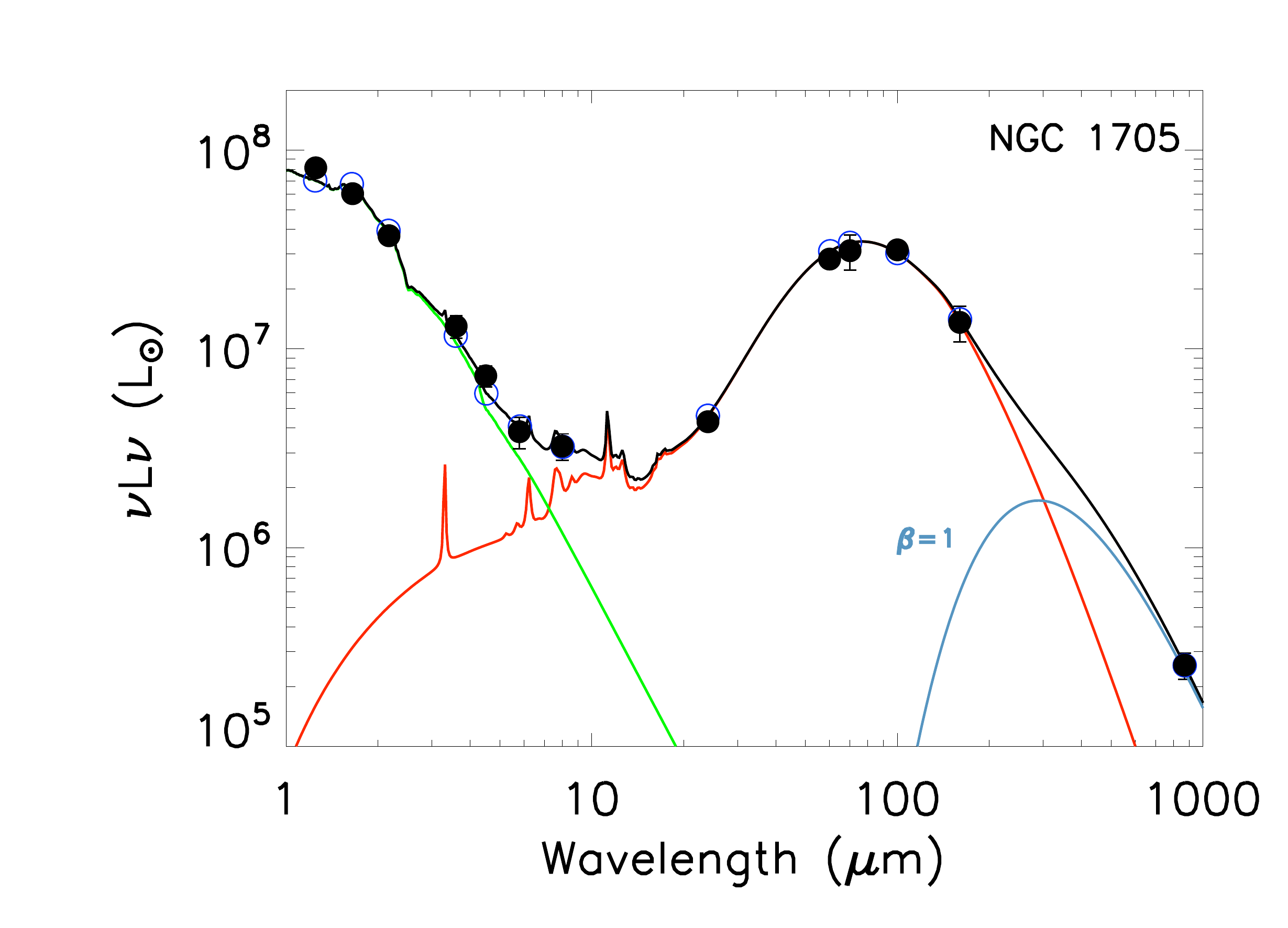}  &
      \includegraphics[width=9cm ,height=6cm]{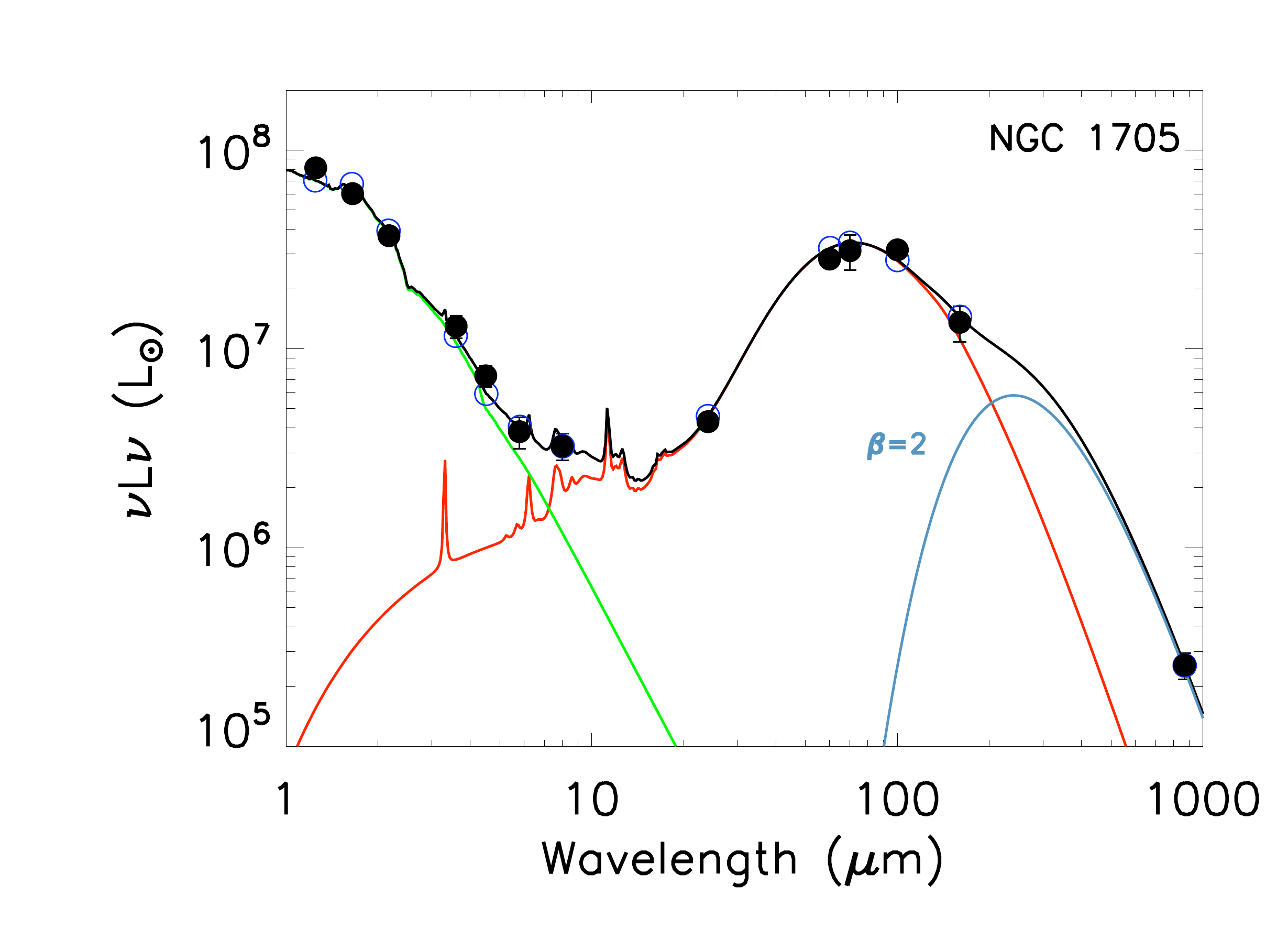}\\
      \includegraphics[width=9cm ,height=6cm]{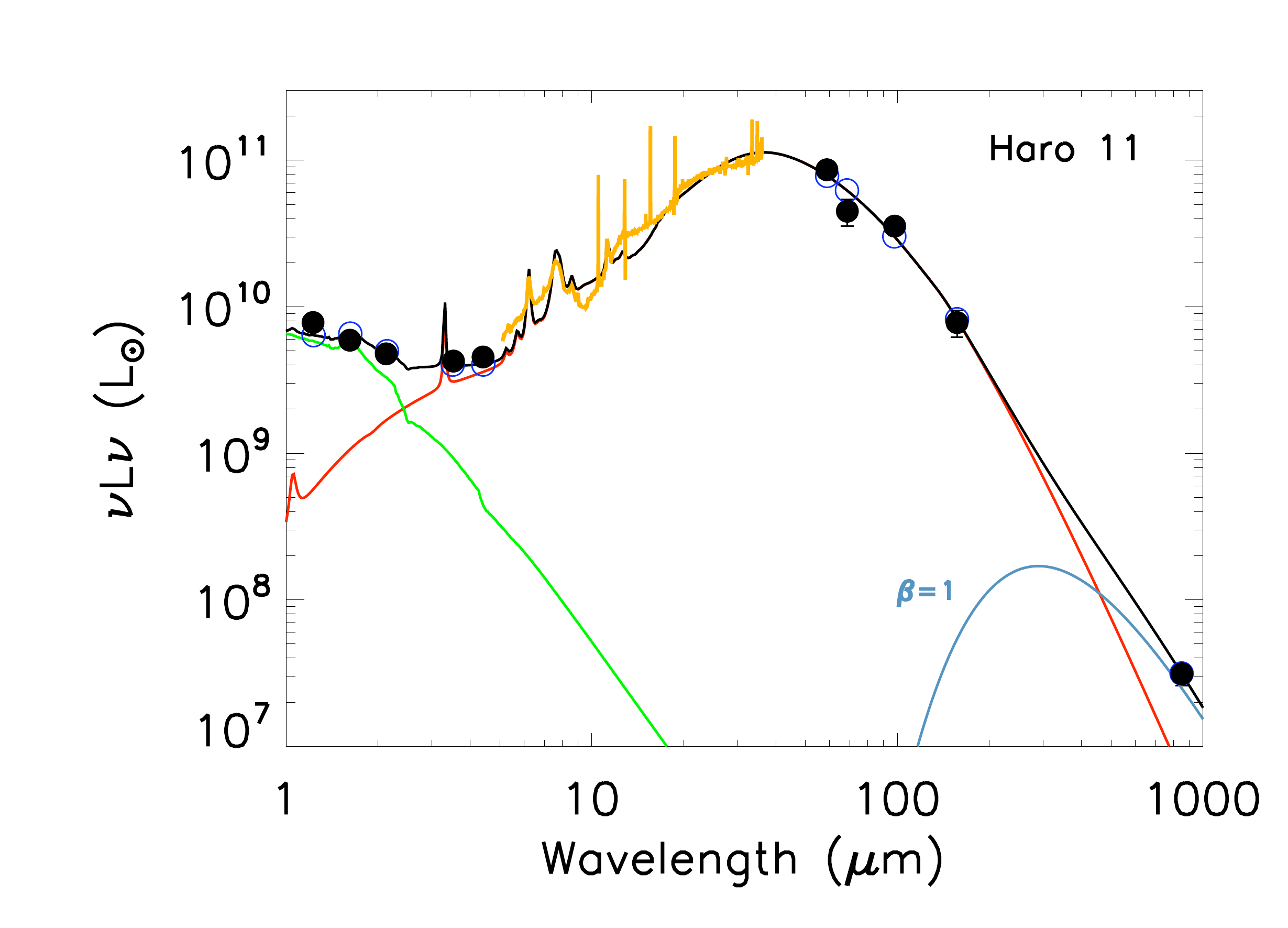} &
       \includegraphics[width=9cm ,height=6cm]{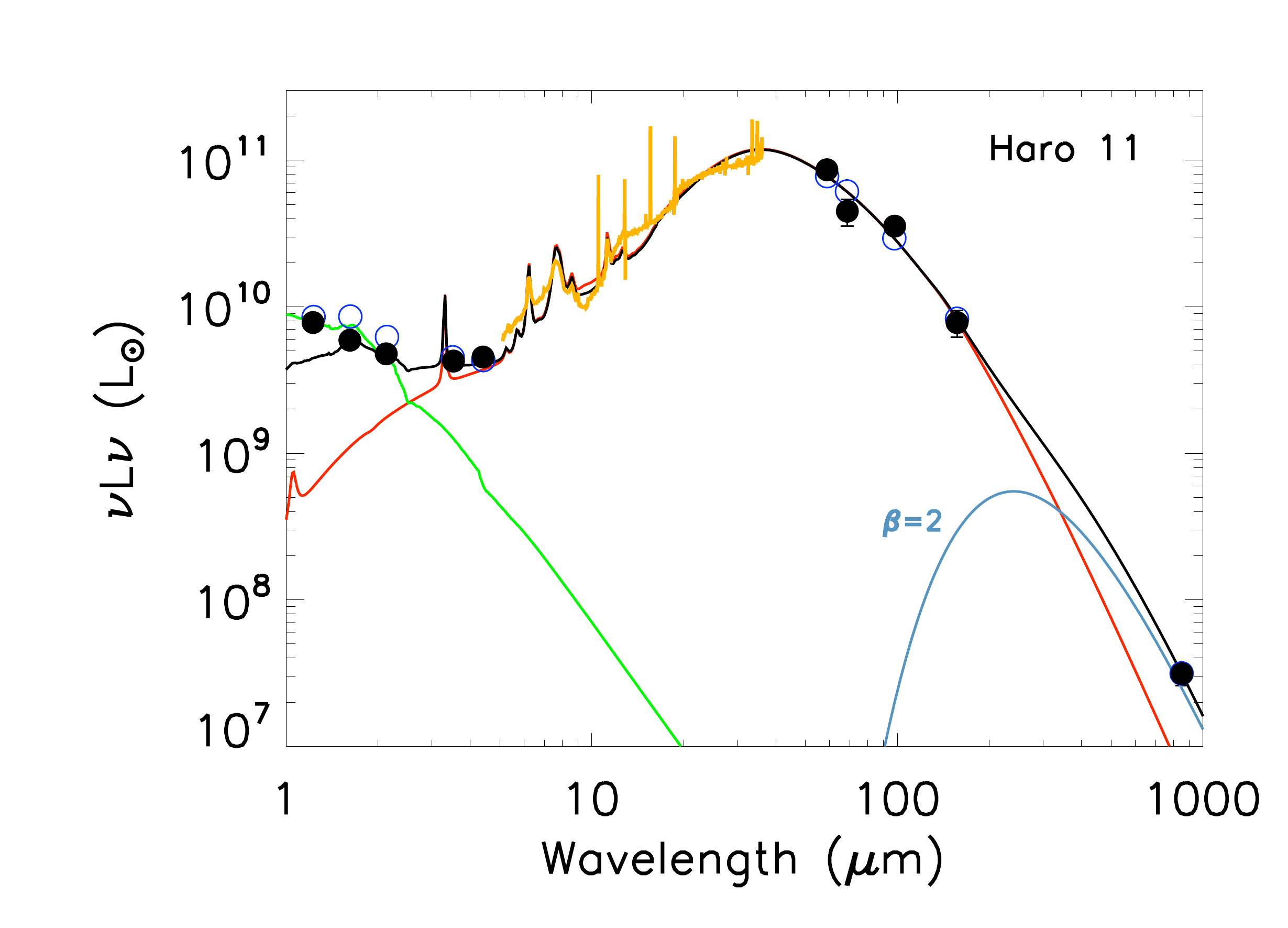} \\
         \end{tabular}
    \caption{SED models of NGC~1705 and Haro~11, adding a cold dust component of 10K with an emissivity coefficient $\beta$ of 1 and 2 to our fiducial model. Observational constraints are superimposed with filled circles. The IRS MIR spectrum is overlaid in orange for Haro~11. The green, red and blue lines represent the contributions of stars, warm and cold dust respectively. The red lines is close to the model generated when submm is omitted in the SED fit. The new models fit well the FIR peak and the 870 \mic\ flux.}
    \label{SEDs_2}
\end{figure*}

\begin{table*}
 \centering
 \begin{tabular}{c|ccc|ccc}
\hline
\hline
 &  \multicolumn{3}{c|}{NGC~1705}& \multicolumn{3}{c}{Haro~11}  \\

& Fiducial modelling & \multicolumn{2}{c|} {Cold dust component} &   Fiducial modelling & \multicolumn{2}{c} {Cold dust component}  \\
& &  $\beta$=1 &   $\beta$=2  &  &    $\beta$=1  &  $\beta$=2      \\
\hline
  & \multicolumn{3}{c|}{} & \multicolumn{3}{c}{}  \\
  
  $\bar{\chi}$$^2$ & 15.7 &  2.4   &   3.3	& 17.2  &  11.3  &  11.5        \\
  
    & \multicolumn{3}{c|}{} & \multicolumn{3}{c}{}  \\

M$_{dust}$ {\footnotesize (\msun)} & 1.3 $\times$ 10$^6$ ($\pm$ 50$\%$)  &  {\bf{1.7 $\times$ 10$^5$} } ($\pm$ 28$\%$)  &  1.3 $\times$ 10$^6$ ($\pm$ 21$\%$) & 1.7 $\times$ 10$^7$ ($\pm$ 40$\%$) &   {\bf{2 $\times$ 10$^7$}} ($\pm$ 17$\%$)&  1.3 $\times$ 10$^8$ ($\pm$ 17$\%$)        \\

  & \multicolumn{3}{c|}{} & \multicolumn{3}{c}{}  \\

M$_{dust}$/M$_{HI}$  & 3.2 $\times$ 10$^{-2}$ &  {\bf4.1 $\times$ 10$^{-3}$} &  2.8 $\times$ 10$^{-2}$ 	& 0.17 & {\bf0.2}  & 1.3   \\

  & \multicolumn{3}{c|}{} & \multicolumn{3}{c}{}  \\

M$_{PAH}$/M$_{dust}$  & 2 $\times$ 10$^{-3}$ &  8.1 $\times$ 10$^{-4}$ &  8 $\times$ 10$^{-5}$ & 1.04 $\times$ 10$^{-2}$ & 3.1 $\times$ 10$^{-3}$  & 6.2 $\times$ 10$^{-4}$    \\

  & \multicolumn{3}{c|}{} & \multicolumn{3}{c}{}  \\
  
\hline
\end{tabular}
\caption{Model parameter results for the SED model introducing a cold dust component at 10K and an emissivity $\beta$ of 1 and 2 to our fiducial model for the galaxies Haro~11 and NGC~1705. Results are compared to the ones of the fiducial modelling. M$_{dust}$ represents the total mass of dust of the galaxy. This mass includes the cold (10K) dust mass for Haro~11 and NGC~1705. M$_{dust}$/M$_{HI}$ represents the dust-to-HI mass ratio and  f$_{PAH}$ the PAH-to-total dust mass ratio. Bold numbers are the dust mass and D/G values  of our prefered SED models.}
  \label{Tableparameters2}
 \end{table*} 
    

\subsection{Refinement: adding a cold dust component}
 
In an attempt to improve the model results, we refine our fiducial SED model for the two galaxies Haro~11 and NGC~1705 by adding a thermal cold dust component with a flux, S$_\nu$, characterised by a modified blackbody:
\begin{equation}
S_{\nu} \propto  \nu^{\beta} B_{\nu}(T_{d}) 
\end{equation}
\noindent where $B_{\nu}$ is the Planck function, T$_{d}$ is the dust temperature and the emissivity coefficient $\beta$. {\revisedbis As we do not have information on the properties of these cold grains, we choose to describe them by extrapolating the properties of graphites grains and assume an emissivity $\beta$ of 1 or 2. We adopted the values $\rho$=2.5 $\times$ 10$^3$ kg m$^{-3}$,  $\lambda$$_0$=100 \mic\ and Q$_0$/a$_0$=150 cm$^{-1}$ for the mass density , the reference wavelength and the absorption coeffcient at $\lambda$$_0$ of the grains respectively. }We note that the absorption opacity chosen may slightly deviate away from the ~\citet{Li_Draine_2001} value when the emissivity is 1 since they usually assume that $\beta$=2. We nevertherless assume that their absorption opacity will still be applicable for $\beta$=1.

Indeed, there is a bend in the SED near 450 \mic. Due to the physical parameters of the grains behaving like a modified black body in the submm, having a natural bend on the Rayleigh-Jeans slope seems complicated unless you suddenly change the emissivity of the grains at 450 \mic. In $\S$ 6.2, we discuss other possible explanations beyond the cold dust hypothesis.

Since the supplemental black body component is only constrained by the 870 \mic\ flux, we cannot solve for the temperature and $\beta$ independently. For this exercise, we fix the temperature to 10K and fit the SED using a  $\beta$ = 1 or 2 to test the hypothesis of a submm excess of cold dust \citep{Galliano2003, Galliano2005, Marleau2006}. The new SED produced is obtained by adding the fiducial SED model and the cold blackbody. Chi-square values are estimated from the deviations of the new SED model to the observational constraints. In this revised model, the warm dust component is described by the observational constraints covering the FIR wavelengths to 160 \mic\ while the cold component is constrained by the 870 \mic\ observations.

The dust masses derived from each SED model for Haro~11 and NGC~1705 using $\beta$=1 and 2, as well as the dust masses derived from the fiducial models, are given in Table~\ref{Tableparameters2}.  Including a 10 K dust component, we find dust masses constrained by $\beta$ = 2 to be 6 and 8 times greater than those with  $\beta$ =1 (Table~\ref{Tableparameters2}) for Haro~11 and NGC~1705 respectively. Indeed, since a 10K blackbody modified with a $\lambda$$^{-1}$ emissivity is flatter on the millimetre tail, the model fits the data with less mass than when a steeper emissivity coefficient of 2 is used. For Haro~11, the best-fit model prefers $\beta$=1, giving a total mass of dust (warm and cold) of 2 $\times$ 10$^7$ \msun, although the success in using $\beta$=1 or 2 is not very different in terms of the reduced chi-square ($\bar{\chi}$$^2$). For both galaxies, the $\bar{\chi}$$^2$ values for the models including a cold component indicate better fits than the fiducial models. The SED models including a cold dust component are presented in Fig~\ref{SEDs_2}. Comparison by eye with the models presented in Fig.~ \ref{SEDs} show obvious improvements in the model fits to the observations. The value of $\beta$ significantly influences the dust masses derived from the SEDs and thus the dust-to-gas mass ratio (D/G) estimate of the galaxies (see section $\S5.3$). We also note that the abundance of PAHs relative to the total dust mass of these galaxies is much lower than the Galactic value of 4.6 $\times$ 10$^{-2}$: a factor of 15 lower for Haro~11 and a factor of 50 times lower for NGC~1705 (Table~\ref{Tableparameters2}). 

For Haro~11, increasing the temperature of the cold dust with $\beta$=1 does not significantly influence the ${\chi}$$^2$ value until the temperature of the cold dust reaches $\sim$20K. For NGC~1705, when we try to increase the temperature of the cold dust component, the 160 \mic\ observations becomes difficult to fit and ${\chi}$$^2$ values increase. Thus, at least for NGC~1705, the very cold dust 10K dust provides a better fit, while for Haro~11, somewhat higher temperatures could also work. However, we need more observational constraints to pin down the precise temperature of the cold dust component.

As mentionned previously, \citet{Galliano2003, Galliano2005} proposed that the submillimeter excess in the dwarf galaxies that they observed could originate from 10 K dust with a dust emissivity index $\beta$=1 and found that dust would account for 40-80$\%$ of the total dust mass in each of these galaxies. We choose to follow the same assumption but other explanations can be proposed. \citet{Lisenfeld2001} proposed that the submillimeter excess would originate from hot ($\sim$100 K) dust with a dust emissivity index $\beta$=1 and the temperature fluctuations of very small grains. \citet{Bendo2006} studied NGC~4631 at 450 and 850 \mic\ with SCUBA and found that the 850-1230 \mic\ emission exceeds what would be expected from thermal emission but the scenario of a very cold dust component which might explain this excess was rejected due to the high dust-to-gas mass ratio it would imply. In fact, submm excess is not always detected in low-metallicity galaxies. {\revisedbis Parkin et al. 2009 (in prep)} investigate the very well sampled SED of the low-metallicity galaxy of the Local Group IC10, using \spitz\ and ISOCAM observations combined with JCMT/SCUBA observations. They model the SED of the 2 main SF regions only and found no submm excess. The observations, however, do not cover the entire galaxy, preventing analysis on the global scale. \citet{Draine2007} present SED models of a large sample of 65 galaxies. For 17 galaxies for which they have SCUBA data at 850 \mic, they fit SED models with and without this submm constraint. Five of their 17 galaxies show an increase in the dust mass when calculated using SCUBA constraints whereas 5 other galaxies show a decrease of this dust mass. In fact, their dust masses obtained with and without SCUBA data agree to within a factor of 1.5 for 11/17 cases and to within a factor of 2.2 for all cases. They concluded that their dust models do not require cold ($\le$10K) dust to account for their submm fluxes. However, they caution that some of the SCUBA galaxies are not completely mapped or are not taken in scan map mode. The difficult data processing might have oversubstracted diffuse, extended emission for these galaxies. Thus assumptions they make for complet submm flux values may require caution in using these data. Their sample studied with submm observations contains only metal-rich galaxies that may show different dust properties and dust temperature distributions than the low-metallicity galaxies we present here. Further studies would be necessary to study the presence of submm excess and the influence of submm constraints on the SED model with respect to the metallicity of the galaxy.

Instead of a cold dust component fitting the submm excess, we also test the formalism presented in \citet{Draine2007} and try to fit the excess with an extra ISM component heated at U=U$_{min}$ for our two galaxies. This addition, in both cases, leads to a very cold component with U$_{min}$$<$10$^{-2}$ due to the fact that the extra component tends to fit the submm excess. The masses of dust derived from the SED models in this case are  3.3 $\times$ 10$^6$ \msun\ for NGC~1705 and 1.2 $\times$ 10$^8$ \msun\ for Haro~11, thus 19 and 6 times higher, respectively, than the mass of dust derived from a SED model using a cold blackbody of 10K with $\beta$=1, unrealistic values due to the D/G mass ratios they imply. 

Additionally, a $\beta$=1.5, which is closer to amorphous carbons \citep{Rouleau_Martin_1991}, might give a better fit and reduce the cold dust mass but it would also imply that the FIR-mm dust mass was dominated by carbons and not silicates , which is contrary to our usual relative dust mass concepts.

Other assumptions can be invoked to account for the submm excess. For example, underestimating the contributions to the 870 \mic\ flux values could have important consequences on the values of the deduced dust mass. Direct measurements of the CO(3-2) line and more mm radio observations would place greater confidence in the dust mass determinations. The submm excess could also be explained as a change in dust emissivity in lieu of the cold dust hypothesis. Models of \citet{Meny2007} modify the dust optical properties to find an effective decrease in the submm emissivity index as the dust temperature increases as suggested by the observations of \citet{Dupac2003}.  However, \citet{Shetty2009} express caution in the inverse temperature - $\beta$ interpretation, showing that flux uncertainties, especially in the Rayleigh-Jeans regime, can affect the results for the SED fits as far as temperature and emissivity are concerned. Although we do not reject the possibility of a modification of the properties of dust with temperature, here we explore the cold dust hypothesis, since it enables us to investigate its consequences on the global properties of the galaxies.

\subsection{Dust-to-gas mass ratios}

We calculate the D/G of our galaxies and compare them to estimates from the chemical evolution model of \citet{Galliano_Dwek_Chanial_2008}. Values are presented in Table~\ref{Tableparameters} for Mrk~1089, in Table~\ref{Tableparameters2} for Haro~11 and NGC~1705. Since our SED for UM~311 contains 3 individual targets, we do not consider it here in the discussion of the D/G.

 {\it NGC 1705 -} This galaxy is surrounded by a very large HI envelope up to 10 times larger than the optical extension (see Fig.~\ref{NGC1705_Halpha_HI}), which is more often the case for dwarf galaxies in comparison with spiral galaxies. From the HI image of NGC~1705, we estimate that only 80$\%$ of the total HI flux is contained in the aperture we choose to derive our dust mass. We then consider that only 80$\%$ of the HI mass, i.e 4.1 $\times$ 10$^7$ \msun, should be used for the D/G estimate. The D/G of NGC~1705 estimated with submm observations is $\sim$ 4.1 $\times$ 10$^{-3}$ with our required very cold dust component using a $\beta$ emissivity factor of 1 (Table~\ref{Tableparameters2}). This value compares to that expected by the chemical evolution model used in \citet{Galliano_Dwek_Chanial_2008} (2 $\times$ 10$^{-3}$ $<$ D/G $<$ 5 $\times$ 10$^{-3}$ ). The solution for $\beta$=2, obtained with a slightly higher $\bar{\chi}$$^2$, leads to a D/G of 2.8 $\times$ 10$^{-2}$, which is a factor of 8 higher than that of $\beta$=1. This high D/G value would be difficult to reconcile with chemical models. We prefer the model with a cold component using $\beta$=1 for NGC~1705. 

 {\it Haro 11 -} The D/G estimated with our SED model is $\sim$ 0.2 when a very cold dust component of $\beta$=1 is used (Table~\ref{Tableparameters2}). If we were to use an emissivity index of 2 for the very cold grains, the D/G derived is even higher and not physical considering the lack of metals in the galaxy. For this reason, in spite of the similar $\bar{\chi}$$^2$ values for the two different fits, we prefer the fit for $\beta$=1 for the cold dust component. 
 The HI mass used for our calculation is an upper limit of 10$^8$ \msun\ given in \citet{Bergvall2000}. \citet{Bergvall_Ostlin_2002} already note from their VLA observations that Haro~11 is dramatically deficient in neutral hydrogen.  For such a low metallicity (12+log(O/H) = 7.9), a D/G of $\sim$ 10$^{-3}$ should be more likely considering chemical evolution models. Our high values for the D/G, constrained by the new submm data, are not easily explained with current models.  For the same galaxy, \citet{Galliano_Dwek_Chanial_2008} estimated the D/G $\sim$ 1.6 $\times$ 10$^{-2}$ (without submm constraints) but this ratio was qualified as uncertain due to the poor HI measurement for this galaxy which could be linked to the difficulty of accurately measuring HI at such distances. Indeed, in \citet{Gordon_Gottesman_1981}, the mean HI mass for BCGs of the size of Haro~11 should be $\sim$ 8.1 $\times$ 10$^{8}$ \msun. If we consider that the existing HI observations provide a reliable upper limit  on the mass of atomic hydrogen \citep{Bergvall2000}, the large estimated dust mass may suggest that a significant fraction of the total gas mass is in forms other than atomic. A major fraction of its gas should be in molecular or ionised form. An upper limit of only $\sim$ 10$^{9}$ \msun\ was estimated from CO observations for the mass of molecular gas in this galaxy~\citep{Bergvall2000}. The total amount of observed molecular + atomic gas (all upper limits) still can not lead to a D/G of 10$^{-3}$ for Haro~11. In fact, CO can be a poor tracer of molecular gas in low metallicity galaxies due to the high excitation and the density of these environments and the small filling factor of the molecular clouds. Self-shielding can be extremely efficient for H$_2$ in regions where CO is photodissociated. Thus, there could be a non negligeable amount of molecular gas not traced by CO~\citep[e.g][]{Poglitsch1995,Madden1997}. In order to reach the D/G value expected by chemical models, a large molecular gas mass on the order of 10$^{10}$ \msun\ would be required to account for the "missing" gas mass, that is to say an order of magnitude more molecular gas than that deduced from current CO upper limits. Molecular hydrogen could be embedded in [CII] emiting envelopes as described in \citet{Madden1997} for the low-metallicity galaxy IC10. The high L$_{CII}$/L$_{CO}$ value ($\>$4 $\times$ 10$^5$) deduced by \citet{Bergvall2000} and also seen in other low metallicity galaxies \citep{Madden2000} is coherent with this theory.
 
  {\it Mrk 1089 -} Our SED model estimate of the D/G leads to a value of 1.9 $\times$ 10$^{-3}$. For a galaxy with the metallicity of Mrk~1089 ($\sim$ 12+log(O/H)= 8.0), the chemical evolution model used in \citet{Galliano_Dwek_Chanial_2008} which links the D/G with metallicity, leads to an estimated D/G of $\sim$ 10$^{-3}$, consistent with the value we obtain to within a factor of 2. 

To summarize, the SED results we will discuss in the following sections are those obtained with our fiducial model for Mrk~1089 (i.e., no additional modified black body) and the SED models which include a cold dust component of 10K with $\beta$=1 for NGC~1705 and Haro~11.

\subsection{Robustness of the results with the assumed radiation field}

The modelling scheme already presented in this paper, does not constrain the form of the global ISRF of the galaxies and the results should be dependent on the shape of the average radiation field to which the dust grains are exposed. The SED results presented here are obtained using the Galactic shape of the ISRF \citep{Mathis1983}, but this may not be accurate in our case for metal-poor galaxies. Modeled ISRFs of dwarf galaxies show that dwarf galaxies can certainly have harder global average ISRFs than that of the Galaxy  \citep{Madden2006} - an effect attributed to the lower dust attenuation in the low metallicity ISM and, as a consequence, the larger mean free path length of the ionising photons. As we are modelling galaxies on the global scale here, we test several ISRFs which have been determined for low metallicity galaxies to quantify this dependence. We use synthesised radiation fields of four dwarf galaxies as a reference: He2-10, IIZw40, NGC1140 and NGC1569. The process to synthesized the ISRFs uses the stellar evolution synthesis code PEGASE \citep{Fioc_Rocca_1997} and the photoionisation model CLOUDY \citep{Ferland1996}. We refer to \citet{Galliano2003,Galliano2005} for more explanations of this process and a description of the synthesized ISRFs of He2-10, IIZw40, NGC1140 and NGC1569. We furthermore produce a grid of dust models for the ISRF of a young and non-ionising cluster, created just after an instantaneous burst, with a Salpeter IMF. Even if the shapes of the resulting SEDs for our galaxies are somewhat modified using different forms of harder radiation fields, we find that the masses of dust derived from these different models and those determined using the Galactic ISRF, differ by less than 10$\%$,  which is within the error bars of the dust masses we estimate (Table~\ref{ISRF_dwarfs}).
Indeed, the shape of the radiation field essentially controls the emission of out-of-equilibrium grains. Increasing the hardness of the radiation field increases the maximum temperature that small grains reach when they fluctutate. Consequently, it produces a short wavelength excess of the grain spectrum as reflected in the resulting modified SEDs. However, this excess is compensated in the global model of $\S$ 5, by lowering the weight of the high intensity regions. Since these hot regions do not contribute significantly to the total dust mass, the latter does not strongly depend on the shape of the ISRF.

\begin{table*}
 \centering
 \begin{tabular}{c|ccc|ccc|cc}
\hline
\hline
Synthesized ISRF  &  \multicolumn{3}{c|}{NGC~1705 $^*$}  &\multicolumn{3}{c|}{Haro~11 $^*$} & \multicolumn{2}{c}{Mrk~1089}  \\
 \multicolumn{1}{c|}{ }  & M$_{warm}$ $_{dust}$ &  M$_{dust}$ &  $\bar{\chi}$$^2$ & M$_{warm}$ $_{dust}$ &  M$_{dust}$ &  $\bar{\chi}$$^2$ & M$_{dust}$ & $\bar{\chi}$$^2$ \\
\hline
 \multicolumn{1}{c|}{ }  & \multicolumn{3}{c|}{} & \multicolumn{3}{c|}{} & \multicolumn{2}{c}{ } \\

Milky Way				 & 2.6 $\times$10$^4$&   	   1.7 $\times$10$^5$	& 2	& 5.9 $\times$10$^6$ &		2 $\times$10$^7$	      & 13 	& 	5.2 $\times$10$^7$  & 5.7\\
Non-ionising cluster		 & 2.5 $\times$10$^4$&	   1.7 $\times$10$^5$  	& 5	& 6.5 $\times$10$^6$  &		2.1 $\times$10$^7$        & 25 	& 	5.0 $\times$10$^7$  	   & 21\\
He 2-10				 & 3.3 $\times$10$^4$&	   1.75 $\times$10$^5$ 	& 20	& 8.7 $\times$10$^6$  &		2.19 $\times$10$^7$      & 45 	& 	7.0 $\times$10$^7$  & 52\\
IIZw40				 & 2.8 $\times$10$^4$&	   1.71 $\times$10$^5$ 	& 12	& 8.7 $\times$10$^6$  &		2.24 $\times$10$^7$      & 36 	& 	4.9 $\times$10$^7$  & 41\\
NGC1140				& 2.8 $\times$10$^4$&	   1.72 $\times$10$^5$ 	& 12	& 8.1 $\times$10$^6$   &		2.2 $\times$10$^7$        & 38 	& 	4.7 $\times$10$^7$  & 46 \\
NGC1569				& 2.6 $\times$10$^4$&   	   1.69 $\times$10$^5$ 	& 2	& 6.6 $\times$10$^6$  &		2.02 $\times$10$^7$      & 14 	& 	5.0 $\times$10$^7$        & 16\\

 \multicolumn{1}{c|}{ }     & \multicolumn{3}{c|}{} & \multicolumn{3}{c|}{} & \multicolumn{2}{c}{ }\\
\hline
\end{tabular}
 \caption{Dust masses derived from SED models using different shapes of ISRFs ($\S$ 6.4): a synthesised ISRF of a non-ionising cluster (created after an instantaneous burst, Salpeter IMF) and the synthesised ISRFs of 4 low metallicity galaxies \citep{Galliano2003,Galliano2005, Madden2006}. 
 \newline
 $^*$ The models used for Haro~11 and NGC~1705 include an independent, very cold dust component at 10K with an emissivity $\beta$=1. M$_{warm}$ $_{dust}$ is the dust mass without including the 10K dust mass (but with cold dust included in the fit) and M$_{dust}$ is the total dust mass of the galaxy, 10K dust mass included.}
  \label{ISRF_dwarfs}
 \end{table*} 

\begin{table*}
 \centering
 \begin{tabular}{cc|ccc}
\hline
\hline
\multicolumn{2}{c|}{}   & NGC~1705 $^*$& Haro~11 $^*$  & Mrk~1089\\
\hline
\multicolumn{2}{c|}{}  & &  & \\
\multicolumn{2}{c|}{T$_{eq}$  $_{(max)}$ (K) }    & 88.2&  125       & 146\\
\multicolumn{2}{c|}{T$_{eq}$  $_{(min)}$ (K)}   & 10&  10      & 16.9  \\
\multicolumn{2}{c|}{ } & &  & \\
\multicolumn{2}{c|}{f$_{mass}$ ( [50 K ; 150 K] )}	 &   0.1 ${\%}$ 	&   0.4 ${\%}$ 		&   0.15 ${\%}$ \\
\multicolumn{2}{c|}{f$_{mass}$ ( [25 K ; 50 K] ) }	 &   7.2 ${\%}$	&   28.9 ${\%}$ 		&   9.5 ${\%}$  \\
\multicolumn{2}{c|}{f$_{mass}$ ( [15 K ; 25 K] ) }	 &   8.3 ${\%}$	&   0 ${\%}$ 	  	&   90.4 ${\%}$\\
\multicolumn{2}{c|}{f$_{mass}$ ( 10 K ) }	 		 &   84.4 ${\%}$&   70.4 ${\%}$   	&   0 ${\%}$\\
\multicolumn{2}{c|}{}  & &  & \\
\hline
\end{tabular}
\caption{Minimum and maximum equilibrium temperature and distribution of the dust mass with dust temperature ranges for the galaxies NGC~1705, Haro~11 and Mrk~1089. 
\newline
$^*$ The models used for Haro~11 and NGC~1705 include an independent, very cold dust component at 10K with an emissivity $\beta$=1. 
}
  \label{Fraction}
 \end{table*} 

\begin{table*}
 \centering
 \begin{tabular}{cc|ccc}
\hline
\hline
\multicolumn{2}{c|}{}   & NGC~1705 $^*$& Haro~11 $^*$  & Mrk~1089 \\
\hline
\multicolumn{2}{c|}{}  & &  & \\
\multicolumn{1}{c}{L$_{TIR}$ = $L_{[3\mu m,1100\mu m]}$ (\lsun)} &  {\small (1)}   &  5.8 $\times$ 10$^7$ &  1.7 $\times$ 10$^{11}$   & 3.9 $\times$ 10$^{10}$ \\ 
& {\small (2)}    & 4.9 $\times$ 10$^7$& 1.4 $\times$ 10$^{11}$  &   3.8 $\times$ 10$^{10}$\\
& {\small (3)}   & 5.9 $\times$ 10$^7$&  1.6 $\times$ 10$^{11}$   &    4.3 $\times$ 10$^{10}$ \\

\multicolumn{2}{c|}{}  & &  & \\

$L_{[3\mu m,50\mu m]}$ (\lsun) 	&   &   1.9 $\times$ 10$^{7}$	&  1.2 $\times$ 10$^{11}$   &  1.4 $\times$ 10$^{10}$\\
$L_{[50\mu m,100\mu m]}$ (\lsun) &	 & 2.2 $\times$ 10$^{7}$& 4.1 $\times$ 10$^{10}$    & 1.3 $\times$ 10$^{10}$  \\
$L_{[100\mu m,1100\mu m]}$ (\lsun) 	&  & 1.6 $\times$ 10$^{7}$&  9.6 $\times$ 10$^{9}$    & 1.1 $\times$ 10$^{10}$ \\
L$_{FIR}$  = $L_{[8\mu m,1100\mu m]}$ (\lsun) &  & 5.1 $\times$ 10$^{7}$&  1.6 $\times$ 10$^{11}$   & 3.7 $\times$ 10$^{10}$  \\

\multicolumn{2}{c|}{ }   &  &  &  \\
$L_{H\alpha}$ (erg s$^{-1}$) 	&   &   6.34 $\times$ 10$^{38}$ $^{(c)}$&  3.2 $\times$ 10$^{42}$ $^{(b)}$   &  4.57 $\times$ 10$^{41}$ $^{(a)}$\\
\multicolumn{2}{c|}{ }   &  &  &  \\
Size &&  0.5' $\times$ 0.5' & 0.25' $\times$ 0.25' & 0.31' $\times$ 0.12' \\
\multicolumn{2}{c|}{ }   &  &  &  \\
SFR (\msun\  yr$^{-1}$  kpc$^{-2}$) & (4)    & 1.6 $\times$ 10$^{-3}$ &5.1 $\times$ 10$^{-2}$  & 5 $\times$ 10$^{-2}$\\
 & (5)    & 9.4 $\times$ 10$^{-4}$& -  & 5.8 $\times$ 10$^{-2}$\\
 & (6)    & 5.2 $\times$ 10$^{-4}$& 2.4 $\times$ 10$^{-2}$  & 1.4 $\times$ 10$^{-2}$ \\
 & (7)    & 5.7 $\times$ 10$^{-4}$& 3 $\times$ 10$^{-2}$  & 1.8 $\times$ 10$^{-2}$\\

\multicolumn{2}{c|}{ }   &  &  &  \\
\hline

\end{tabular}
\caption{Luminosities, size and SFR of NGC~1705,  Haro~11 and Mrk~1089.
\newline
- Total infrared luminosities (L$_{TIR}$) are  estimated: 1) by integrating our SED models with a submm constraint from 3 to 1100 \mic, 2) from the \citet{Dale_Helou_2002} formula, 3) from the \citet{Draine_Li_2007} formula. We also give the distribution of the luminosities within different wavelength windows.  
\newline
- H$\alpha$ luminosities ($L_{H\alpha}$) are given in a)~\citet{Cannon_NGC1705_2006}, b) ~\citet{Bergvall2000} and c)~\citet{Iglesias-Paramo_Vilchez_1997}. 
\newline
- Star formation rates (SFR) are estimated: 4) using the relation of \citet{Kennicutt1998} based on L$_{FIR}$, 5) using the relation of \citet{Zhu2008} based on $\nu$ L$_{\nu}$[24 $\mu$m]. (This last relation is only valid for galaxies with a L$_{TIR}$ $<$ 10$^{11}$ L$_{\odot}$  and can not be used for Haro~11), 6) using the relation of \citet{Calzetti2007} based on the 24 \mic\ luminosity (L$_{24}$) alone, 7) using the relation of \citet{Calzetti2007} based on L$_{24}$ and L$_{H\alpha}$, .
\newline
$^*$ The models used for Haro~11 and NGC~1705 include an independent, very cold dust component at 10K with an emissivity $\beta$=1.
}
  \label{LTIR}
 \end{table*} 

 \subsection{Distribution of the dust temperatures}

To get an idea of how the dust mass is distributed as a function of the temperature, we calculate the fraction of the dust mass in several ranges of temperatures: above 50K (hot), between 25 and 50K, between 15 and 25K and a cold 10K component for the two galaxies NGC~1705 and Haro~11. We estimate the fraction of dust between two temperatures as the mass of dust exposed to a radiation field such that the large silicates, which are at thermal equilibrium, have temperatures between these two temperatures.  

The energy absorbed by a dust grain is given by :

\begin{equation}
\Gamma_{abs}\propto \int_{0}^{\infty} Q_{abs}(\lambda)U_{\lambda}(\lambda)d\lambda
 \end{equation}
 \noindent where Q$_{abs}$ is the absorption efficiency  and U the mean intensity of the radiation field.
 
 The energy emitted by a dust grain is given by :
 
 \begin{equation}
\Gamma_{em}\propto \int_{0}^{\infty} \nu^{\beta}B_{\nu}(T_{eq})d\nu \propto T_{eq}^{4+\beta}
 \end{equation}
  \noindent where B$_{\nu}$ is the Planck function and T$_{eq}$ the equilibrium temperature.
  
{\revisedbis As $\Gamma$$_{abs}$=$\Gamma$$_{em}$, we can equalize these two expressions. We thus obtain a relation of proportionnality between the radiation intensity U and the equilibrium temperature T$_{eq}$. Grains are assumed to possess an emissivity $\beta$=2:}
\begin{equation}
{\it U}(T_{eq})\propto\left(\frac{T_{eq}}{17.5}\right)^{4+{\beta}}\propto\left(\frac{T_{eq}}{17.5}\right)^6{}
 \end{equation}
Note that in this equation, the equilibrium temperature is normalised to the equilibrium dust temperature of the Galaxy of 17.5K \citep{Boulanger1996}. 

{\revisedbis Finally, from the prescription of \citet{Dale2001} (Eq.~\ref{Dale}), we can derive the heated dust mass associated to each radiation intensity U and thus to each equilibrium temperature. The fraction of the total mass exposed to radiation between temperatures T{\footnotesize 1} and T{\footnotesize 2} is given by:
\begin{equation}
f_{mass}([T_{1} , T_{2}]) = \frac{U_{2}^{1-\alpha}-U_{1}^{1-\alpha}}{U_{max}^{1-\alpha}-U_{min}^{1-\alpha}} =  \frac{T_{2}^{6(1-\alpha)}-T_{1}^{6(1-\alpha)}}{T_{max}^{6(1-\alpha)}-T_{min}^{6(1-\alpha)}}
 \end{equation}
 \noindent where U$_{min}$ and U$_{max}$ are the boundaries of the intensity range obtained from the SED modelling.}
 
For  Haro~11 and NGC~1705, the fraction of dust at 10K is given by the ratio between the dust mass of the cold dust component and the total dust mass of the galaxy . The results are summarized in Table~\ref{Fraction}. The cool phase ($\le$ 25K) constitues the major part of the dust in our galaxies, at least 70\% for Haro~11 and up to 90\% for Mrk~1089 and NGC~1705. The SED of Haro~11 peaks at usually short wavelengths ( 36 \mic), with a significant fraction ($\sim$ 30$\%$) of the dust mass at a temperature $>$ 25K while this same dust mass fraction does not exceed 10$\%$ for NGC~1705 and Mrk~1089. Thus their global SEDs reflect the different levels of SF activity (and/or morphologies) and the consequences on the dust heating.

\begin{figure*}

   \begin{tabular}{ p{0mm} m{8.5cm} }
     a) &  \includegraphics[width=16cm ,height=10.cm]{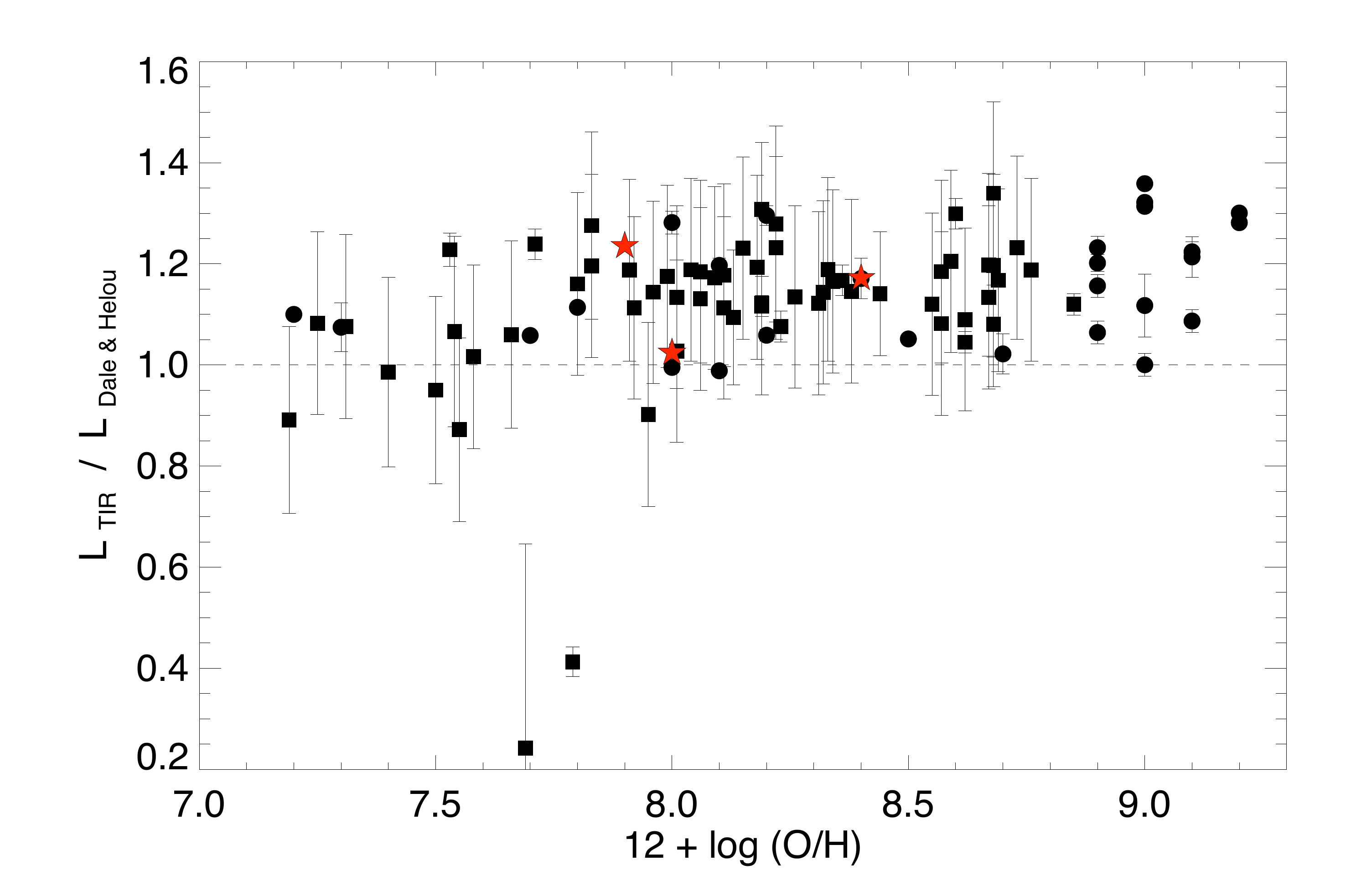} \\
     b) &  \includegraphics[width=16cm ,height=10.cm]{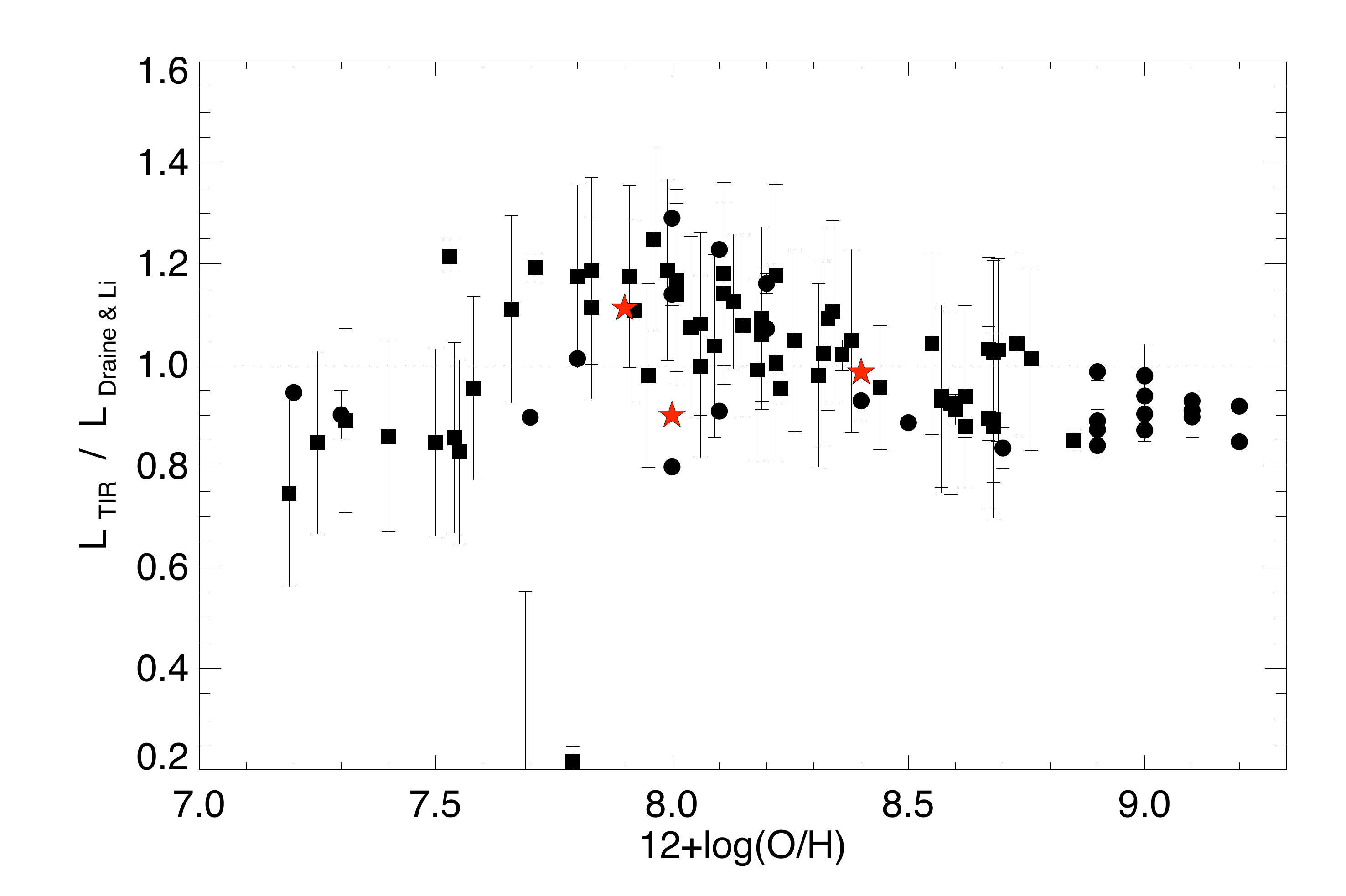} \\
         \end{tabular}
    \caption{{\it a)} Ratio between the total infrared luminosty (L$_{\footnotesize TIR}$) derived by integrating directly the SED models from 3 to 1100 \mic\ and L$_{TIR}$ given by the \citet{Dale_Helou_2002} formula. The squares indicate the \citet{Engelbracht2008} galaxies whereas the circles indicate the \citet{Galliano_Dwek_Chanial_2008} sample. Stars represent our three galaxies Haro~11, Mrk~1089 and NGC~1705 in metallicity increasing order .  {\it b)} Same as {\it a)} but performed with L$_{TIR}$ given by the \citet{Draine_Li_2007} formula.}
    \label{Dale_Draine}
\end{figure*}

\subsection{Total Infrared luminosity}

We calculate the total IR luminosity (L$_{TIR}$) for our galaxies by integrating our modeled SEDs from 3 \mic\ to 1100 \mic\ and compare our L$_{TIR}$ values with prescriptions in the literature using IR broadbands. For example, \citet{Dale_Helou_2002} have made phenomenological SED models to derive L$_{TIR}$ based  on \iras\, \iso\ and some 850 \mic\ data and provide a recipe to estimate the 3 \mic\ to 1100 \mic\ luminosity using the \spitz\ MIPS bands:

\begin{equation}
L_{TIR}=1.6~L_{24} + 0.77~L_{70} + 1.35~L_{160} 
\end{equation}

In the same fashion, \citet{Draine_Li_2007} expand this L$_{TIR}$ relation to include the IRAC 8 \mic\ flux. 
\begin{equation}
L_{TIR}=0.95~L_{8} + 1.15~L_{24} + L_{70} + L_{160} 
\end{equation}

We estimate the L$_{TIR}$ of our galaxies using these two formulas and compare to the L$_{TIR}$  we determine from our SED model. For the galaxy Mrk~1089, the 160 \mic\ observation is not available but the flux density at 160 \mic\ can be extrapolated from our SED models. We estimate this flux density to be 5334 ($\pm$ 266) mJy when using the submm constraint. For Haro~11 and NGC~1705, the L$_{TIR}$ was calculated from the revised models which include a cold dust component with an emissivity $\beta$=1. Results are presented in Table~\ref{LTIR}. For the UM~311 system, we derive, from our SED model with a 160 \mic\ constraint equal to our upper limit, a L$_{TIR}$ which should not exceed 1.4 $\times$ 10$^9$ \lsun.

The values given by the \citet{Dale_Helou_2002} formula for the L$_{TIR}$ seem to be lower for our galaxies than the L$_{TIR}$ derived directly from our SED models while the \citet{Draine_Li_2007} formula leads to closer results. To know if our overestimation of L$_{TIR}$ compared to the \citet{Dale_Helou_2002} formula is a systematic effect observed in other galaxies, we derive the L$_{TIR}$ of the galaxies of \citet{Galliano_Dwek_Chanial_2008} by integrating their SEDs from 3 \mic\ to 1100 \mic\ and compare the results to the L$_{TIR}$ obtained by both \citet{Dale_Helou_2002} and \citet{Draine_Li_2007} relations. 
\citet{Engelbracht2008} also calculate the L$_{TIR}$ of their galaxies from their SED modelling. We calculated the L$_{TIR}$ derived from the \citet{Dale_Helou_2002} and the \citet{Draine_Li_2007} relations for their galaxies (from the fluxes available in the paper) and add these galaxies to the sample. The ratio between these different estimates are plotted in Fig.~\ref{Dale_Draine} as a function of metallicity. We clearly observe a systematic underestimation of the L$_{TIR}$  when using the \citet{Dale_Helou_2002} relation, with a shift of about 20$\%$ and a scatter of $\sim$ 20$\%$. The \citet{Draine_Li_2007} formula seems to better estimate the L$_{TIR}$ of the galaxies to within $\sim$ 20$\%$. For the lowest metallicity galaxies used in this comparison (12+log(O/H)$<$7.5), we note a potentially systematic underestimation but larger samples would be required to investigate this point.

Table~\ref{LTIR} also presents the distribution of the IR luminosity according to wavelength windows: from 3 to 50 \mic, from 50 to 100 \mic\ and from 100 to 1100 \mic. For Mrk~1089 and NGC~1705, the luminosity is roughly evenly distributed over the wavelength windows while for Haro~11, 70$\%$ of the L$_{TIR}$ comes out at wavelengths shorter than 50 \mic. While the longer wavelengths account for the major fraction of the dust mass, not more than 6$\%$ of the L$_{TIR}$ comes out from wavelength superior to 100 \mic.

\begin{figure*}
    \centering
      \includegraphics[width=10cm ,height=10cm]{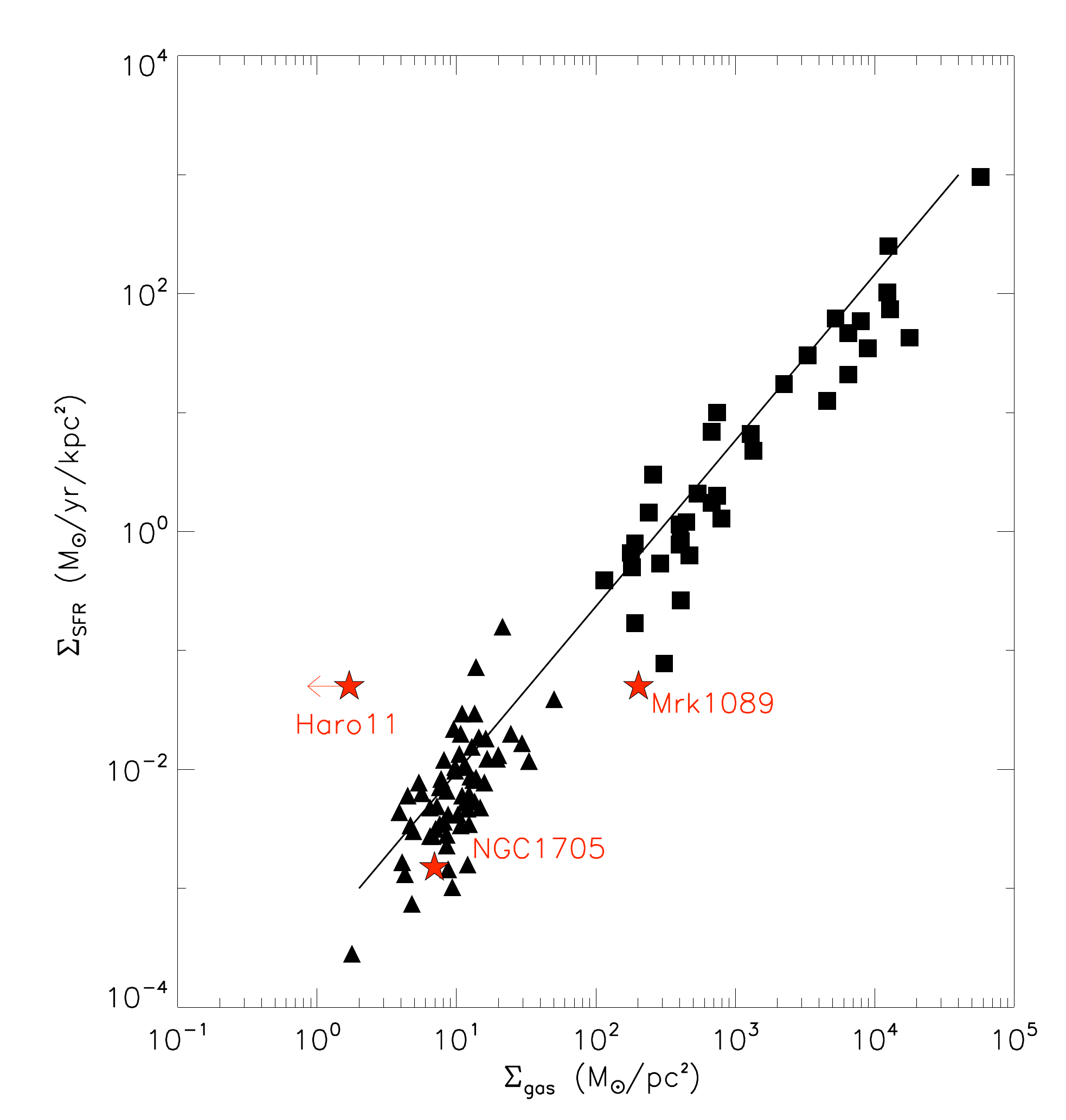} \\
    \caption{ Positions of Haro~11, NGC~1705 and Mrk~1089 in the Schmidt-Kennicutt diagram. The triangles represent normal spirals and squares represent starburst galaxies \citep[see][ for details on the galaxies represented by triangles]{Kennicutt1998}. Red stars indicate our sources. The position of Haro~11 in this diagram accounts for the HI+H$_2$ upper limit of total gas mass ($\sim$ 10$^9$\msun) of the galaxy. }
    \label{Schmidt_Kennicutt}
\end{figure*}

 \subsection{Star Formation Rates}
 
To investigate how these galaxies behave with respect to the Schmidt law, which describes the tight relationship between global SFR and gas density, originally formulated for normal spirals, we derive the star formation rates (SFRs) for our galaxies using the relation of \citet{Kennicutt1998}: 
\begin{equation}
SFR~( M_{\odot} ~yr^{-1}) =  4.5 \times 10^{44} ~L_{FIR}~ (erg~s^{-1})
\end{equation}
 \noindent where L$_{FIR}$ is the IR luminosity obtained integrating our SED models between 8 and 1000 \mic. 
 
 We compare these values with the SFRs obtained by \citet{Zhu2008} for star forming galaxies, formula which is only valid when L$_{TIR}$$<$ 10$^{11}$ \lsun\ and thus can not be used for the LIRG Haro~11:
 
 \begin{equation}
SFR~_{24\mu m}~( M_{\odot} ~yr^{-1}) =  \frac{\nu L_{\nu}[24 \mu m]}{7.79 \times 10^8 L_{\odot}}
\end{equation}
 
We finally use the relations of \citet{Calzetti2007} to derive SFR estimates from the 24 \mic\ luminosity L$_{24}$ and the H$\alpha$ luminosity (L$_{H\alpha}$).

 \begin{equation}
SFR~( M_{\odot} ~yr^{-1}) =  1.24 \times 10^{-38} ~[L_{24} ~(erg~s^{-1})]^{0.88}
\end{equation}

 \begin{equation}
SFR~( M_{\odot} ~yr^{-1}) =  5.3 \times 10^{-42} ~[L_{H\alpha}+0.031~L_{24} ]
\end{equation}

The SFR estimates are presented on Table~\ref{LTIR}. The values given by the \citet{Kennicutt1998} relation are higher than the one obtained with the \citet{Calzetti2007} formula but compare to the ones of \citet{Zhu2008}.

We investigate the behaviour of our galaxies with respect to the Schmidt law~\citep{Kennicutt1998}. We plot the SFR estimated with the relation of \citet{Kennicutt1998} as a function of the gas density for our three galaxies. We overplot a large sample of spirals and starburst galaxies presented in ~\citep{Kennicutt1998}; (Fig.~\ref{Schmidt_Kennicutt}). The two galaxies Mrk~1089 and NGC~1705 seem to follow the Schmidt law. For Haro~11, we take the large upper limit of the HI  + H$_2$ mass ~\citep[$\sim$ 10$^9$ \msun\ - ][]{Bergvall2000} into account in the gas mass estimate. This galaxy does not fall close to the Schmidt law. 
Its location in Fig.~\ref{Schmidt_Kennicutt} might be due to the fact that it is a dwarf galaxy, which can explain why the gas seems to be consumed at a much higher rate than in normal spiral galaxies. Undergoing a merger, Haro~11 should also have a high star-formation rate with a lower gas consumption time than less luminous galaxies ~\citep[see][for studies on normal nearby and submillimiter galaxies respectively]{Kennicutt1994,Tacconi2006}. Nevertherless, \citet{Kennicutt1998} suggests that the SFR in LIRGs do follow the Schmidt Law dependance on gas surface density, which means that the relation linking the SFR and L$_{FIR}$ of starburst systems should not differ significantly from normal disk galaxies. In conclusion, the location of Haro~11 as an outlier confirms that some gas is missing and that the current HI and CO observations do not lead to an amount of gas coherent with what is usually expected. {\revisedbis The fact that Haro~11 falls far from the relation favors the cold dust explanation over the change in the emissivity of the grains for this galaxy.} To follow the Schmidt law, Haro~11 would require 10$^{10}$ \msun\ of gas, which would imply an order of magnitude more than that suggested by CO and HI measurements and which is coherent with the "missing" gas mass required to obtain a D/G expected by the chemical evolution models (see $\S$ 6.3).


\section{Conclusion and summary}

The quantification of the dust mass of a galaxy aids our understanding of its evolution and star formation history. Larger dust masses are sometimes found in low metallicity galaxies when using submm constraints in the SED modelling. In this context, submm observations are clearly necessary to lead a more complete description of the distribution and properties of dust. 

We focused our paper on the dust modelling of four low metallicity galaxies observed with \APEX/\lab. \\

In this paper:

\begin{enumerate}

 \item We present the first images of four dwarf galaxies carried out with the \APEX/\lab\ instrument observing at 870 \mic. 
  
 \item We construct the SEDs with \spitz\ IRAC and MIPS bands as well as the IRS spectra for Haro~11. We apply our SED model and determine  the dust properties of these galaxies.
  
  \item We find that the mass of PAHs accounts for 0.08 to 0.8 $\%$ of the total dust mass of the galaxies, which is a factor of 5 to 50 lower than that of the Galaxy. 
  
  \item To investigate the influence of the submm constraints on the interpretation, we test the effect on the SED model results when submm 870 \mic\ observations are taken into account and compare with SED models not taking into account the 870 \mic\ flux, but with observational constraints at wavelengths only as long as 160 \mic. We find that the use of submm observational constraints always leads to an increase of  the total dust mass derived for our low metallicity galaxies. 
  
  \item We choose to include an additional component to account for the excess submm emission of NGC~1705 and Haro~11. A cold dust component ($\sim$ 10K) with a $\beta$ emissivity index of 1 substantially improves the fit. We find at least 70$\%$ of the total dust mass residing in a cold ($\sim$ 10K) dust component for these two galaxies. We note that describing a cold component of $\beta$=2 does not give very different $\bar{\chi}$$^2$ values, but would give unrealistically larger D/G. Our results however do not rule out the hypothesis of a change in dust emissivity as a function of wavelength proposed in recent studies \citep[e.g.][]{Dupac2003,Meny2007}.
  
\item While Haro~11 has a substantial ($\sim$ 70$\%$) cold dust component, it also harbours a significant fraction of dust mass (30$\%$) in a warmer dust component ($>$ 25K). The SED peaks at unusually short wavelengts (36 \mic), highlighting the importance of the warm dust.

 \item We determine the D/G for Mrk~1089, NGC~1705 and Haro~11 to be 1.9 $\times$ 10$^{-3}$,  4.1 $\times$ 10$^{-3}$ and 0.2, respectively. For Mrk~1089 and NGC~1705, these D/G are consistent with current chemical evolution models. On the contrary, Haro~11 has an excessively high D/G considering the upper limits detected in HI and CO. Haro~11 also falls far from the  Schmidt law, perhaps due to the observed deficit of gas in this galaxy. This could suggest the presence of a large amount of molecular gas. 10 times more molecular gas, compared to that deduced from CO measurements, may be present but not necessarily traced by CO observations.
   
  \item From our SED models, we determine the total infrared luminosity of our galaxy sample to range from 5.8 $\times$ 10$^{7}$ for NGC~1705 to 1.7 $\times$ 10$^{11}$ for the LIRG Haro~11. These values of L$_{TIR}$ are systematically higher than those obtained using the \citet{Dale_Helou_2002} formula but compare better to the \citet{Draine_Li_2007} formula. While $\sim$90$\%$ of the dust mass is residing in the FIR to submm regime, not more than 6$\%$ of the total IR luminosity in Haro~11 emerges from the FIR to submm (100 to 1100 \mic), while most of the luminosity (70$\%$) emerges in the NIR to MIR (3 \mic\ to 50 \mic) window. This is in contrast to Mrk~1089 and NGC~1705 which distribute their luminosities more equally in these two wavelength windows.

 \end{enumerate}

Better observational coverage of the Rayleigh Jeans side of the SEDs should help us to disentangle possible scenarios to explain the excess detected in some galaxies in the submm: dominant diffuse ISM dust, modifications of the dust optical properties at submm wavelengths, very cold dust component etc. The \hersc\  guaranteed time key program of 50 dwarf galaxies (PI: S.Madden), which covers a wide range of metallicity values, will provide a broader coverage of wavelengths between 60 \mic\ to 600 \mic\ and will enable us to better sample the warm and cold dust. The observations will especially enable us to study the slope of the Rayleigh Jeans side of the SED and to learn if an independant very cold dust component does indeed exist. A following paper will discuss the systematic increase in the dust mass estimate with or without submm constraints and the implications for the D/G ratios for a broader sample of galaxies and study how it could possibly be influenced by metallicity. \\


\begin{acknowledgements}
We thank Xander Tielens, Ant Jones and Eli Dwek for very stimulating discussion which improved the content of this paper. We would also like to thank Gerhardt R. Meurer for his high quality HI map of NGC~1705. This publication is based on data acquired with the Atacama Pathfinder Experiment (APEX). APEX is a collaboration between the Max-Planck-Institut fur Radioastronomie, the European Southern Observatory, and the Onsala Space Observatory.  
\end{acknowledgements}


\newpage
\bibliographystyle{apj}
\bibliography{./mybiblio.bib}

\end{document}